\newcommand{\kms}{{km~s$^{-1}$}}

\newcommand{\msol}{\mathcal{M}_\odot}
\newcommand{\lsol}{\mathcal{L}_\odot}

\newcommand{\sdu}{$\msol$ pc$^{-2}$}

\newcommand{\halp}{H$\alpha$}

\newcommand{\oiii}{[O{\scshape iii}]}
\newcommand{\hone}{H{\scshape i}}

\newcommand{\pa}{\phi_0}

\newcommand{\vsys}{V_{\rm sys}}

\newcommand{\vrot}{V_{\rm rot}}

\newcommand{\mhi}{\mathcal{M}_{\rm HI}}

\newcommand{\hr}{h_{\rm R}}

\newcommand{\vmax}{V_{\rm max}}
\newcommand{\rHI}{R_{\rm HI}}

\newcommand{\itf}{i_{\rm TF}}

\newcommand{\arcdeg}{\mbox{$^\circ$}}

\newcommand{\lk}{\mathcal{L}_K}

\documentclass[traditabstract]{aa} % for the abstract without structuration 
\usepackage{graphicx}
\usepackage[varg]{txfonts}
\usepackage{longtable}
\usepackage{lscape}
\usepackage{subfigure}
\usepackage{rotating}
\usepackage[spanish,english]{babel}
\usepackage{natbib}
\bibpunct{(}{)}{;}{a}{}{,}
\usepackage{hyperref}

\defcitealias{bershady2010a}{Paper~I}
\defcitealias{westfall2011b}{Paper~IV}
\defcitealias{martinsson2013a}{Paper~VI}
\defcitealias{martinsson2013b}{Paper~VII}
\defcitealias{westfall2014}{Paper~VIII}
\defcitealias{swaters2014}{Paper~IX}

\begin{document}

% ================================== Title & Authors =====================================
\title{The DiskMass Survey. X. \\Radio synthesis imaging of spiral galaxies}

\author{Thomas P. K. Martinsson\inst{1,2}
   \and Marc A. W. Verheijen\inst{3}
   \and Matthew A. Bershady\inst{4}
   \and \\Kyle B. Westfall\inst{5}
   \and David R. Andersen\inst{6}
   \and Rob A. Swaters\inst{7}}

\institute{
  Instituto de Astrof\'{i}sica de Canarias (IAC), E-38205 La Laguna, Tenerife, Spain\\
  \email{tmartinsson@iac.es}
\and
  Departamento de Astrof\'{i}sica, Universidad de La Laguna, E-38206 La Laguna, Tenerife, Spain
\and
  Kapteyn Astronomical Institute, University of Groningen, PO Box 800, 9700 AV Groningen, 
  The Netherlands\\
  \email{verheyen@astro.rug.nl}
\and
  Department of Astronomy, University of Wisconsin, 475 N. Charter St., Madison, WI 53706,
  USA\\
  \email{mab@astro.wisc.edu}
\and
  Institute of Cosmology and Gravitation, Univ.\ of Portsmouth, Dennis Sciama
  Building, Burnaby Road, Portsmouth PO1 3FX, UK\\
  \email{kyle.westfall@port.ac.uk}
\and
  NRC Herzberg Astronomy and Astrophysics, 5071 West Saanich Road, Victoria, 
  British Columbia, V9E 2E7, Canada \\
  \email{david.andersen@nrc-cnrc.gc.ca}
\and
  Department of Astronomy, University of Maryland, College Park, MD 20742, USA\\
  \email{rob@swaters.net}
}
% ========================================================================================

%\date{\today}
\date{Received 27 July 2015 / Accepted 13 October 2015}

% ===================================== Abstract =========================================
\abstract{
We present results from 21 cm radio synthesis imaging of 28 spiral galaxies from the
DiskMass~Survey obtained with the VLA, WSRT, and GMRT facilities.
We detail the observations and data reduction procedures and present a brief analysis of
the radio data.
We construct 21 cm continuum images, global \hone\ emission-line profiles, column-density
maps, velocity fields, and position-velocity diagrams. From these we determine
star formation rates (SFRs), \hone\ line widths, total \hone\ masses, rotation curves, and
azimuthally-averaged radial \hone\ column-density profiles.
All galaxies have an \hone\ disk that extends beyond the readily observable stellar disk,
with an average ratio and scatter of $\rHI/R_{25}=1.35\pm0.22$, and a majority of the
galaxies appear to have a warped \hone\ disk.
A tight correlation exists between total \hone\ mass and \hone\ diameter, with the largest
disks having a slightly lower average column density.
Galaxies with relatively large \hone\ disks tend to exhibit an enhanced stellar velocity
dispersion at larger radii, suggesting the influence of the gas disk on the stellar
dynamics in the outer regions of disk galaxies.
We find a striking similarity among the radial \hone\ surface density profiles, where the
average, normalized radial profile of the late-type spirals is described surprisingly well
with a Gaussian profile. These results can be used to estimate \hone\ surface density
profiles in galaxies that only have a total \hone\ flux measurement.
We compare our 21 cm radio continuum luminosities with 60 $\mu$m luminosities from $IRAS$
observations for a subsample of 15 galaxies and find that these follow a tight
radio-infrared relation, with a hint of a deviation from this relation at low luminosities.
We also find a strong correlation between the average SFR surface density and the $K$-band
surface brightness of the stellar disk.
}
% ========================================================================================


   \titlerunning{DMS-X. Radio synthesis imaging of spiral galaxies}
   \authorrunning{Thomas P. K. Martinsson et al.}
   \maketitle

% =============================== Introduction ===========================================
\section{Introduction}
\label{sec:HIintro}
The observed distribution and kinematics of atomic hydrogen (\hone) gas in galaxies
provide important information about their baryonic composition and dynamical state, which
is difficult to obtain from optical observations. Generally, \hone\ gas in disk galaxies
has a radial extent greater than the optical disk \citep[e.g.,][]{broeils1997}.
Consequently, since the rotation speed at large radii is the most important constraint on
the total mass of the dark matter halo, \hone\ observations are critical for decomposing
rotation curves into dark and luminous components and establishing the dark matter density
profile.
In addition, \hone\ observations are critical for inferring the disk stellar mass density
from observed stellar velocity dispersion by providing a direct measurement of the atomic
gas component (\citet[][hereafter \citetalias{westfall2011b}]{westfall2011b};
\citet[][hereafter \citetalias{martinsson2013b}]{martinsson2013b}). Furthermore, the
atomic gas is the reservoir for the molecular gas and future star formation. These
considerations make resolved \hone\ observations essential for understanding galaxy
dynamics, formation, and evolution.

Since the mid 1960s \citep[e.g.,][]{burbidge1964} it has been known that the rotation
curves of disk galaxies often display a non-Keplarian decline or even no decline at all
(see the reviews of \cite{kruit1978} and \cite{faber1979}).
The ground-breaking studies of \cite{bosma1978,bosma1981a,bosma1981b} showed that \hone\
rotation curves, in general, remain flat out to the last measured point, many optical disk
scale lengths ($\hr$) from the center. Comparable studies using optical tracers of the
ionized gas showed similar results  \citep[e.g.,][]{rubin1978}.
Extended flat rotation curves were also found by \cite{begeman1987,begeman1989} who, with
better data and an improved fitting algorithm, demonstrated that the eight galaxies in his
sample all showed flat rotation curves.  The rotation curve of one of the galaxies in his
sample (NGC~3198) even remained flat to within 5~\kms\ out to the last measured point at
11 $\hr$.  This generic flatness of extended \hone\ rotation curves
\citep[e.g.,][]{sofue2001} is now commonly interpreted as evidence for the existence of an
extended distribution of dark matter that surrounds the exponential stellar disk.
The flatness of the rotation curve suggests an $R^{-2}$ radial decline in the density of
the dark matter at radii where the dark matter dominates the total gravitational
potential.

One complication in using extended velocity fields for dynamical inference is the presence
of warps in the \hone\ gas distribution. For more than half a century, it has been known
that the \hone\ disk of our own Galaxy is warped in the outer parts
\citep{burke1957,kerr1957}, and it later became clear that many spiral galaxies have warps
\citep[e.g.,][]{sancisi1976,bosma1981b}. We now believe that most  \hone\ disks are
warped. \Citet{kruit2007} found that the \hone~warp starts at around 1.1 times the
truncation radius of the optical disk, and \cite{garcia2002} even claim that, whenever a
galaxy has an extended \hone\ disk with respect to the stellar disk, it has a warp.

The most common approach to derive a rotation curve in the presence of a warp is to model
the observed velocity field with a set of nested tilted rings, allowing
the inclination and position angle of the rings to vary with radius. Usually, the position
angles of the rings can be readily measured from the velocity field.
\citet[][]{begeman1989} demonstrated that below inclinations of $\sim$40$\arcdeg$, a
strong degeneracy exists between the inclination and the rotational velocity of a ring,
even for symmetric velocity fields with random velocity errors of $\sim$5 \kms.
For low-inclination disks, this degeneracy may yield prohibitively large
inclination errors based on the measured \hone\ velocity fields.  However, high-quality
optical IFU kinematic data can yield accurate and precise kinematic inclinations for the
optical, non-warped disk down to $\sim$15\arcdeg\ when using a different approach that
models the entire velocity field as a single, inclined disk \citep{andersen2013}.
Nonetheless, the presence of non-axisymmetric motions can lead to significant errors at
low inclination in all but the most regular velocity fields; solid-body rotation also
precludes accurate inclinations derived from kinematics.
For the nearly face-on galaxies in this paper, we therefore take advantage of the small
scatter in the Tully-Fisher relation \citep{tullyfisher1977,verheyen2001b} to calculate
robust inverse Tully-Fisher inclinations. This is done by comparing the circular velocity
of the gas as predicted from the galaxies absolute $K$-band magnitudes to our measurements
of their projected rotation speeds
\citep[see][hereafter \citetalias{martinsson2013a}]{martinsson2013a}.

The measurement of the azimuthally-averaged radial \hone\ mass surface density profile
($\Sigma_{\rm HI}(R)$) is, on the other hand, much less affected by uncertainties in the
inclination. Some studies have noted the similarities in $\Sigma_{\rm HI}(R)$ among
galaxies, especially within the same morphological type
\citep{rogstad1972,cayatte1994,wang2014}. Others point out the diversity in the radial
behavior of $\Sigma_{\rm HI}(R)$ in galaxies with a wide range of global properties 
\citep[e.g.,][]{verheyen2001a}. \cite{swaters2002} found that the outer part of the 
$\Sigma_{\rm HI}(R)$ profile in many late-type dwarf galaxies can be well fitted with an
exponential decrease, and it can be argued that spiral galaxies in general should display
a similar exponential profile at the outer radii.
Here, we parameterize the typical radial behavior of $\Sigma_{\rm HI}(R)$ and study its
dependence on global photometric and kinematic properties of the galaxies.

Our observations also allow us to construct 21 cm continuum images and derive total 
radio continuum luminosities at 1.4~GHz, which we use to estimate global star formation 
rates (SFRs). We use these data to investigate the correlation between SFR and other
global properties of the galaxies in our sample. We also use literature values from
$IRAS$ far-infrared (FIR) fluxes to investigate the FIR-radio correlation
\citep[e.g.,][]{condon1992,yun2001}.

This paper outlines the data reduction and observational results from 21 cm radio
synthesis observations of 28 spiral galaxies from the DiskMass~Survey
\citep[DMS;][hereafter \citetalias{bershady2010a}]{bershady2010a}.
The main observational goals of the DMS are to obtain rotation curves and
velocity dispersion profiles of the stars and ionized gas in a sample of $\sim$40 nearly
face-on spiral galaxies, taking advantage of the two custom-built integral field units
(IFUs) SparsePak \citep{bershady2004,bershady2005} and PPak \citep{verheyen2004,kelz2006}.
The kinematics of the stars and ionized gas were measured with the PPak IFU
following \cite{westfall2011a} and presented in \citetalias{martinsson2013a}.
Using these data, together with the results from the \hone\ observations presented in this
paper, we have shown that, in general, spiral galaxies are submaximal
(\cite{bershady2011}; \citetalias{martinsson2013b};
\citet[][hereafter \citetalias{swaters2014}]{swaters2014});
at a radius of 2.2 $\hr$, the baryons contribute $\lesssim$50\% to the total potential in
the disk.

The paper is organized in the following way: Sections~\ref{sec:HIsample} and 
\ref{sec:observations} discuss the sample and the observations carried out with the
Westerbork Synthesis Radio Telescope (WSRT)\footnote{The Westerbork Synthesis Radio
Telescope is operated by the ASTRON (Netherlands Foundation for Research in
Astronomy) with support from the Netherlands Foundation for Scientific Research NWO.},
the Very Large Array (VLA)\footnote{The Very Large Array is operated by the The 
National Radio Astronomy Observatory (NRAO). NRAO is a facility of the National Science
Foundation operated under cooperative agreement by Associated Universities, Inc.}, and
the Giant Metrewave Radio Telescope (GMRT)\footnote{The Giant Metrewave Radio Telescope
is run by the National Centre for Radio Astrophysics of the Tata Institute of
Fundamental Research. We thank the staff of the GMRT who have made these observations
possible.}.
In Sect.~\ref{sec:Reduction_HI}, we describe the data reduction procedures.
Observational results such as disk geometry, \hone\ column-density maps, velocity fields,
and rotation curves are described in Sect.~\ref{sec:Products} and presented in an Atlas
(Appendix~\ref{sec:Atlas}).
Section~\ref{sec:HIproperties} presents the \hone\ properties of the galaxies in the
sample, with an investigation of the radial distribution of the \hone\ gas and an
inspection of warps. Section~\ref{sec:SFR} presents results from our measured 
21 cm radio continuum fluxes, from which we estimate global star formation rates.
Finally, Sect.~\ref{sec:Summary_HI} summarizes this work.
% ========================================================================================

% ========================================================================================
\section{The reduced HI sample}
\label{sec:HIsample}
The complete DMS sample selection procedure has been described in detail in 
\citetalias{bershady2010a} and an abridged summary is given in 
\citetalias{martinsson2013a}. All 43 galaxies of the Phase-B sample for which
stellar-kinematic measurements were obtained with the IFUs have been imaged at 21 cm using
WSRT, VLA, and GMRT.
This sample was augmented with UGC~6869 for which stellar-kinematic measurements were
obtained with SparsePak during a pilot study. In this paper, the \hone\ data for 28 of the
44 galaxies are presented; 24 galaxies from the PPak sample presented in
\citetalias{martinsson2013a}, and 4 additional galaxies for which stellar-kinematic data
were obtained with SparsePak. We refer to the galaxy sample studied here as the ``reduced
\hone\ sample''; the galaxies in this sample are listed in
Table~\ref{tab:HIobs}. Properties of these galaxies, such as distances, colors and
coordinates, can be found in \citetalias{bershady2010a} and \citetalias{martinsson2013a}.
The scope of this paper is limited to a description of the data reduction and a concise 
analysis of the reduced \hone\ sample. These radio data products have already been used
for analysis in \citetalias{westfall2011b}, \citetalias{martinsson2013b},
\citet[][hereafter \citetalias{westfall2014}]{westfall2014} and
\citetalias{swaters2014}.

% ------------------------------------------------------------------------------
% Table of observed Galaxies

\begin{table*}
\caption{\label{tab:HIobs}
Observed galaxies in the reduced \hone\ sample.
}
\centering
{\footnotesize 
\renewcommand{\tabcolsep}{1.2mm}
\begin{tabular}{|l l c l r c c c c|}
\hline
\multicolumn{1}{|c}{UGC} &
\multicolumn{1}{c}{Array} &
\multicolumn{1}{c}{Obs.Date} &
\multicolumn{1}{c}{Calibrators} &
\multicolumn{1}{c}{T$_{\rm obs}$} &
\multicolumn{1}{c}{$\nu_{\rm c}$} &
\multicolumn{1}{c}{beam size} &
\multicolumn{1}{c}{$\sigma_{\rm chan}$} &
\multicolumn{1}{c|}{$\sigma_{\rm cont}$}\\
\multicolumn{1}{|c}{} &
\multicolumn{1}{c}{}  &
\multicolumn{1}{c}{}  &
\multicolumn{1}{c}{}  &
\multicolumn{1}{c}{(hrs)}  &
\multicolumn{1}{c}{(MHz)}  &
\multicolumn{1}{c}{(arcsec$^{\rm 2}$)}  &
\multicolumn{1}{c}{(mJy/bm)}  &
\multicolumn{1}{c|}{(mJy/bm)} \\
\multicolumn{1}{|c}{(1)} &
\multicolumn{1}{c}{(2)}  &
\multicolumn{1}{c}{(3)}  &
\multicolumn{1}{c}{(4)}  &
\multicolumn{1}{c}{(5)}  &
\multicolumn{1}{c}{(6)}  &
\multicolumn{1}{c}{(7)}  &
\multicolumn{1}{c}{(8)}  &
\multicolumn{1}{c|}{(9)} \\
%
% UGC                        Array          Obs.Date    Used Calibrators   Tobs     f_c      resolution          rms    rms
%
\hline
 \phantom{11}448           & WSRT        & 2007-09-03 & 3C286; 3C48      & 12.0  & 1397.9  & $29.9\times13.6$  & 0.47 & 0.20 \\
 \phantom{11}463           & VLA         & 2005-09-22 & 3C48; 0119+084   &  2.3  & 1399.7  & $14.7\times12.9$  & 0.54 & 0.34 \\
                           &             & 2009-09-23 & 3C48; 0119+084   &       &         &                   &      &      \\
 \phantom{1}1087           & VLA         & 2005-09-27 & 3C48; 0204+152   &  2.0  & 1401.1  & $16.2\times13.2$  & 0.58 & 0.28 \\
 \phantom{1}1635           & VLA         & 2005-09-24 & 3C48; 0204+152   &  2.1  & 1404.3  & $14.4\times13.5$  & 0.51 & 0.24 \\
                           &             & 2009-09-25 & 3C48; 0204+152   &       &         &                   &      &      \\
 \phantom{1}3140           & VLA         & 2005-09-27 & 3C48; 0459+024   &  2.0  & 1400.5  & $15.9\times13.9$  & 0.59 & 0.76 \\
 \phantom{1}3701           & WSRT$^{*}$ & 2007-12-11 & 3C48; 3C286      & 12.0  & 1409.8  & $15.9\times14.6$  & 0.43 & 0.08 \\
 \phantom{1}3997           & WSRT        & 2007-12-10 & 3C48; 3C286      & 12.0  & 1393.0  & $23.6\times14.6$  & 0.44 & 0.08 \\
 \phantom{1}4036           & WSRT        & 2007-12-19 & 3C48; 3C286      & 12.0  & 1404.2  & $15.8\times14.9$  & 0.40 & 0.09 \\
 \phantom{1}4107           & WSRT        & 2008-01-02 & 3C48; 3C286      & 11.0  & 1404.0  & $20.5\times14.3$  & 0.41 & 0.06 \\
 \phantom{1}4256           & WSRT        & 2008-01-03 & 3C48; 3C286      &  9.8  & 1396.0  & $35.5\times14.9$  & 0.45 & 0.08 \\
 \phantom{1}4368           & GMRT        & 2008-11-15 & 3C48; 3C286;     &  6.8  & 1402.0  & $16.3\times13.7$  & 0.58 & 0.64 \\
                           &             &            & 0834+555         &       &         &                   &      &      \\
 \phantom{1}4380           & WSRT        & 2008-01-14 & 3C48; 3C286      & 10.3  & 1385.8  & $20.1\times14.0$  & 0.44 & 0.07 \\
 \phantom{1}4458$^\dagger$ & GMRT        & 2009-11-17 & 3C48; 3C286;     &  9.7  & 1398.3  & $13.9\times10.5$  & 0.57 & 0.38 \\
                           &             & 2009-11-18 & 0834+555         &       &         &                   &      &      \\
 \phantom{1}4555           & WSRT        & 2008-05-18 & 3C48; 3C286      & 12.0  & 1400.5  & $29.8\times15.5$  & 0.41 & 0.09 \\
 \phantom{1}4622           & WSRT$^{*}$ & 2007-12-31 & 3C48; 3C286      & 12.0  & 1362.1  & $26.4\times16.2$  & 0.41 & 0.07 \\
 \phantom{1}6463           & WSRT        & 2008-05-19 & 3C147; 3C286     & 12.0  & 1408.5  & $31.3\times13.9$  & 0.44 & 0.16 \\
 \phantom{1}6869           & WSRT        & 2008-05-21 & 3C147; 3C286     & 12.0  & 1416.5  & $21.6\times15.7$  & 0.55 & 0.10 \\
 \phantom{1}6903           & GMRT        & 2009-05-30 & 3C286; 1347+122; &  6.8  & 1411.5  & $14.6\times10.5$  & 0.63 & 0.24 \\
                           &             &            & 1130-148         &       &         &                   &      &      \\
 \phantom{1}6918           & GMRT        & 2008-05-29 & 3C286; 1400+621; &  5.6  & 1415.0  & $16.6\times12.6$  & 0.59 & 0.74 \\
                           &             & 2008-05-31 & 1219+484         &       &         &                   &      &      \\
 \phantom{1}7244           & WSRT        & 2008-05-08 & 3C147; CTD93     & 12.0  & 1400.0  & $17.1\times14.4$  & 0.43 & 0.10 \\
 \phantom{1}7416           & WSRT        & 2008-11-28 & 3C147; CTD93     & 12.0  & 1388.6  & $23.4\times14.6$  & 0.44 & 0.15 \\
 \phantom{1}7917           & WSRT        & 2008-11-29 & 3C147; CTD93     & 12.0  & 1388.2  & $25.3\times14.5$  & 0.43 & 0.11 \\
 \phantom{1}8196           & WSRT        & 2008-07-09 & 3C147; CTD93     & 12.0  & 1382.0  & $18.0\times14.7$  & 0.47 & 0.12 \\
 \phantom{1}8230           & WSRT        & 2008-06-29 & 3C147; CTD93     & 12.0  & 1387.2  & $18.8\times14.6$  & 0.42 & 0.06 \\
 \phantom{1}9177           & GMRT        & 2008-05-31 & 3C286; 1347+122; & 10.8  & 1379.6  & $18.2\times12.2$  & 0.42 & 0.38 \\
                           &             & 2008-06-06 & 1445+099; 1609+266; &$^{**}$&     &                   &      &      \\
                           &             & 2008-09-15 & 2130+050         &       &         &                   &      &      \\
 \phantom{1}9837           & WSRT        & 2007-07-21 & 3C286; 3C48      & 12.0  & 1407.9  & $17.5\times14.5$  & 0.43 & 0.09 \\
 \phantom{1}9965           & GMRT        & 2008-05-29 & 3C286; 3C48;     & 10.3  & 1399.3  & $18.8\times17.7$  & 0.43 & 0.25 \\
                           &             & 2008-06-07 & 1609+266         &$^{**}$&        &                   &      &      \\
                           &             & 2008-09-16 &                  &       &         &                   &      &      \\
           11318           & WSRT$^{*}$ & 2007-07-20 & 3C286; 3C48      & 12.0  & 1393.1  & $17.3\times13.8$  & 0.48 & 0.08 \\
\hline
\end{tabular}
}
\tablefoot{
Columns show: (1) UGC number; (2) array used for observation; (3) starting date of the
observation; (4) observed flux and phase calibrators; (5) total integration on source;
(6) central frequency of the bandpass; (7) FWHM of the synthesized beam; (8) average rms
noise in a single channel map; (9) rms noise in the continuum map.
\newline
%Notes:
$(^{*})$ Also observed by the VLA in 2005.\\
$(^{**})$ Many antennas non-operational due to thunderstorm. UGC~9177 and UGC~9965 have
3.3 and 3.5 hours on source, respectively.\\
$(^\dagger)$ For UGC~4458, we use \hone\ data from the WHISP survey. It was observed with
WSRT for 12 hours, with a synthesized beam of $45.6\arcsec \times 15.5\arcsec$, a velocity
resolution of 16.5~\kms\ after Hanning smoothing, and with a noise in the channel maps of
0.43 mJy/beam \citep{noordermeer2005}.}
\end{table*}

% ------------------------------------------------------------------------------

% =========================================================================================

% ========================================================================================
\section{Observational strategy and configurations}
\label{sec:observations}
Collecting \hone\ imaging data for 44 galaxies comprises a substantial observational
program. Therefore, in order to collect these data, the observations were distributed
over the three largest aperture synthesis imaging arrays that operate at 1.4~GHz, and over
multiple observing semesters and cycles.
Our strategy was to use the WSRT only for galaxies with a declination ($\delta$) above
30$\arcdeg$. Because of the east-west configuration of the WSRT antennas, the elliptical
synthesized beam of the WSRT is elongated in the north-south direction on the sky,
approximately proportional to 1/sin($\delta$) such that the synthesized beam is
$\sim$15$\arcsec$ and circular at the North Celestial Pole, while it is
$\sim$15$\arcsec$$\times$30$\arcsec$ at $\delta = 30\arcdeg$.
At lower declinations, the WSRT beam becomes too large compared to the diameters of the
target galaxies. The Y-shape along which the VLA and GMRT antennas are laid out allows for
a more or less circular synthesized beam at lower declinations, and these arrays
were used to mainly target galaxies at $\delta<30\arcdeg$. Of the galaxies for which the
\hone\ data are presented here, 7 were observed with the VLA, 18 with the WSRT, including
3 galaxies previously observed with the VLA, and 6 galaxies were observed with the GMRT.
Details of the observational setups depend on the array that was used and are described
below. A summary of the observational setups is provided in Tables~\ref{tab:HIobs} and
\ref{tab:TelPar}.

\subsection{Observations}
Observations with the WSRT were carried out in its maxi-short configuration between July
2007 and November 2008. A total of 234 hours was allocated, spread over semesters 07B and
08A with some observations delayed to semester 08B. As an east-west array, the WSRT takes
advantage of the Earth's rotation to sample the UV-plane. Each galaxy was observed during
a 12-hour track, preceded and followed by observations of a flux calibrator. The excellent
phase stability of the WSRT at 1.4~GHz does not require observations of phase calibrators
during the 12-hour track. 

Observations with the VLA were carried out in its C-configuration in September and October
2005. A total of 20 hours was allocated, spread over five observing tracks. Each galaxy
was observed during 3--5 scans, with each scan lasting $\sim$30 minutes.
Every 30-minute scan was bracketed by short observations of a nearby phase calibrator with
the same correlator settings. A flux calibrator was observed once during each observing
track, with correlator settings that were relevant for the galaxies observed
during that track.

Observations with the GMRT were carried out between May 2008 and November 2009. A total
of 193 hours was allocated, spread over four observing cycles.
In each cycle, the allocated time was split up over several tracks that each lasted 
10--19 hours. During each track, 1--3 galaxies were observed and most galaxies 
were observed during multiple tracks. Galaxy observations in any given track consist of
several scans, each lasting 40--60 minutes. Similar to the VLA observing strategy, every
scan was bracketed by short observations of a nearby phase calibrator with the same
correlator settings. A flux calibrator was observed once or twice during each observing
track with correlator settings that were relevant for the galaxies observed
during that track.

The flux and phase calibrators used for each galaxy are provided in Table~\ref{tab:HIobs}.

\subsection{Telescope configurations and correlator settings}
For the WSRT, the longest and shortest baselines in its maxi-short configuration, which
provides optimum imaging performance for extended sources within a single 12-hour
observation, are 2700 and 36 meters respectively. This allows for an angular resolution of
$\sim$15$\arcsec$ in the east-west direction, while the largest observable structures are
$\sim$20$\arcmin$ in size.
The longest and shortest baselines of the VLA in its C-configuration are 3400 and
35 meters respectively, which allows for an angular resolution of $\sim$13$\arcsec$.
The largest observable structures are about 20$\arcmin$.
The GMRT consists of 30 dishes in a fixed configuration with 14 dishes located in a 
central region of about 1~km$^2$ and 16 dishes distributed along three arms of the overall
Y-shaped configuration.
Its longest baseline is 26 km and the shortest about 100 meters without
foreshortening, which allows for an angular resolution of $\sim$2$\arcsec$ at 1.4~GHz with
the largest observable structures of $\sim$7$\arcmin$.
The high angular resolution comes at the expense of \hone\ column density sensitivity and
therefore, during the data reduction, the distribution of baselines was tapered such that
the angular resolution of the GMRT observations was  similar to the WSRT and VLA
observations. 
In all observations from the three telescopes, the largest observable structures are much
larger than our target galaxies (see Table~\ref{tab:properties}).

% ------------------------------------------------------------------------------
% Table with telescope parameters

\begin{table}
\caption{\label{tab:TelPar}
Configurations of the interferometric observations.}
\centering
{\footnotesize
\renewcommand{\tabcolsep}{1.05mm}
\begin{tabular}{|l c c c|}
\hline
                                       & VLA              & WSRT              & GMRT             \\
\hline
No.\ of galaxies observed \hfill       &  7$^{*}$         & 18                & 20$^{**}$         \\
Allocated observing time \hfill   (hr) & 20               & 234               & 193              \\
Configuration \hfill                   & C                & maxi-short        & fixed            \\
Freq.\ of observations \hfill (MHz)    & 1362$-$1407      & 1362$-$1417       & 1380$-$1415      \\
Bandwidth \hfill                 (MHz) & 3.125            & 10.0              & 8.0              \\
Number of channels \hfill              & 128              & 1024              & 256              \\
Channel width \hfill      (kHz [\kms]) & 24.4 [5.23]      & 9.77 [2.09]       & 31.3 [6.69]      \\
Integration time \hfill          (sec) & 30               & 60                & 16.9             \\
No.\ of antennas used \hfill           & 23               & $11-13$           & $18-21$          \\
Maximum baseline \hfill           (km) &   3.4            &   2.7             &  26              \\
Minimum baseline \hfill            (m) &  35              &  36               & 100              \\
Primary beam \hfill           (arcmin) & 30               & 36                & 24               \\
Synthesized beam \hfill       (arcsec) & $13.4\times15.3$ & $14.6\times22.7$  & $12.9\times16.4$ \\
rms noise \hfill            (mJy/beam) & 0.71             & 0.55              & 0.64             \\
\hline
\end{tabular}
}
\tablefoot{
The stated frequency of observations gives the range of central frequencies. The channel
width in \kms\ is for a frequency of 1400 MHz. No.\ of antennas used is the number used in
the Fourier transforms.
Maximum and minimum baseline are valid for a complete array without projection effects.
Synthesized beam sizes and rms noise levels are average values, where the rms noise is
measured in individual channel maps. All observations were carried out in
dual-polarization.\\
$(^{*})$ Three of these galaxies were also observed with the WSRT, providing higher-quality
data.\\
$(^{**})$ Data for 6 galaxies observed with the GMRT are presented here.}
\end{table}

% ------------------------------------------------------------------------------

For all galaxies observed with the same telescope, the same correlator settings were used,
but the observing frequencies were tuned to match the recession velocity of each galaxy.
Observations were carried out in dual-polarization mode to increase the signal-to-noise
(S/N) in the unpolarized 21 cm \hone\ emission line by a factor $\sqrt{2}$.
The signal from the correlator was integrated in time intervals of 60, 30, and 16.9
seconds for the WSRT, VLA and GMRT, respectively, after which the visibilities were
recorded.
For the WSRT, an observing bandwidth of 10~MHz, or 2110~\kms at the rest frequency of
1420.405 MHz, was split into 1024 channels of each 9.77~kHz, or 2.06~\kms.
For the VLA, the observing bandwidth of 3.125 MHz (660~\kms) was split into 128 channels
of each 24.4~kHz (5.15~\kms), and for the GMRT, the observing bandwidth of 8~MHz
(1690~\kms) was split into 256 channels of each 31.25~kHz (6.60~\kms).
Doppler tracking to heliocentric velocities was enabled for the WSRT and VLA to compensate
for the drifting observing frequencies, but not for the GMRT.
At these observed frequencies, the Full Width Half Maximum (FWHM) of the primary beams of
the WSRT, VLA and GMRT are 36$\arcmin$, 30$\arcmin$ and 24$\arcmin$, respectively.
Because of the large primary beams, one or more satellites or companion galaxies were
often detected within the field of view (FOV) and frequency range covered by the
observations.

A summary of the various telescope and correlator settings is provided in
Table~\ref{tab:TelPar}.
% ========================================================================================

% ========================================================================================
\section{Data reduction}
\label{sec:Reduction_HI}
The reduction and analysis of standard spectral-line aperture synthesis imaging data, such
as obtained here, by and large take place in two different domains. Flagging, calibrating,
and Fourier transforming the recorded visibilities, including gridding, weighting and
tapering, is done in the UV domain. For this purpose we used the Astronomical Image
Processing System ({\it AIPS}\footnote{\url{www.aips.nrao.edu/}}) software package.
Post-imaging reduction and analysis of the data cubes are performed in the image domain
with the Groningen Image Processing SYstem
({\it GIPSY}\footnote{\url{www.astro.rug.nl/~gipsy/}}) software package 
\citep{hulst1992,vogelaar2001}. Some details of the data reduction procedures are
described below.

\subsection{Flagging, calibrating, and Fourier imaging the visibilities}
The raw visibilities of a particular galaxy and its associated calibrators were extracted
from the recorded data sets of each track, loaded into {\it AIPS}, and concatenated if the
collected data were distributed over multiple files. Obviously bad data from defunct
antennas or data affected by radio frequency interference (RFI) or correlator glitches
were flagged. Telescope-based gain, phase, and bandpass corrections were determined using
the observed fluxes from the calibrators. Calibrated visibilities of the galaxy scans were
closely inspected and additional flags were applied when necessary. Subsequently, the
calibrated visibility data of a galaxy were Fourier transformed to the image domain.
During the map-making process no ``cleaning'' or deconvolution was applied to remove the
sidelobes of the synthesized beam. 
For each made data cube, the resulting size of the synthesized beam is listed in
Table~\ref{tab:HIobs} and indicated in the maps in the Atlas.

Although the calibration procedures in principle are very similar for data from the 
different arrays, they are rather different in practice. Below, we provide some details of
the calibration and imaging procedures separately for the three different arrays.

\subsubsection{The WSRT data}
\label{sec:AIPS-WSRT}
The WSRT observations were carried out with Antenna 5 missing from the standard array of
14 antennas as it was equipped with a prototype receiver for the APERTIF system.
The collected visibility data were processed and calibrated following standard
procedures.
Upon loading and concatenating the raw visibility data into {\it AIPS}, they were weighted
according to the elevation-dependent behavior of the system temperature. Antennas with an
anomalous behavior of their system temperature were flagged; three tracks suffered from
one dysfunctional antenna, and one track from two dysfunctional antennas. Although the
WSRT observations were scheduled as to maximize night-time observing, many galaxies were
still partly observed during the day, and solar RFI clearly affected the shortest baseline
on several runs. Therefore, in every observation, we removed this RFI by blindly flagging
the four shortest baselines (antenna pairs 9A, 9B, AB, and CD) whenever the sun was above
the horizon. Before proceeding with the calibration, the visibility amplitudes were
visually inspected for both the XX and YY polarizations. Additional data affected by RFI
was noted and manually flagged.

A continuum data set was produced by averaging the central 75\% of the channels. The 
expected flux levels of the calibrators 3C48 and 3C147 were calculated for the observed
frequencies based on the known fluxes and spectral indices for these sources. For
calibrator CTD93, which lacks an {\it AIPS} model, the Stokes-I flux was set manually to
4.83~Jy, based on the VLA Calibrator
Manual\footnote{http://www.vla.nrao.edu/astro/calib/manual}. For calibrator 3C286, which
is approximately 10\% linear polarized, the Stokes parameters were set manually to
(I,Q,U,V)=(14.65,0.56,1.26,0.00)~Jy, taken from the ASTRON 
homepage\footnote{http://www.astron.nl/radio-observatory/astronomers/analysis-wsrt-data/analysis-wsrt-dzb-data-classic-aips/analysis-wsrt-d}.
For every track, antenna-based complex gain and phase corrections were determined for both
calibrator scans separately by comparing the observed complex visibilities to the expected
fluxes and phases of the calibrators. The corrections were then linearly interpolated in
time between the two calibrator scans that bracket the 12-hour scan of the galaxy.
For each observation, the interpolated gain and phase corrections were transferred from
the continuum to the line data set. The shapes of the complex bandpasses of the antennas
were determined on the basis of the frequency-dependent complex visibilities for both
calibrators separately, and subsequently interpolated in time. The gain-, phase- and
bandpass-calibrated visibility data of the galaxies were cleaned from remaining RFI by
flagging visibilities with an amplitude above 5$\sigma$ of the root-mean-square (rms)
noise (0.24$\pm$0.02~Jy) in the UV data, after subtracting a continuum baseline to the
line-free channels. We verified that the 5$\sigma$ clip level was high enough not to
affect the \hone\ signal itself.

For every galaxy, an initial data cube was made by Fourier transforming the calibrated
UV data, including the continuum flux from the galaxy and other sources in the field. From
this data cube, the channels that are free of line emission were determined, and these
channels were used to fit the continuum baseline for the purpose of flagging RFI as
mentioned above. The line-free channels were also averaged to produce a single-channel
continuum UV data set for each galaxy. The UV data sets were then Fourier transformed to
produce the final line-free continuum image and the \hone+continuum data cube for every
galaxy, as well as maps and cubes of the corresponding antenna patterns.

% ------------------------------------------------------------------------------
% Figure of robustness parameter

\begin{figure}
\centering
\includegraphics[width=0.5\textwidth]{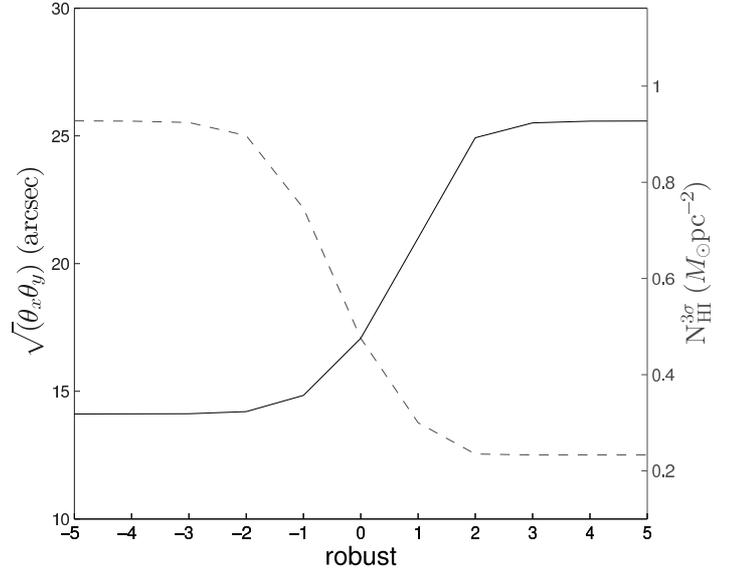}
\caption{Illustration of the trade-off between angular resolution and column-density
sensitivity for different robust weightings of the WSRT visibility data, ranging from
uniform to natural weighting. The solid curve indicates the size of the synthesized beam,
and the dashed curve shows the 3$\sigma$ column-density sensitivity in a single channel.
}
\label{fig:robust}
\end{figure}
% ------------------------------------------------------------------------------

The WSRT visibilities were given a ``Robust=0'' weighting. Figure~\ref{fig:robust}
illustrates the trade-off between angular resolution and column density sensitivity for
the WSRT data of UGC~4107. As a function of the `Robust' parameter, it shows the effective
angular resolution ($\sqrt{\Theta_x\Theta_y}$) as a solid line, and the 3$\sigma$ \hone\
column density sensitivity per channel as a dashed line. A value of Robust=$-$5 
corresponds to a uniform weighting of the UV data, yielding the smallest beam 
($\sim$14$\arcsec$) and the worst column density sensitivity limit 
($\sim$0.93~$\msol$pc$^{-2}$). Robust=+5 corresponds to a natural weighting of the
UV data, yielding the best column density sensitivity ($\sim$0.23~$\msol$pc$^{-2}$), but
the largest beam ($\sim$26$\arcsec$) with significant sidelobes. A Robust=0 weighting
provides the best compromise for our purpose, yielding a synthesized beam of
14.3$\arcsec$$\times$20.5$\arcsec$ for the case of UGC~4107 at $\delta$$=$+49.5$\arcdeg$.
All data cubes and continuum maps constructed from the WSRT data have a pixel size of 
5$\arcsec$ in right ascension and 5$\arcsec$/sin($\delta$) in declination to properly
sample the beam. The channel maps and continuum images have a dimension of 512$\times$512
pixels and the antenna patterns a dimension of 1024$\times$1024 pixels.

\subsubsection{The VLA data}
\label{sec:AIPS-VLA}
The VLA observations were performed with 4 of the 27 antennas missing from the array as
they were being refurbished for the EVLA expansion. The reduction of the VLA data was
performed in a way similar to what was described in the previous section, except that
antenna-based complex gain corrections and bandpass solutions were derived for
each scan of both the flux and the phase calibrators and interpolated in time over the
galaxy scans to account for temporal variations.

The calibrated and partially flagged visibilities of an observed galaxy were Fourier
transformed into data cubes containing the 21 cm continuum and \hone\ line emission, as
well as cubes with the frequency-dependent antenna patterns. As for the WSRT data,
the visibilities were given a Robust=0 weighting. All data cubes constructed from the
VLA data have a pixel size of 4.5\arcsec$\times$4.5\arcsec\ and the same dimensions as the
WSRT data cubes.

\subsubsection{The GMRT data}
\label{sec:AIPS-GMRT}
The reduction and calibration of the UV data from the GMRT follows the same overall
strategy as applied to the WSRT and VLA data, but there are a few notable differences in
practice.

First of all, identifying and flagging RFI is much more tedious and time consuming for
GMRT data compared to the VLA and WSRT data. This is mainly due to higher levels of RFI at
the GMRT site and the larger data volumes generated by the GMRT; the typical data volume
per galaxy from the GMRT is $\sim$3 times that from the WSRT and $\sim$14 times that from
the VLA. Several remote antennas and long baselines were flagged blindly upfront,
motivated by the fact that the highest angular resolution provided by the longest GMRT
baselines is not required to obtain a synthesized beam similar to that provided by the VLA
and WSRT. Therefore, we flagged a) nine of the outermost antennas that provide the longest
baselines; b) baselines longer than 25 k$\lambda$ between the remaining inner antennas of
the array; and c) the nine shortest baselines ($<$ 300 meters) whenever the sun was above
the horizon in order to remove solar RFI. All these antennas and baselines were excluded
from the calibration and imaging process.
Subsequently, for each track the preliminary-calibrated remaining GMRT visibilities were
prepared for visual inspection by arranging them for each polarization in
three-dimensional data cubes with baseline number, time and frequency channel as their
axes. The visually identified presence of RFI was recorded and flagged manually.

Second, due to steep phase gradients across the 8 MHz bandpass, a continuum data set
for determining the gain and phase corrections was made by averaging only $\sim$10
RFI-free channels near the center of the bandpass to avoid effective de-correlation of the
signal that would have occurred when the default 75\% of the channels would have been
averaged.

Third, because the shape of the bandpass of the GMRT is quite stable in time, a single
average bandpass was determined for each galaxy by making use of all the scans of the flux
and phase calibrators observed for that galaxy.

After some trails and consultation with the staff at the GMRT in Khodad and the NCRA in
Pune, the calibrated UV data were Fourier transformed to the image domain with a 
``Robust=0'' weighting, a Gaussian UV taper with its width at 30\% set to 16~k$\lambda$ in
both U and V, and excluding baselines longer than 25~k$\lambda$ (5.3~km). This yielded 
angular resolutions that are similar to what was obtained with the VLA and the WSRT as
listed in Table~\ref{tab:HIobs}. All data cubes constructed from the GMRT data have a
pixel size of 4\arcsec$\times$4\arcsec\ and the same dimensions as the WSRT and VLA data
cubes.

\subsection{Post-imaging deconvolution and signal definition}
\label{sec:GIPSY}
The data cubes produced with {\it AIPS} for each galaxy, containing the 21 cm sky signals
and corresponding frequency-dependent beam patterns or ``dirty'' beams (the
radio equivalent of a point spread function), were further processed with {\it GIPSY}.
Before deriving the various data products, the continuum and line emission must be
separated and, after defining the regions that contain the continuum and \hone\ signal in
each channel map, the images need to be deconvolved, or ``{\tt CLEAN}ed'', to remove the
sidelobes of the synthesized beams. The post-imaging processing of the data cubes is
basically identical for all three arrays and described in detail in the following
subsections.

\subsubsection{Velocity smoothing}
All observations were carried out with a uniform frequency taper which means that the
spectral response to an infinitely narrow emission line is a sync function with a FWHM of
1.2 times the width of a frequency channel. Furthermore, strong continuum sources may
produce a Gibbs ripple that can affect a large part of the bandpass. To suppress the 
sidelobes of the sync function and the Gibbs phenomenon, the data cubes were first 
convolved along the frequency axis with a Hanning smoothing kernel which extends over 3 
channels with relative weights of (1/4,1/2,1/4). As a consequence, the spectral response
function becomes triangular in shape with a FWHM of twice the channel width. At the rest
frequency of the \hone\ line, this corresponds to a velocity resolution of 4.12, 10.3, and
13.4 \kms\ for the WSRT, VLA, and GMRT data, respectively.

To increase the S/N and to obtain a similar velocity resolution as the VLA and GMRT data,
the data cubes from the WSRT were smoothed further in velocity to a near-Gaussian response
function with a FWHM of 4 channels, or 8.3~\kms. Subsequently, every other channel in the
WSRT data cubes were discarded such that the FWHM of the spectral response function was
sampled by 2 channels, reducing the number of channels from 1024 to 512 while each
retained channel was doubled in width to 19.5~kHz (4.12~\kms), and preserved its
observed flux density. All operations were performed on both the data cubes and the cubes
containing the beam patterns.

\subsubsection{Continuum subtraction}
From each spectrum in the data cubes, the continuum emission was subtracted using an
iterative rejection scheme. A second-order polynomial was fit as a baseline to each
spectrum separately and subtracted. Subsequently, the rms noise was calculated in each
continuum-subtracted channel map, and pixels above and below 2$\sigma$ were masked. This 
mask was transferred to the input data cube and baselines were refitted, ignoring the
masked pixels. After subtracting the refitted baselines, the rms was recalculated and the
pixel mask was adjusted. This process was repeated until the pixel mask no longer changed
significantly. This iterative rejection method maximizes the number of line-free channels
in the fitting, while still ensuring that most of the \hone\ signal was rejected before
fitting the final continuum baseline. For the VLA and GMRT data, a continuum map was 
created by calculating the values of the fitted baselines at the center of the observed 
bandpass. For the WSRT data, the continuum UV-data sets were already prepared 
in the UV domain by averaging line-free channels and Fourier transforming that data set to
the image domain (Sect.~\ref{sec:AIPS-WSRT}).

For every galaxy, this iterative procedure resulted in a continuum-free data cube that
only contains the signal from the \hone-emission line, as well as a line-free image that
only contains the continuum flux of the galaxy and other radio sources in the field.
In these cubes and images, the line and continuum signal is still convolved with the
corresponding beam patterns.

\subsubsection{Signal definition and {\tt CLEAN}ing}
\label{sec:signal}
The signal in the continuum images and \hone\ data cubes was deconvolved with the 
{\tt CLEAN} algorithm as developed by \cite{hogbom1974} and implemented in {\it GIPSY}.
To avoid mistaking noise peaks for signal and to speed up the search for 
{\tt CLEAN} components, we defined masks or search areas that contain the continuum and 
\hone\ signals. For the \hone\ data cubes, the shapes of these masks vary from channel to
channel as different parts of the rotating \hone\ disks are seen at different frequencies
or recession velocities. 

For the continuum images, the masks were made in an iterative way.
First, the brightest continuum sources were identified visually and search areas enclosing
these sources were created manually. The overall rms noise in the continuum maps was
calculated and the maps were {\tt CLEAN}ed down to 1$\sigma$. This removed the sidelobes
of the brightest sources, reduced the rms noise in the maps, and revealed fainter
continuum sources for which enclosing search areas were added to the pre-existing ones.
The lower rms noise was recalculated and the continuum image was {\tt CLEAN}ed again down
to 1$\sigma$ with the updated map containing the search areas. This was repeated until all
sidelobes were removed, the noise no longer decreased, and no more fainter sources were
revealed. The {\tt CLEAN} components found within the final set of search areas were
restored into the map with a Gaussian beam of the same FWHM and position angle as the 
dirty beam pattern.

Constructing masks or search areas for the \hone\ channel maps is more elaborate as the
spatially extended \hone\ signal occurs in many channels and at different locations as a 
function of frequency. Also, extended emission at lower column densities may disappear
below the noise level, which makes it difficult to identify this emission and include it
in the search areas. The procedure we adopted was as follows. First, all channel maps in a
data cube were {\tt CLEAN}ed blindly down to four times the rms noise in a channel map and
the detected {\tt CLEAN} components were restored with a Gaussian beam similar in size and
orientation as the dirty beam pattern. This removed most of the sidelobes from the
brightest \hone\ sources, including possible companion galaxies near the velocity extremes
of the data cubes. Second, the channel maps were spatially smoothed to a beam that is
twice as large as the Gaussian beam with which the {\tt CLEAN} components were restored 
for the WSRT and VLA data, and to a $30\arcsec\times30\arcsec$ beam for the GMRT data. 
This enhances the S/N of \hone\ emission at lower column densities. Third, we calculated
the rms noise in the spatially smoothed channel maps and selected only those pixels with 
values above the 2$\sigma$ level. This resulted in frequency-dependent masks that contain
the \hone\ signal from the galaxies, as well as some $>$2$\sigma$ noise peaks. As the
final fourth step, we visually inspected the continuation of the search areas in all three
dimensions of the data cubes and manually removed all noise peaks. Naturally, there is 
some risk that the faintest dwarf satellites were misjudged to be noise peaks. Also, the
edges of the search areas are affected by noise at the 2$\sigma$ level and emission from
the lowest column density gas, below 2$\sigma$ in the smoothed channel maps, is still
excluded from the search areas. 

The channel-dependent search areas based on the smoothed data cubes were used to 
{\tt CLEAN} the high-resolution data cubes down to 1$\sigma$ of the rms noise level. The 
{\tt CLEAN} components that were found within the search areas were restored with a
Gaussian beam of similar FWHM as the antenna pattern. In the {\tt CLEAN}ed channel maps 
the sidelobes of the  beam pattern were fully removed.

It is noteworthy that for 15 of the 28 observed target galaxies we also detected
\hone\ emission from one or more satellites or companion galaxies, but these are not
considered in any further analysis. In total, 34 companion galaxies were detected.
% ========================================================================================

\section{Data products}
\label{sec:Products}
In this section, we discuss the primary data products derived from the continuum maps and
the \hone\ data cubes, including flux densities, global \hone\ lines and their widths and
integrated fluxes, \hone\ maps and radial \hone\ column density profiles, \hone\ velocity
fields, position-velocity (PV) diagrams, and rotation curves. For every galaxy, the
results are collected and presented in the appended Atlas.

The \hone\ data of UGC~4458 were taken from the WHISP survey \citep{hulst2001} and
have already been presented by \cite{noordermeer2005}. These data were of higher quality
than our GMRT observations. For consistency within the set of derived data products, we
have taken the data cube that was obtained with the WSRT and re-processed it in the same
manner as the other galaxies we observed.

% ------------------------------------------------------------------------------
% Comparisons of with literature values

\begin{figure*}[t]
\centering
\includegraphics[width=1.0\textwidth]{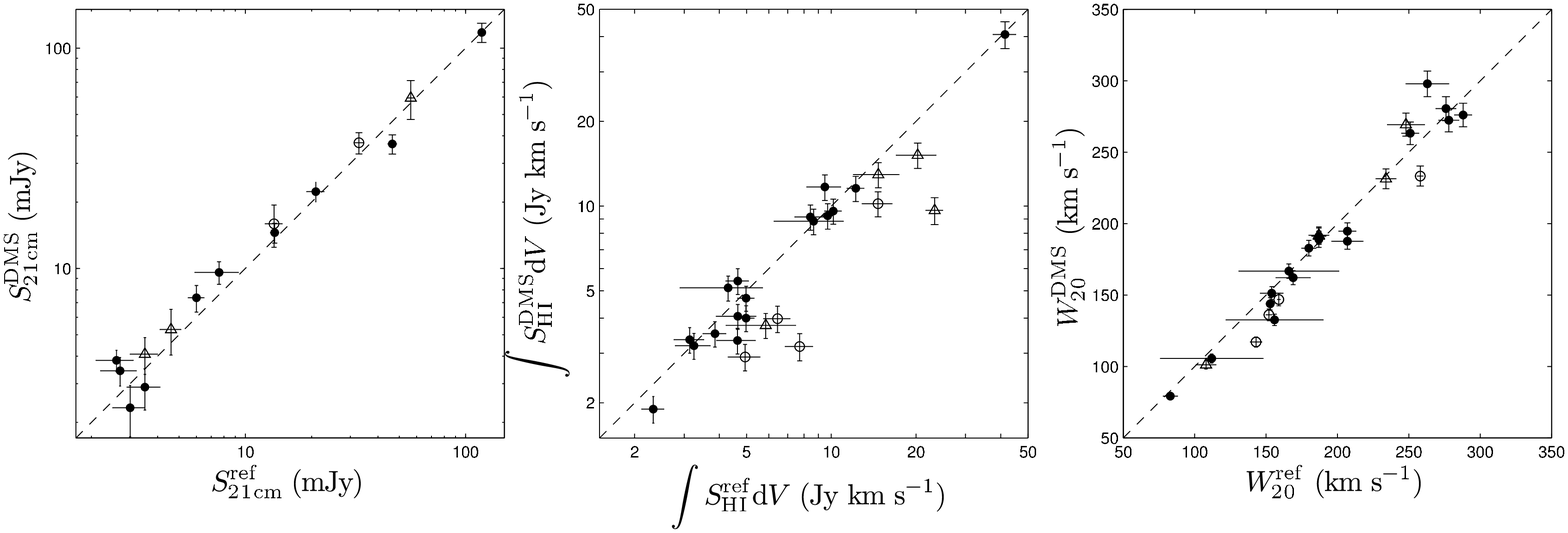}
\caption{Comparison of our measurements with data from the literature. 
{\bf Left:} Total 21 cm radio continuum flux densities.
{\bf Middle:} Integrated \hone\ line fluxes.
{\bf Right:} \hone\ line widths at the 20\% level of peak intensity.
Solid symbols represent data from the WSRT, open circles from the VLA, and open triangles
from the GMRT. The data are listed in Table~\ref{tab:Fluxes} with references to the
literature values.
}
\label{fig:fluxcomp}
\end{figure*}
% ------------------------------------------------------------------------------

\subsection{The 21 cm continuum maps and flux densities}
\label{sec:S21}
The 21 cm continuum flux densities ($S_{\rm 21 cm}$) were measured from the
{\tt CLEAN}ed continuum maps as presented in the Atlas. Only the flux within the search
area of the galaxy used by the {\tt CLEAN} algorithm was considered. If the galaxy was
not clearly detected in radio continuum, $S_{\rm 21 cm}$ was calculated in an area that
corresponds to the optical disk corrected for the effect of beam smearing.
We estimated the formal errors on $S_{\rm 21 cm}$ from the rms noise in the maps.
All galaxies with a measured $S_{\rm 21 cm}$ at least three times higher than the
error were considered to have a significant detection. 
However, four galaxies (UGC~3997, UGC~4380, UGC~4458, UGC~7244) had measured
$S_{\rm 21 cm}$ lower than three times the error. We have excluded these four galaxies
in the following analysis that uses $S_{\rm 21 cm}$ measurements.
Measured fluxes were corrected for the attenuation of the primary beams, but this is only 
a minute correction given that the galaxies are relatively small and located near the 
centers of the FOVs.

The radio continuum maps obtained with the GMRT show strong imaging artefacts from 
imperfections in the calibration procedure.
A proper self-calibration procedure can correct for these effects, but has not
been carried out as it is not required for the \hone\ data in which we are primarily
interested\footnote{However, in a future publication, we will include an
additional 14 DMS galaxies observed with the GMRT, and we intend to then redo and extend
the GMRT calibrations to also include self-calibration.}.
Because of these artefacts, galaxies observed with the GMRT might have underestimated
errors on their $S_{\rm 21 cm}$ measurements.
We estimate an additional 10\% systematic error on $S_{\rm 21 cm}$ to account for
uncertainties in the flux calibration. The continuum flux densities are presented
in Table~\ref{tab:Fluxes}.

The left panel of Fig.~\ref{fig:fluxcomp} shows a comparison of our measured 
$S_{\rm 21 cm}$ for 15 galaxies with literature values reported by the NRAO VLA Sky Survey
\citep[NVSS;][]{condon1998}. We find a reasonable agreement, indicating that our flux
calibrations were successful and that we are not suffering from missing short spacings in
the UV plane.

% ------------------------------------------------------------------------------
% Table with S21, W20, W50, R, etc

\begin{table*}
%\begin{sidewaystable}
\caption{\label{tab:Fluxes}
Obtained measurements and literature values.}
\centering
{\tiny
\renewcommand{\tabcolsep}{1.2mm}
\begin{tabular}{|crrrcrr|rccrlc|}
\hline
\multicolumn{1}{|c}{}  &
\multicolumn{6}{c}{}   &
\multicolumn{6}{|c|}{} \\
\multicolumn{1}{|c}{}                   &
\multicolumn{6}{c}{This study}          &
\multicolumn{6}{|c|}{Literature values} \\
\multicolumn{1}{|c}{}  &
\multicolumn{6}{c}{}   &
\multicolumn{6}{|c|}{} \\
\hline
\multicolumn{1}{|c}{UGC}                 & 
\multicolumn{1}{c}{$S_{\rm 21cm}$}       & 
\multicolumn{1}{c}{$W_{\rm 20}$}         & 
\multicolumn{1}{c}{$W_{\rm 50}$}         & 
\multicolumn{1}{c}{R}                    & 
\multicolumn{1}{c}{$\int S_V$d$V$}       & 
\multicolumn{1}{c}{$\vsys^{\rm prof}$}      & 
\multicolumn{1}{|c}{$S_{\rm 21cm}$}      & 
\multicolumn{1}{c}{$W_{\rm 20}$}         & 
\multicolumn{1}{c}{$W_{\rm 50}$}         & 
\multicolumn{1}{c}{$\int S_V$d$V$}       & 
\multicolumn{1}{c}{$\vsys^{\rm lit}$}       & 
\multicolumn{1}{c|}{ref.}                \\
\multicolumn{1}{|c}{}                    &
\multicolumn{1}{c}{(mJy)}                &
\multicolumn{1}{c}{(\kms)}               &
\multicolumn{1}{c}{(\kms)}               &
\multicolumn{1}{c}{(\kms)}               &
\multicolumn{1}{c}{(Jy \kms)}            &
\multicolumn{1}{c}{(\kms)}               &
\multicolumn{1}{|c}{(mJy)}               &
\multicolumn{1}{c}{(\kms)}               &
\multicolumn{1}{c}{(\kms)}               &
\multicolumn{1}{c}{(Jy \kms)}            &
\multicolumn{1}{c}{(\kms)}               &
\multicolumn{1}{c|}{}                   \\
\multicolumn{1}{|c}{(1)}  &
\multicolumn{1}{c}{(2)}   &
\multicolumn{1}{c}{(3)}   &
\multicolumn{1}{c}{(4)}   &
\multicolumn{1}{c}{(5)}   &
\multicolumn{1}{c}{(6)}   &
\multicolumn{1}{c}{(7)}   &
\multicolumn{1}{|c}{(8)}  &
\multicolumn{1}{c}{(9)}   &
\multicolumn{1}{c}{(10)}  &
\multicolumn{1}{c}{(11)}  &
\multicolumn{1}{c}{(12)}  &
\multicolumn{1}{c|}{(13)} \\
%
% UGC        S_1.4GHz              W20                 W50     Res         /Sdv                  Vsys           S_21(NVSS)           W20           W50              /Sdv               Vsys      ref
%
\hline
  448 & \phantom{00}3.4 $\pm$ \phantom{0}0.7 & 193.2 $\pm$ 0.7           & 179.1 $\pm$ 1.6           & \phantom{0}8.3 &  5.42 $\pm$ 0.57 &  4855.3 $\pm$ 0.3 &                 & 187 $\pm$ \phantom{0}5           & 176 $\pm$ \phantom{0}5           &  4.65 $\pm$ 0.43 &  4857 $\pm$  5 & a \\
  463 & \phantom{0}37.2 $\pm$ \phantom{0}4.1 & 236.8 $\pm$ 0.8           & 220.5 $\pm$ 2.8           & 10.3           &  3.17 $\pm$ 0.35 &  4457.1 $\pm$ 0.2 &  32.8 $\pm$ 1.7 & 258 $\pm$ \phantom{0}2           & 221 $\pm$ \phantom{0}2           &  7.72 $\pm$ 0.86 &  4462          & b \\
 1087 & \phantom{00}6.5 $\pm$ \phantom{0}1.2 & 150.3 $\pm$ 0.8           & 136.1 $\pm$ 1.7           & 10.3           &  3.98 $\pm$ 0.43 &  4485.5 $\pm$ 0.6 &                 & 159 $\pm$ \phantom{0}1           & 136 $\pm$ \phantom{0}1           &  6.44 $\pm$ 0.69 &  4484          & b \\
 1635 & \phantom{00}2.4 $\pm$ \phantom{0}0.7 & 120.6 $\pm$ 1.1           & 109.6 $\pm$ 2.0           & 10.3           &  2.91 $\pm$ 0.31 &  3517.6 $\pm$ 0.2 &                 & 143 $\pm$ \phantom{0}4           & 127 $\pm$ \phantom{0}4           &  4.94 $\pm$ 0.64 &  3516          & b \\
 3140 & \phantom{0}15.9 $\pm$ \phantom{0}3.5 & 139.6 $\pm$ 0.5           & 120.5 $\pm$ 1.0           & 10.3           & 10.20 $\pm$ 1.04 &  4623.3 $\pm$ 0.9 &  13.5 $\pm$ 1.2 & 152 $\pm$ \phantom{0}2           & 119 $\pm$ \phantom{0}2           & 14.66 $\pm$ 1.79 &  4620          & b \\
 3701 & \phantom{00}1.2 $\pm$ \phantom{0}0.3 & 153.4 $\pm$ 0.7           & 131.9 $\pm$ 1.3           & \phantom{0}8.3 &  9.15 $\pm$ 0.94 &  2917.6 $\pm$ 0.6 &                 & 154 $\pm$ \phantom{0}8           & 135 $\pm$ \phantom{0}8           &  8.43 $\pm$ 1.01 &  2915          & b \\
 3997 & {\it \phantom{00}0.7 $\pm$ \phantom{0}0.3} & 168.8 $\pm$ 1.6           & 154.7 $\pm$ 1.4           & \phantom{0}8.3 &  3.33 $\pm$ 0.35 &  5941.6 $\pm$ 0.8 &                 & 166 $\pm$ 35                     & \phantom{0}95 $\pm$ 23           &  4.64 $\pm$ 0.73 &  5915 $\pm$ 12 & c \\
 4036 & \phantom{00}9.6 $\pm$ \phantom{0}1.1 & 134.7 $\pm$ 0.5           & 119.9 $\pm$ 0.6           & \phantom{0}8.3 &  8.83 $\pm$ 0.92 &  3468.3 $\pm$ 0.3 &   7.6 $\pm$ 1.7 & 156 $\pm$ 34                     & 128 $\pm$ 25                     &  8.65 $\pm$ 2.39 &  3470 $\pm$ 17 & a \\
 4107 & \phantom{00}3.8 $\pm$ \phantom{0}0.4 & 164.2 $\pm$ 0.9           & 152.4 $\pm$ 1.2           & \phantom{0}8.3 &  3.19 $\pm$ 0.34 &  3509.1 $\pm$ 0.2 &   2.6 $\pm$ 0.5 & 169 $\pm$ 12                     & 146 $\pm$ 12                     &  3.25 $\pm$ 0.46 &  3512          & b \\
 4256 & \phantom{0}36.7 $\pm$ \phantom{0}3.7 & 196.8 $\pm$ 0.7           & 171.3 $\pm$ 1.4           & \phantom{0}8.3 & 11.57 $\pm$ 1.19 &  5248.1 $\pm$ 0.3 &  46.5 $\pm$ 2.1 & 207 $\pm$ \phantom{0}6           & 188 $\pm$ \phantom{0}6           & 12.22 $\pm$ 0.84 &  5259 $\pm$  3 & a \\
 4368 & \phantom{0}13.2 $\pm$ \phantom{0}3.0 & 274.7 $\pm$ 1.4           & 243.5 $\pm$ 1.4           & 13.4           & 12.94 $\pm$ 1.33 &  3864.4 $\pm$ 3.2 &                 & 248 $\pm$ 13                     & 243 $\pm$ \phantom{0}6           & 14.69 $\pm$ 2.71 &  3870 $\pm$  6 & a \\
 4380 & {\it \phantom{00}0.9 $\pm$ \phantom{0}0.3} & 145.9 $\pm$ 1.0           & 120.7 $\pm$ 2.0           & \phantom{0}8.3 &  3.35 $\pm$ 0.35 &  7480.7 $\pm$ 2.1 &                 & 153 $\pm$ \phantom{0}3           & 116 $\pm$ \phantom{0}3           &  3.14 $\pm$ 0.38 &  7483          & b \\
 4458 & {\it \phantom{0}10.1 $\pm$ \phantom{0}3.6} & 284.0 $\pm$ 0.8           & 240.9 $\pm$ 1.2           & 13.4           & 11.69 $\pm$ 1.22 &  4758.5 $\pm$ 2.2 &   6.2 $\pm$ 0.5 & 288 $\pm$ \phantom{0}6           & 257 $\pm$ \phantom{0}5           &  9.49 $\pm$ 1.31 &  4749 $\pm$  5 & a \\
 4555 & \phantom{00}3.4 $\pm$ \phantom{0}0.5 & 265.3 $\pm$ 0.5           & 243.4 $\pm$ 1.3           & \phantom{0}8.3 &  4.06 $\pm$ 0.41 &  4239.0 $\pm$ 1.2 &   2.7 $\pm$ 0.5 & 251 $\pm$ \phantom{0}6           & 247 $\pm$ \phantom{0}7           &  4.65 $\pm$ 0.75 &  4244 $\pm$  6 & a \\
 4622 & \phantom{00}1.0 $\pm$ \phantom{0}0.3 & 221.1 $\pm$ 0.8           & 194.6 $\pm$ 1.8           & \phantom{0}8.3 &  3.38 $\pm$ 0.35 & 12826.3 $\pm$ 1.5 &                 &                                  &                                  &                  &                &   \\
 6463 & \phantom{00}2.3 $\pm$ \phantom{0}0.6 & 191.0 $\pm$ 0.5           & 175.3 $\pm$ 0.7           & \phantom{0}8.3 &  9.60 $\pm$ 0.98 &  2503.1 $\pm$ 0.3 &   3.0 $\pm$ 0.5 & 187 $\pm$ \phantom{0}4           & 178 $\pm$ \phantom{0}4           & 10.16 $\pm$ 0.70 &  2507 $\pm$  5 & a \\
 6869 & 117.9           $\pm$ 11.8           & 282.6 $\pm$ 0.9           & 254.4 $\pm$ 1.4           & \phantom{0}8.3 & 40.71 $\pm$ 4.45 &   798.8 $\pm$ 0.3 & 118.4 $\pm$ 4.2 & 276 $\pm$ \phantom{0}7           & 260 $\pm$ \phantom{0}5           & 41.41 $\pm$ 3.81 &\phantom{0}798 $\pm$  5 & a \\
 6903 & \phantom{00}5.3 $\pm$ \phantom{0}1.2 & 197.4 $\pm$ 1.0           & 178.0 $\pm$ 1.8           & 13.4           &  9.65 $\pm$ 1.06 &  1888.8 $\pm$ 1.4 &   4.6 $\pm$ 0.5 & 187 $\pm$ \phantom{0}7           & 180 $\pm$ 12                     & 23.29 $\pm$ 1.61  &  1892 $\pm$  5 & a \\
 6918 & \phantom{0}59.4 $\pm$ 11.9           & 236.9 $\pm$ 1.6           & 204.6 $\pm$ 4.3           & 13.4           & 15.17 $\pm$ 1.56 &  1103.7 $\pm$ 2.9 &  56.4 $\pm$ 2.3 & 234 $\pm$ \phantom{0}7           & 214 $\pm$ \phantom{0}5           & 20.28 $\pm$ 3.27 &  1108 $\pm$  5 & a \\
 7244 & {\it \phantom{00}1.1 $\pm$ \phantom{0}0.4} & 107.6 $\pm$ 0.7           & \phantom{0}86.7 $\pm$ 1.0 & \phantom{0}8.3 &  5.12 $\pm$ 0.52 &  4356.7 $\pm$ 0.7 &                 & 112 $\pm$ 36                     & \phantom{0}72 $\pm$ 24           &  4.3  $\pm$ 1.4\phantom{0}&  4350 $\pm$ 12 & d \\
 7416 & \phantom{00}7.4 $\pm$ \phantom{0}1.0 & 189.7 $\pm$ 1.6           & 166.5 $\pm$ 1.2           & \phantom{0}8.3 &  3.52 $\pm$ 0.36 &  6900.5 $\pm$ 0.4 &   6.0 $\pm$ 0.5 & 207 $\pm$ 11                     & 170 $\pm$ 11                     &  3.86 $\pm$ 0.36  &  6901 $\pm$ 10 & a \\
 7917 & \phantom{00}2.9 $\pm$ \phantom{0}0.6 & 274.4 $\pm$ 0.8           & 259.4 $\pm$ 1.8           & \phantom{0}8.3 &  4.00 $\pm$ 0.42 &  6988.5 $\pm$ 1.2 &   3.5 $\pm$ 0.6 & 278 $\pm$ \phantom{0}7           &                                  &  4.98 $\pm$ 0.34  &  6985 $\pm$  5 & a \\
 8196 & \phantom{0}14.5 $\pm$ \phantom{0}1.5 & 163.0 $\pm$ 1.3           & 126.9 $\pm$ 3.3           & \phantom{0}8.3 &  5.92 $\pm$ 0.61 &  8329.6 $\pm$ 1.6 &  13.6 $\pm$ 0.6 &                                  &                                  &                   &                &   \\
 8230 & \phantom{00}2.1 $\pm$ \phantom{0}0.3 & 299.9 $\pm$ 2.0           & 286.9 $\pm$ 5.4           & \phantom{0}8.3 &  1.90 $\pm$ 0.21 &  7163.3 $\pm$ 1.2 &                 & 263 $\pm$ 15                     &                                  &  2.33 $\pm$ 0.21  &  7152 $\pm$ 10 & a \\
 9177 & \phantom{00}6.3 $\pm$ \phantom{0}1.7 & 304.6 $\pm$ 1.5           & 279.8 $\pm$ 3.9           & 13.4           &  2.48 $\pm$ 0.29 &  8861.3 $\pm$ 1.8 &                 &                                  &                                  &                   &  8860 $\pm$ 10 & e \\
 9837 & \phantom{00}3.3 $\pm$ \phantom{0}0.6 & 184.9 $\pm$ 0.5           & 166.8 $\pm$ 0.9           & \phantom{0}8.3 &  9.24 $\pm$ 0.96 &  2656.5 $\pm$ 0.0 &                 & 180 $\pm$ \phantom{0}5           & 161 $\pm$ \phantom{0}5           &  9.71 $\pm$ 0.45 &  2657 $\pm$  4 & a \\
 9965 & \phantom{00}4.1 $\pm$ \phantom{0}0.8 & 106.6 $\pm$ 0.7           & \phantom{0}83.8 $\pm$ 1.4 & 13.4           &  3.77 $\pm$ 0.38 &  4525.3 $\pm$ 0.3 &   3.5 $\pm$ 0.5 & 108 $\pm$ \phantom{0}7           & \phantom{0}87 $\pm$ \phantom{0}7 &  5.85 $\pm$ 1.62 &  4528 $\pm$  8 & a \\
11318 & \phantom{0}22.3 $\pm$ \phantom{0}2.3 & \phantom{0}81.3 $\pm$ 0.5 & \phantom{0}59.2 $\pm$ 0.7 & \phantom{0}8.3 &  4.71 $\pm$ 0.48 &  5884.4 $\pm$ 0.1 &  20.9 $\pm$ 1.9 & \phantom{0}83 $\pm$ \phantom{0}5 & \phantom{0}57 $\pm$ \phantom{0}8 &  4.98 $\pm$ 0.34 &  5886 $\pm$  5 & a \\
\hline
\end{tabular}
}
\tablefoot{
(1) UGC number; (2) Radio continuum flux denisty. Galaxies with $< 3 \sigma$ detection
(Sect.~\ref{sec:S21}) have measured values in italic; (3) Width of the global \hone\
profile at 20\% of the peak flux density; (4) Width of the global \hone\ profile at 50\%
of the peak flux density; (5) Velocity resolution of the data cube; (6) Integrated \hone\
flux; (7) Systemic velocity derived from the global \hone\ profile. Columns (8)$-$(12)
represent similar measurements collected from the literature.
Column (13) provides references to the literature: a) \cite{bottinelli1990}; b)
\cite{springob2005}; c) \cite{schneider1992}; d) \cite{theureau1998}; e) \cite{paturel2003}.}
%
%\end{sidewaystable}
\end{table*}
%
%
% a: Bottinelli et al 1990, A&A Suppl., 82, 391
% b: Springob et al 2005, ApJ Suppl., 160, 149
% c: Schneider et al 1992, ApJ Suppl., 81, 5
% d: Theureau et al 1998, A&A Suppl., 130, 333
% e: Paturel et al 2003, A&A, 412, 57
%
% From HyperLEDA-II (Paturel et al 2003, A&A, 412, 57)
%
%                                         RA      DEC  Log(2Vsini)      m21                Vhel
%
% PGC0051237  UGC9177                J142030.5+102555 2.404+/-0.028 0  16.39+/- 0.17 0    8860.+/- 10. 0
%
% m21 = -2.5log(0.2366F)+15.84 with F in Jy*km/s

% ------------------------------------------------------------------------------

\subsection{The global HI profile}
\label{sec:globHIprof}
The global \hone\ profile was constructed by calculating the flux density in each channel
map within the search area that defines the signal from the galaxy. The error on the flux
density measurement was calculated empirically, and we took advantage of the fact that
the signal of the galaxy only occupies a small fraction of the data cube.
We extracted the smallest data cube that encloses the search areas defining the
\hone\ signal. This small ``signal'' cube was replicated 27 times in an orthogonal
3$\times$3$\times$3 configuration within the full data cube, and centered on the target
galaxy. Consequently, 26 of the small cubes were centered on line-free areas of the main
data cube, modulo the presence of a possible companion galaxy. Subsequently, for every
channel in the ``signal'' cube the flux density was measured within the replicated search
area in each of the 26 surrounding ``noise'' cubes. The error on the flux-density
measurement was then calculated as the rms scatter in the 26 flux density measurements from
the ``noise'' cubes. The measured flux densities and their errors were corrected for the
minute effect of the primary-beam attenuation.
The global \hone\ profiles are presented in the accompanying Atlas.

\subsubsection{Integrated line flux}
The integrated \hone\ line flux ($\int S_{\rm HI} {\rm d}V$) was calculated by summing the
primary-beam-corrected flux densities in all channel maps and multiplying it with the
channel width (d$V$) defined to be the velocity width of the central frequency channel.
The integrated \hone\ fluxes are listed in Table~\ref{tab:Fluxes}. The middle panel of 
Fig.~\ref{fig:fluxcomp} shows a comparison between our integrated \hone\ fluxes and 
literature values taken mainly from \cite{bottinelli1990} and \cite{springob2005}, who
collected measurements that were mainly obtained with single-dish telescopes. The 
correlation between our measurements and the literature values is less strong than for the
continuum fluxes; the integrated fluxes reported in the literature tend to be higher than
our measurements obtained with the VLA and GMRT arrays (open symbols in 
Fig.~\ref{fig:fluxcomp}) while there is good correspondence with the WSRT measurements 
(solid symbols).

To investigate whether this offset is caused by a systematic error in the flux calibration
of the VLA and GMRT observations we recall the left panel of Fig.~\ref{fig:fluxcomp},
demonstrating no systematic offset in the continuum fluxes of the galaxies. To confirm
this, we have compared the continuum flux densities of other continuum point sources in 
the field with those reported by the NVSS. 
Again, no systematic offset was found. Furthermore, as pointed out earlier, the shortest
baselines in our interferometric observations allow us to detect structures that are 
larger than the dimensions of the targeted galaxies. Therefore, it is unlikely that some
of the \hone\ flux is ``resolved out'' by the interferometers. 

Three of our target galaxies have been observed by both the VLA and WSRT arrays, and for 
two of those galaxies (UGC~3701 and UGC~11318) the VLA provides a useful integrated \hone\
flux measurement. These galaxies were also observed with the Green Bank Telescope
(GBT) 91m telescope, with single-dish measurements reported by \cite{springob2005}. There
is reasonable correspondence among the integrated fluxes from the VLA, WSRT and
GBT measurements: 8.0, 9.2 and 8.4 Jy~\kms\ for UGC~3701, and 4.4, 4.7 and 3.1
Jy~\kms\ for UGC~11318 for the VLA, WSRT and GBT respectively.

We note that the literature values with which our four VLA measurements are compared in 
Fig.~\ref{fig:fluxcomp} all come from \cite{springob2005}, who observed these galaxies
with the Arecibo telescope. No other galaxies observed by us have literature values from 
this combination of reference and telescope so there is an exclusive VLA-Arecibo
correspondence. We also note that the fluxes measured with Arecibo were corrected by
\cite{springob2005} for pointing offsets and beam-filling effects, following a model that
describes the presumed radial extent of the \hone\ gas in the galaxies. We suspect that 
these applied beam corrections are systematically too large for the Arecibo observations.
Finally, we note that the fluxes as measured by the single-dish observations may be 
contaminated by contributions from companion galaxies at similar recession velocities as
the target galaxies (see Sect.~\ref{sec:signal}). Indeed, we detect at least one
more \hone\ source in the nearby field at similar recession velocity as the target
galaxies for three (UGC~463, UGC~1087, UGC~1635) of the four galaxies observed with the
VLA and Arecibo. For UGC~1635, we also detect two more companions, a bit further away from
UGC~1635, and just outside the frequency range of that galaxy's \hone\ detection. However,
these companion galaxies are at a distance of $\sim$10$\arcmin$ from UGC~463 and UGC~1087,
and $\sim$5$\arcmin$ from UGC~1635, so should not have contributed enough flux to explain
the discrepancy in the flux measurements from the Arecibo telescope, which has a FWHM beam
of $\sim$3$\arcmin$.

Given the facts that: a) there is no offset in the continuum fluxes; b) the
interferometers can detect the largest structures in our target galaxies; c) there is a 
reasonable correspondence between the VLA, WSRT and GBT fluxes of 2 galaxies
observed by both arrays, and d) significant and uncertain beam corrections were
applied to the Arecibo measurements, we conclude that there is no reason to question the
integrated \hone\ fluxes that we have measured.

\subsubsection{Line width and rotation speed}
The width of the \hone\ line is traditionally defined to be the full width (in \kms) 
between the outer edges of the profile where the flux density is 20\% of the maximum 
observed flux density of the emission line. Variations on this 
definition include considering the 25\% or 50\% of the peak flux density or of the mean 
flux density. In the nominal case of a double-horned profile, the peak flux densities of 
the two horns may be considered separately as well. Here, we limit ourselves to measuring
$W_{20}$ and $W_{50}$, the full widths of the profile at 20\% or 50\% of the absolute peak
flux density observed over the entire profile.

We proceed by calculating the velocities at the 20\% level ($V_{20}^{\rm low}$ \& 
$V_{20}^{\rm high}$) and the 50\% level ($V_{50}^{\rm low}$ \& $V_{50}^{\rm high}$) by 
linear interpolation between the two channels that are just above and below these 
thresholds. The widths follow from $W_{20}=V_{20}^{\rm high}-V_{20}^{\rm low}$ and 
$W_{50}=V_{50}^{\rm high}-V_{50}^{\rm low}$. The systemic velocities based on the global
profiles ($\vsys^{\rm prof}$) are calculated as $\left(V^{\rm high}+V^{\rm low}\right)/2$
for both the 20\% and 50\% levels and then averaged. The error on $\vsys^{\rm prof}$ is
taken as half the difference between $V_{\rm sys,20}$ and $V_{\rm sys,50}$. The measured
line widths and their formal errors, as well as the systemic velocities and their errors,
are listed in Table~\ref{tab:Fluxes}. The global \hone\ profiles are presented on the
Atlas pages, with $W_{20}$ indicated by a horizontal dotted line.

The shape of the global \hone\ profile results from the convolution between the radial 
distribution of the \hone\ gas, the rotation curve of the galaxy, and the orientation of
the gas disk in terms of its inclination and the possible presence of a warp. Relatively 
shallow edges of an \hone\ profile (e.g., UGC~4458 and UGC~8196) hint at the presence of a 
declining rotation curve or a warp towards an edge-on orientation. The absence of a clear
double-horned signature may indicate that the flat part of the rotation curve is not
traced by the \hone\ gas over an extended radial range (e.g., UGC~463) or may be due to a
nearly face-on orientation of the disk (UGC~9965 and UGC~11318). In an ideal situation
(flat rotation curve, extended \hone\ disk, no warp), the width of the \hone\ line is a
measure of the maximum rotation speed of the galaxy after application of several 
corrections to the observed width. Typically, these corrections account for the finite
velocity resolution of the instrument, the turbulent motions of the \hone\ gas, and 
the inclination of the gas disk.

We correct $W_{20}$ for instrumental broadening using the expression motivated and
provided by \cite{verheyen2001b},

\begin{equation}
W_{20}^{\rm R} = W_{20} - 35.8 \left[ \sqrt{1+\left( \frac{\rm R}{23.5} \right)^2}
 -1\right],
\label{eq:W20inst}
\end{equation}

\noindent
where $W_{20}^{\rm R}$ is the line width corrected for instrumental broadening, $W_{20}$
is the observed width, and R is the velocity resolution in \kms\ as listed in 
Table~\ref{tab:Fluxes}. Given the range of velocity resolutions of our observations,
R=8.3--13.4 \kms, this correction amounts to a modest\footnote{The {\it WHISP} data of
UGC~4458 have R=16.5 \kms, with a correction of $\sim$8~\kms} 2--5 \kms.
The right panel of Fig.~\ref{fig:fluxcomp} shows a comparison of our corrected line widths
($W_{20}^{\rm DMS}=W_{20}^{\rm R}$) with values from the literature, most often obtained
from single-dish observations. The correspondence is quite good, keeping in mind that the
single-dish observations are often of low S/N while it is not always clear from the
literature if and how the correction for instrumental broadening was applied. It should
also be noted that global \hone\ profiles measured with single-dish telescopes may be
broadened by \hone\ emission from satellite galaxies at similar recession velocities.

\begin{figure}
\centering
\includegraphics[width=0.5\textwidth]{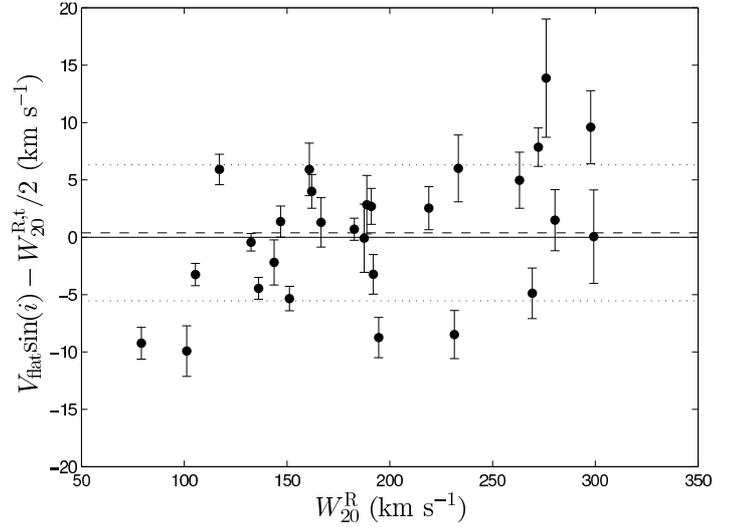}
\caption{Difference between the weighted-average, projected rotational velocities of the
\halp, \oiii\ and \hone\ PV diagrams and half the corrected widths of the \hone\ global
profiles. The dashed line shows the average difference while the rms scatter is indicated
by the two dotted lines.
}
\label{fig:W20vsVmax}
\end{figure}

Subsequently, we correct our profile widths for the turbulent motion of the gas with the 
expression given by \cite{tully1985}

\begin{eqnarray}
\left(W_{20}^{\rm R,t}\right)^2 & = & 
\left(W_{20}^{\rm R}\right)^2 + \left(W_{\rm t,20}\right)^2 
\left[1-2\rm e^{-\left(\frac{W_{20}^{\rm R}}{W_{\rm c,20}}\right)^2}\right] \nonumber \\ 
& & {} - 2 W_{20}^{\rm R} W_{\rm t,20} 
\left[1-\rm e^{-\left(\frac{W_{20}^{\rm R}}{W_{\rm c,20}}\right)^2} \right] ,
\label{eq:W20rand}
\end{eqnarray}

\noindent
where the value of the transition parameter is set to $W_{\rm c,20}=120$ \kms. The
turbulence parameter is set to $W_{\rm t,20}=2 k_{20} \sigma_{\rm ran}$. For a purely
Gaussian profile $k_{20}=1.80$, but we follow \cite{bottinelli1983} and adopt
$k_{20}=1.89$, accounting for a somewhat narrower core and broader wings than a Gaussian
profile.
The value of the velocity dispersion is determined by comparing the half-width of the
corrected \hone\ line ($W_{20}^{\rm R,t}$/2) with the amplitude of the projected flat part
of the rotation curve ($V_{\rm flat}$sin($i$)) as the average of the \oiii, \halp, and
\hone\ PV-diagrams (\citetalias{martinsson2013a}).
Figure~\ref{fig:W20vsVmax} shows the best agreement is found for
$\sigma_{\rm ran}$$=$$10$ \kms, which results in an average difference of +0.4 \kms and a
scatter of 5.9 \kms, excluding UGC~4458 which has a declining rotation curve with a flat
part much lower than the maximum velocity. The value of $\sigma_{\rm ran}$$=$$10$ \kms\ is
higher than observed by \cite{obrien2010}, who found that most of their gas-rich dwarf
systems display velocity dispersions of 6.5--7.5~\kms, and it is at the high end of the
characteristic value of 8--10~\kms\ found by \cite{tamburro2009} for galaxies more similar
to those in our sample. However, compared to \cite{caldu-primo2013} who found a median
velocity dispersion of $11.9\pm3.1$~\kms, our value is in the lower end.
The correction of the $W_{20}^{\rm R}$ line widths for the turbulent motion of the gas in
our galaxies amounts to 18--38~\kms.

% ------------------------------------------------------------------------------
% Figure of W20 vs Vmax

% ------------------------------------------------------------------------------

Finally, we correct the width of the \hone\ line for the inclination ($\itf$; see 
Sect.~\ref{sec:geometry}) of the galaxy:

\begin{equation}
W_{20}^{\rm cor} = W_{20}^{\rm R,t} / \sin(\itf) .
\label{eq:W20incl}
\end{equation}

\noindent The derived $W_{20}^{\rm cor}/2$ is indicated as a dashed line in the Atlas
rotation-curve panel and can there be compared to the observed rotational velocities.

\subsection{The HI column-density map}
A total \hone\ column-density map was made by adding together all channel maps containing
\hone\ emission. Before doing so, the pixels in every channel map that are located outside
the search area were set to zero. In this way, we avoid adding noise to areas of the total
\hone\ map from channel maps that do not contain \hone\ emission from the galaxy at that
particular location on the sky. This procedure allows us to obtain a higher S/N, but has
the disadvantage that the noise level varies across the total \hone\ map because
a different number of channels were added at each position.
The pixel-dependent noise in the \hone\ map was determined empirically and analogous to 
the procedure with which the noise in the global \hone\ profile was determined 
(Sect.~\ref{sec:globHIprof}).

The pixel values of the \hone\ map were converted to column densities according to

\begin{equation}
 N_{\rm HI}= 1.823\times 10^{18} 
 \int T_{\rm b}\;{\rm d}V {\rm \;\;\;\;\;[atoms\;cm}^{-2}{\rm ]} ,
\label{eq:NHI}
\end{equation}

\noindent
where d$V$ is the velocity range in \kms\ over which the emission line was integrated and
$T_{\rm b}$ is the brightness temperature in Kelvin (K). The conversion from mJy/beam to K
for an elliptical Gaussian beam is calculated according to 

\begin{equation}
 T_{\rm b} = \frac{605.7}{\Theta_x\;\Theta_y}\;S\;\left(\frac{\nu_0}{\nu}\right)^2 
 \;\;\;\;\; {\rm [K]} ,
\label{eq:Tb}
\end{equation}

\noindent
where $S$ is the flux density in mJy/beam, $\Theta_x$ and $\Theta_y$ are the major and
minor FHWM of the Gaussian beam in arcseconds, $\nu_0$ is the rest frequency of the \hone\
line (1420.40575177 MHz), and $\nu$ is the frequency at which the redshifted \hone\ line is
observed. The conversion from [atoms cm$^{-2}$] to [$\msol$pc$^{-2}$] is 

\begin{equation}
1\;\;[\msol\;{\rm pc}^{-2}] = 1.249\times10^{20} \;\;
[{\rm atoms}\;{\rm cm}^{-2}] .
\label{eq:NHIconv}
\end{equation}

\noindent
The \hone\ column-density maps can be found in the Atlas with all maps having the same
contour levels in terms of $\msol$ pc$^{-2}$.

\subsection{The observed HI velocity field}
The observed velocity field of the rotating \hone\ disk was created by fitting a single
Gaussian to the \hone\ emission line in the spectrum at every position on the sky.
The fitting routine makes use of initial estimates for the amplitude, centroid, and 
dispersion of the Gaussian function. For each spectrum, the initial estimate for the 
amplitude is simply the peak flux, while the initial estimates for the centroid and 
dispersion were calculated as the first and second moment of the spectrum in which pixels
outside the search areas were set to zero. The Gaussian fit was accepted if the following
four conditions were met: (1) the amplitude of the fitted Gaussian exceeds 2.5 times the 
rms noise in the spectrum; (2) the centroid of the fitted Gaussian lies within 300~\kms\ 
of the systemic velocity (from Table~2 in \citetalias{bershady2010a}); (3) the formal 
error in the centroid is smaller than 5 \kms; and (4) the velocity dispersion lies in the 
range 5--100~\kms. The lower limit in the last condition is motivated by the velocity
resolution of the observations, and we expect the velocity dispersion of the \hone\ gas to
be well below the upper limit \citep[e.g.,][]{tamburro2009}.

The observed velocity fields are presented in the Atlas. For presentation purposes only,
the values of pixels within the velocity field for which the Gaussian fit failed were
calculated by interpolating the values from neighboring pixels, weighted by the 
Gaussian beam. For subsequent analysis of the kinematics, only pixels with accepted 
Gaussian fits were considered.

\subsubsection{Modeling the HI kinematics}
\label{sec:geometry}
We model the axisymmetric behavior of each observed velocity field with a set of nested,
concentric tilted rings following \cite{begeman1989}. Along the major axis, each tilted 
ring is $10\arcsec$ wide, or $\sim$2/3 of the size of the synthesized beam, and the radii 
of the centers of the rings are $R_{j}=5\arcsec + j\times10\arcsec$, where $j$=0,1,...,20.
The tilted-ring fitting procedure is carried out in four steps using the {\tt ROTCUR} 
program within {\it GIPSY}. In these four steps, for each galaxy, we determine: 
(1) the systemic recession velocity ($\vsys$);
(2) the position angle of the receding kinematic major axis of the gas disk ($\phi$);
(3) the inclination of the disk ($i$); and
(4) the rotational velocities ($\vrot$).
We fit $\vsys$ and $\phi$ using uniform weighting, while $i$ and $\vrot$ are fit with a
$\cos(\phi)$ weighting. We always use all the data points.
The four steps are here described in detail.

{\bf Step 1: Determine $\vsys$.}
We keep the location of the dynamical center of all rings fixed at the morphological
center of the galaxy. For the 24 galaxies observed with PPak, we refer to
\citetalias{martinsson2013a} for a discussion on how the morphological centers were
determined. For the four galaxies observed with SparsePak, we adopt the centers as
reported by the Sloan Digital Sky Survey (SDSS)\footnote{\url{http://www.sdss.org/}}.
The adopted dynamical centers of the galaxies are provided in the Atlas Tables
(Appendix~\ref{sec:Atlas}).
The inclination of all rings was fixed to be the value derived from the inverse
Tully-Fisher relation ($i$$=$$\itf$; see \citetalias{martinsson2013a}). For the four
galaxies in the reduced \hone\ sample that were not discussed in
\citetalias{martinsson2013a}, we have calculated their $\itf$ according to the absolute
$K$-band magnitudes ($M_K$) as reported in \citetalias{bershady2010a}. The inferred
$\itf$ can be found for each galaxy in the Atlas Table.

After imposing the above restrictions, we fit $\vsys$, $\phi$ and $\vrot$ for each ring
such as to reproduce the cosine behavior of the recession velocities as a function of
azimuthal angle in the plane of the galaxy. For each galaxy, the radial trend of $\vsys$
is presented in the upper geometry panel on the Atlas page. The $\vsys$ measurement of
each galaxy, listed in Table~\ref{tab:Geometry}, is calculated as the weighted average of
all rings and indicated by a horizontal solid line in the Atlas.

% ------------------------------------------------------------------------------
% Figure of Delta-PA & Delta-Vsys

\begin{figure}[t]
\centering
\includegraphics[width=0.5\textwidth]{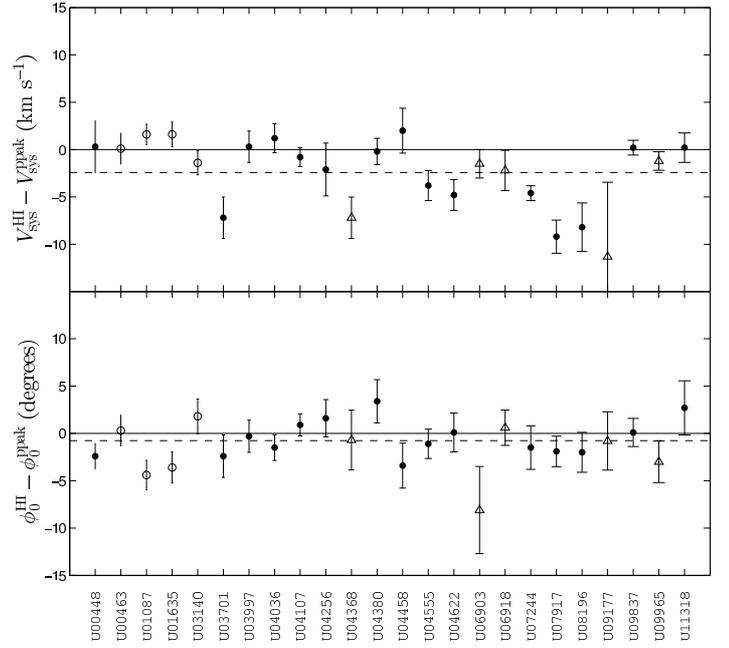}
\caption{Comparison of results from \hone\ observations with those from optical PPak
observations. The figure shows differences between the systemic velocities (upper panel)
and between position angles of the kinematic major axis in the inner regions of the
galaxies (lower panel), measured from the \hone\ velocity fields and those measured with
the PPak IFU from \oiii\ and stellar velocity fields \citepalias{martinsson2013a}.
The dashed horizontal lines show the average differences. Solid points correspond to WSRT
data, open circles to VLA data, and open triangles to GMRT data.
}
\label{fig:PAdiff}
\end{figure}
% ------------------------------------------------------------------------------

In the upper panel of Fig.~\ref{fig:PAdiff} we compare $\vsys$ derived using our \hone\
velocity fields ($\vsys^{\rm HI}$) with those derived from the stellar and \oiii\ velocity
fields ($\vsys^{\rm PPak}$) obtained with the PPak IFU \citepalias[see][]{martinsson2013a}.
Overall, there is good agreement between the PPak and \hone\ measurements, with a minor
systematic offset of $\Delta \vsys = -2.5$~\kms\ or 1/4$^{\rm th}$ of the typical
velocity resolution of the \hone\ observations.

{\bf Step 2: Determine $\phi$.}
We now fix $\vsys$ to the value determined in Step~1 for all rings and refit $\phi$ and
$\vrot$ for each ring. In the observed velocity fields of several galaxies, a twisting of
the isovelocity contours in the outer disks often reveals the presence of a warp, and this
is generally confirmed by the radial behavior of $\phi$ of the fitted tilted rings.
The radial trend of $\phi$ is presented for every galaxy in the middle geometry panel on
the Atlas page.
For many galaxies, a $\phi$~warp starting outside the FOV of the IFUs ($R>35\arcsec$) is
clearly detected, e.g., for UGC~4036, UGC~6869, and UGC~9965. The representative inner
kinematic position angle ($\pa$) of the gas disk is calculated as the weighted-average
position angle of the four inner rings, covering $R<40\arcsec$, and only including rings
containing more than 10 non-blank pixels. Subsequently, we fit a straight line to the
measured $\phi$ values at $R>40\arcsec$, forcing $\phi(R)=\pa$ at $R=40\arcsec$ for
continuity. If the slope $k$ of this fitted line is significantly non-zero ($|k|$ more
than three times larger than the uncertainty in $k$) we then allow for the presence of a
$\phi$~warp in subsequent fitting steps. Otherwise we adopt $\pa$ to be valid for all
outer rings as well. Whenever appropriate, we also adjust the radius at which the
$\phi$~warp starts, or adjust by eye the radial behavior of the $\phi$~warp to better
represent the data.

For 14 galaxies in our sample, the shapes of the outer isovelocity contours are best
described by introducing a $\phi$~warp. The finally adopted radial behavior of the
kinematic position angle of the gas disk is indicated by the solid line in the middle
geometry panels shown in the Atlas. The position angles representative for the inner gas
disks ($\pa$) are listed in Table~\ref{tab:Geometry}.

In the lower panel of Fig.~\ref{fig:PAdiff}, we compare the representative position angles
derived for the inner \hone\ velocity fields ($\pa^{\rm HI}$) with those derived for
stellar and \oiii\ velocity fields ($\pa^{\rm PPak}$) obtained with the PPak IFU
\citepalias{martinsson2013a}.
The average difference is $-0\fdg76 \pm 0\fdg43$, with an rms scatter around this mean of
$2\fdg0$. UGC~6903 was excluded from the calculation as it has poor
PPak data. The systematic offset with a significance of 1.8$\sigma$ may indicate a small
north-south misalignment of the PPak IFU module.

{\bf Step 3: Attempt to fit $i$.}
While keeping the dynamical center, $\vsys$ and $\phi$ (with its possible radial
dependence) fixed, we fit $i$ and $\vrot$ for each ring. Solutions for these two
parameters are highly degenerate for nearly face-on galaxies \citep{begeman1989}, such
that our results are primarily used as a consistency check against the expected $\itf$
values. For each galaxy, the bottom geometry panel in the Atlas shows the best-fit
inclination and its formal error for every ring. Indeed, except for a few cases, the
formal uncertainties in $i$ are very large. Notable exceptions are UGC~4368
($\itf$=45\degr), UGC~6869 ($\itf$=55\degr), UGC~6918 ($\itf$=38\degr),
and UGC~9837 ($\itf$=31\degr). The first two galaxies are the two most inclined
galaxies in the sample, UGC~6918 is in the top-five of the most inclined disks, and the
\hone\ velocity field of UGC~9837 is exceptionally regular and axisymmetric. In all four
cases, the average inclinations found from the tilted-ring fits are consistent with the
TF-based inclination. Therefore, instead of the results from the tilted-ring fits, we
adopt $i=\itf$ for the inner parts of the \hone\ disks.

An inclination ($i$) warp with its straight line of nodes is much more difficult to detect
than a warp in $\phi$. We have seen, however, that $\phi$~warps occur frequently. Given
the random orientation of the galaxies in space, $i$~warps should be equally common among
our target galaxies.
Therefore, to characterize these $i$~warps in our galaxies, we first consider the shape of
the deprojected rotation curve under the assumption that $i=\itf$ for the entire disk. Any
slope in the rotation curve beyond $R=35\arcsec$ is removed by adopting a linear $i$~warp
that forces the rotation curve to be roughly flat.
For ten galaxies in our sample, we judged that introducing such an $i$~warp is warranted.
Six of these ten galaxies also display a $\phi$~warp.

{\bf Step 4: Determine $\vrot$.}
We fix the dynamical centers and $\vsys$, as well as the radial behavior of $\phi$
and $i$ of the rings as determined in the second and third step.
The resulting rotation curves, sampled every 10$\arcsec$ in radius, are indicated for each
galaxy by the crosses in the ``Rotation Curve'' panel, and projected onto the PV diagrams
presented in the Atlas.

% ------------------------------------------------------------------------------
% Table with galaxy geometry

\begin{table}
%\begin{sidewaystable}
\caption{\label{tab:Geometry}
Properties of the observed \hone\ velocity fields.}
\centering
{\tiny
\renewcommand{\tabcolsep}{1.2mm}
\begin{tabular}{|rrrcccc|}
\hline
\multicolumn{1}{|c}{UGC}           &
\multicolumn{1}{c}{$\vsys$}        &
\multicolumn{1}{c}{$\pa$}          &
\multicolumn{2}{c}{$R_{\phi}$}     &
\multicolumn{2}{c|}{$R_{\rm i}$}   \\
\multicolumn{1}{|c}{}                &
\multicolumn{1}{c}{(\kms)}           &
\multicolumn{1}{c}{(deg)}            &
\multicolumn{1}{c}{(arcsec)}         &
\multicolumn{1}{c}{($R/R_{\rm 25}$)} &
\multicolumn{1}{c}{(arcsec)}         &
\multicolumn{1}{c|}{($R/R_{\rm 25}$)} \\
\multicolumn{1}{|c}{(1)} &
\multicolumn{1}{c}{(2)} &
\multicolumn{1}{c}{(3)} &
\multicolumn{1}{c}{(4)} &
\multicolumn{1}{c}{(5)} &
\multicolumn{1}{c}{(6)} &
\multicolumn{1}{c|}{(7)} \\
%
% UGC    Vsys            PA_kin      R_pa  R_pa  R_i  R_i
% 
\hline
  448 &  4857.1 $\pm$ 0.9 & 303.7 $\pm$ 1.2 &  - &    - & 50 & 0.99 \\
  463 &  4459.5 $\pm$ 0.5 &  68.8 $\pm$ 1.5 & 35 & 0.70 &  - &    - \\
 1087 &  4482.5 $\pm$ 0.3 &  80.4 $\pm$ 1.2 &  - &    - & 45 & 0.97 \\
 1635 &  3518.1 $\pm$ 0.4 & 141.4 $\pm$ 1.3 &  - &    - &  - &    - \\
 3140 &  4621.0 $\pm$ 0.4 & 354.3 $\pm$ 1.6 & 50 & 0.85 & 65 & 1.10 \\
 3701 &  2918.2 $\pm$ 0.7 &  88.6 $\pm$ 1.9 & 65 & 1.12 & 75 & 1.30 \\
 3997 &  5943.6 $\pm$ 0.5 &  32.2 $\pm$ 1.2 & 15 & 0.40 &  - &    - \\
 4036 &  3467.4 $\pm$ 0.5 & 193.9 $\pm$ 1.1 & 65 & 1.05 &  - &    - \\
 4107 &  3506.8 $\pm$ 0.3 & 291.0 $\pm$ 1.0 &  - &    - &  - &    - \\
 4256 &  5246.1 $\pm$ 0.9 & 292.1 $\pm$ 1.8 &  - &    - &  - &    - \\
 4368 &  3869.0 $\pm$ 0.7 & 127.4 $\pm$ 3.1 & 35 & 0.50 &  - &    - \\
 4380 &  7481.1 $\pm$ 0.4 &  32.5 $\pm$ 1.4 &  - &    - & 35 & 0.97 \\
 4458 &  4753.7 $\pm$ 0.7 & 286.3 $\pm$ 2.2 &  - &    - &  - &    - \\
 4555 &  4240.1 $\pm$ 0.5 &  91.8 $\pm$ 1.5 & 35 & 0.71 &  - &    - \\
 4622 & 12827.1 $\pm$ 0.4 & 118.4 $\pm$ 1.5 &  - &    - & 35 & 0.82 \\
 6463 &  2505.2 $\pm$ 0.8 & 145.2 $\pm$ 1.4 &  - &    - &  - &    - \\
 6869 &   793.6 $\pm$ 0.7 & 298.5 $\pm$ 1.3 & 45 & 0.52 &  - &    - \\
 6903 &  1890.2 $\pm$ 0.3 & 134.9 $\pm$ 3.4 & 35 & 0.44 & 75 & 0.94 \\
 6918 &  1110.0 $\pm$ 0.7 & 192.2 $\pm$ 1.8 & 55 & 0.78 & 75 & 1.06 \\
 7244 &  4356.5 $\pm$ 0.2 & 148.3 $\pm$ 1.3 &  - &    - &  - &    - \\
 7416 &  6902.5 $\pm$ 0.9 &  67.0 $\pm$ 1.0 & 35 & 0.74 &  - &    - \\
 7917 &  6988.9 $\pm$ 0.5 & 218.6 $\pm$ 1.4 &  - &    - &  - &    - \\
 8196 &  8328.7 $\pm$ 0.8 &  88.5 $\pm$ 1.8 & 45 & 0.96 & 65 & 1.39 \\
 8230 &  7163.8 $\pm$ 2.2 &  69.6 $\pm$ 3.6 &  - &    - &  - &    - \\
 9177 &  8855.3 $\pm$ 2.6 & 243.7 $\pm$ 3.0 &  - &    - &  - &    - \\
 9837 &  2655.1 $\pm$ 0.2 & 311.0 $\pm$ 1.1 &  - &    - &  - &    - \\
 9965 &  4525.9 $\pm$ 0.3 & 200.2 $\pm$ 1.6 & 35 & 0.94 & 35 & 0.94 \\
11318 &  5884.7 $\pm$ 0.5 & 351.4 $\pm$ 2.3 & 35 & 0.70 &  - &    - \\
\hline
\end{tabular}
}
\tablefoot{
(1) UGC number; (2) systemic velocity of the rotating
\hone\ disk; (3) position angle of the receding side of the kinematic major axis of the
inner \hone\ disk; (4) and (5) onset radius of the position-angle warp in arcseconds and
scaled by the optical radius $R_{\rm 25}$; (6) and (7) onset radius of the inclination
warp in arcseconds and scaled by $R_{\rm 25}$.}
%\end{sidewaystable}
\end{table}
% ------------------------------------------------------------------------------

\subsubsection{Model and residual velocity fields}
Based on the (warped) tilted-ring models derived above, we used the program {\tt VELFI} in
{\it GIPSY} to construct model velocity fields. These are presented in the Atlas with the
same isovelocity contours as the observed velocity fields. The differences between the
observed and modeled velocity fields are shown in the Atlas as residual maps.

When inspecting the observed, modeled, and residual velocity fields, there are several 
things to keep in mind.
First, the model velocity fields are axisymmetric by definition, whereas the observed
velocity fields can be significantly asymmetric and affected by non-circular streaming
motions.
Second, irregular and small-scale structure in the shapes of the isovelocity contours of
the observed velocity fields may be caused by the interpolation of observed recession
velocities across blank pixels.
Third, the model velocity fields are based on inclinations suggested by inverting the
Tully-Fisher relation, which may deviate from the formally best-fitting tilted-ring
inclinations.
Fourth, several observed velocity fields suffer from significant beam-smearing effects
which tends to remove curvature from the intrinsic isovelocity contours, making them run
more parallel to each other in the inner regions. This results in systematically
underestimating the inclination of the tilted rings, worsening the degeneracy between
inclination and rotation velocity and increasing the errors on the inclination.
An illustrative example is the case of UGC~4555 with its kinematic minor axis aligned
along the elongated synthesized beam. The characteristic pattern in its residual velocity
field is the tell-tale signature of an inclination mismatch.

\subsection{Position-velocity diagram and HI surface density profile}
\label{sec:PVsigmaHI}
From the cleaned data cubes, we have extracted two-dimensional position-velocity (PV)
slices along the inner kinematic major ($\pa$) and minor ($\pa$+90) axes of the galaxies,
centered on the adopted dynamical centers. These slices do not follow any $\phi$~warp and,
consequently, the contours of the major-axis PV diagrams do not necessarily indicate the
projected rotation curves. The PV diagrams are presented in the Atlas, where the vertical
dashed line indicates the position of the dynamical center, while the horizontal dashed
line corresponds to $\vsys$. The small crosses in the PV diagrams indicate the rotation
curves as derived in the previous section projected onto the PV diagrams, accounting for
possible warps.

The orientation of the tilted rings was also followed when extracting radial \hone\
surface density profiles from the \hone\ column density maps. The \hone\ surface densities
were azimuthally averaged in the $10\arcsec$-wide rings, not only for the entire ring, but
also for the receding and approaching sides of the galaxy separately.
In the case of a warp, adjacent tilted rings do overlap and the total signal in the
overlapping regions was appropriately divided among both rings to conserve the total mass.
Finally, the azimuthally-averaged \hone\ column densities were corrected for the
line-of-sight integration through the projected disk, assuming the \hone\ gas is optically
thin along each line of sight. The face-on \hone\ column-density profiles are also
presented on the Atlas pages.

\subsection{Correcting rotation curves for beam smearing}
\label{sec:RCbeamcor}
Because of beam smearing, the shape of a velocity profile along the line of sight may
deviate strongly from a Gaussian, especially in the central region of a galaxy where the
velocity gradient across the beam is largest. The velocity profiles are generally skewed
by tails towards the systemic velocity, leading to an underestimate of the rotational
velocities.
This effect also compromises the observed velocity fields in terms of the shapes of the 
inner isovelocity contours which become less ``pinched'' towards the dynamical center.
For our tilted-ring fitting, we expect beam smearing to yield rotational velocities that
may be significantly underestimated in the fourth step of our procedure. This effect can
be seen in the PV diagrams for, e.g., UGC~4256 and UGC~9837.

Several methods have been suggested to correct for the effects of beam smearing, e.g., the
envelope-tracing method \citep{sofue2001} which makes use of the terminal velocity in a
PV diagram along the major axis, or by correcting the velocity field itself on the basis
of local velocity gradients \citep{begeman1989}. Here, we adopt an interactive variant of
the envelope-tracing method following \cite{verheyen2001a}. That is, we correct the
rotation curves manually by inspecting the PV diagrams along the kinematic major axis. We
have considered each cross on the receding and approaching sides of every galaxy
and shifted it to the most extreme velocity, while keeping in mind the contour levels, the
three-dimensional size of the beam, and the S/N in the data cube. 
The corrected, but still projected, rotational velocities are indicated in the major-axis
PV diagrams, with solid symbols on the receding side and with open symbols on the
approaching side of the galaxies. These points are also indicated with the same symbols in
the Rotation Curve panels of the Atlas, with a solid line connecting the midpoints.
We note that our beam-smearing corrections contribute to kinematic asymmetries in the
rotational velocities between the receding and approaching halves of the galaxy.

The errors on the velocities in the beam-corrected rotation curve come from adding the
estimated measurement errors (ranging between 1--4~\kms\ on the projected velocities
depending on the amplitude of the projected rotation curve, or 4--11~\kms\ on the 
deprojected velocities depending on the inclination of the galaxy) to half the difference
between the measured velocity on the receding and approaching sides. In the few cases 
where we only have a measurement on one side of the galaxy, we adopt an error based on the
average difference at all other radii for which we have measured rotational velocities on
both sides.
% =========================================================================================

% =========================================================================================
\section{Physical properties of the gas disks}
\label{sec:HIproperties}
This section summarizes some of the physical properties of the \hone\ gas disks in our
target galaxies such as their absolute and relative \hone\ masses, diameters, and
radial column-density profiles. We also discuss their kinematic properties such as their 
rotation curves and warps. When considering the following it should be kept in mind that
our reduced \hone\ sample is by no means statistically complete, but merely representative
of regular, disk-dominated galaxies. The main purpose of the data collected here is to
support the analysis of the baryonic and dark matter mass distributions in the galaxies as
presented in \citetalias{martinsson2013b}.

% ------------------------------------------------------------------------------
% Figure of HI vs Total mass

\begin{figure*}[]
\centering
\includegraphics[width=1.0\textwidth]{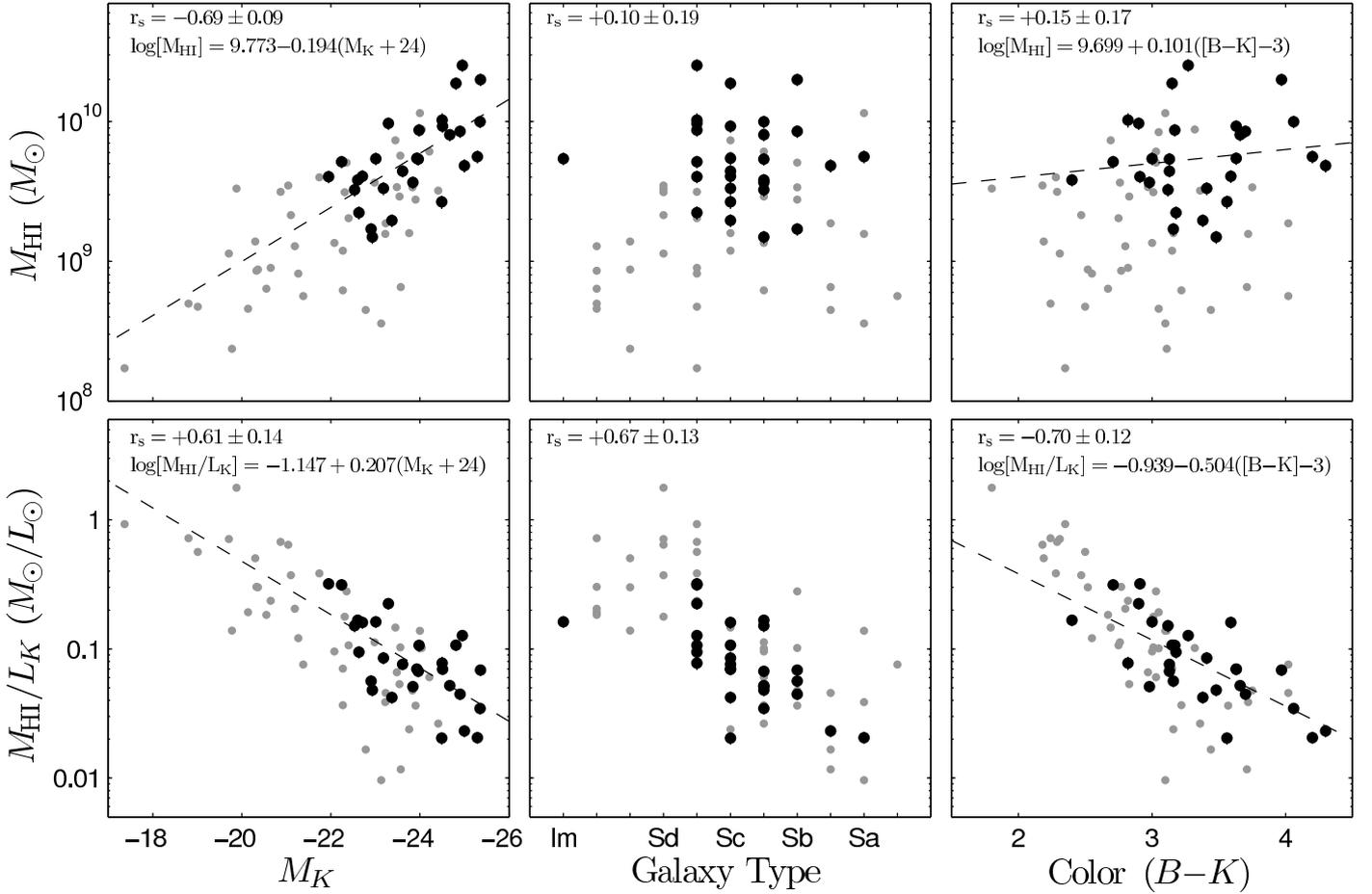}
\caption{Total and relative \hone\ masses as a function of $M_K$ (left),
morphological type (middle), and $B-K$ color (right). The dashed lines
are fits to the DMS galaxies (black symbols). The gray symbols shows the galaxies from
\cite{verheyen2001a}.
}
\label{fig:M_rel}
\end{figure*}
% ------------------------------------------------------------------------------

\subsection{HI masses}
The total \hone\ mass of each galaxy is derived by calculating
\begin{equation}
\mhi\;[\msol] = 2.36\times10^5\; D^2\;[{\rm Mpc}] 
\int S_{\rm HI}\;[{\rm Jy}]\; {\rm d}V \; [{\rm km~s^{-1}}] ,
\label{eq:totHI}
\end{equation}
where $\int S_{\rm HI} {\rm d}V$ is the integrated \hone\ line flux listed in 
Table~\ref{tab:Fluxes}. The \hone\ masses of our target galaxies, listed in 
Table~\ref{tab:properties}, range between 1.5$\times$10$^9$ $\msol$ and 
2.5$\times$10$^{10}$ $\msol$. The upper panels of Fig.~\ref{fig:M_rel} show the total
\hone\ masses against absolute $K$-band magnitude ($M_K$), morphological type, and $B-K$
color. The total \hone\ mass correlates well with the total $K$-band luminosity, but
trends with morphological type and color are absent. The most gas-rich galaxies are among
the most luminous galaxies that are of type Sb--Scd, but these are not the bluest galaxies
in our sample.

When considering the \hone\ gas content of galaxies per unit luminosity, clear
correlations emerge with all three global properties as shown in the lower panels of 
Fig.~\ref{fig:M_rel}. The trend with $M_K$ is inverted; the most luminous galaxies contain
the largest amount of gas, but the lowest relative gas content. Correlations with total
luminosity, morphological type, and color are consistent with overall trends along the
Hubble sequence and these trends are well known from previous studies
\citep{roberts1994,broeils1997,verheyen2001a,swaters2002}.
The linear fits to the correlations are presented in the figure panels, as well as the
Spearman rank-order coefficient ($r_s$) of the relations,
calculated following \citetalias{westfall2014}.
The relative gas-mass fractions in these galaxies were further discussed in
\citetalias{martinsson2013b}.

We compare our results with the galaxies in the Ursa Major sample of \cite{verheyen2001a},
plotted as gray symbols in Fig.~\ref{fig:M_rel}. We use the same absolute $K$-band
magnitude of the sun, $\mathcal{M}_{\odot,K}=3.30$ \citepalias[see][]{westfall2011b},
for both samples when calculating $\lk$.
Although the galaxies in \cite{verheyen2001a} have typically much lower total gas masses,
as is evident in the top panels, they seem to follow the same relations as the DMS
galaxies except for a possible steeper relation between $\mhi/\lk$ and color. However,
this difference is not statistically significant.

\subsection{Sizes of HI disks}
\label{sec:HIsize}
Next, we investigate the sizes ($D_{\rm HI}$) of the \hone\ disks, which we define as
twice the radius ($\rHI$) at which the azimuthally-averaged \hone\ column density has
dropped to 1~\sdu. In the total \hone\ maps in the Atlas, this column density level is
indicated with a thicker contour. In the optical images of the galaxies on the Atlas
pages, shown on the same scale as the \hone\ maps, the inclination and
extinction-corrected $D_{25}$ diameters are indicated by solid ellipses. 
These $D_{25}$ values are equal to $2 \times R_{25}$, the isophotal radius at a blue
surface brightness level of 25 mag/arcsec$^2$ as reported by the
NASA/IPAC~Extragalactic~Database\footnote{Operated by the Jet Propulsion Laboratory, 
California Institute of Technology, under contract with the National Aeronautics and Space
Administration.} (NED).

We do not measure $\rHI$ from the total \hone\ maps, but instead from the
azimuthally-averaged radial \hone\ column-density profiles as presented in the Atlas
(see Sect.~\ref{sec:HIprofiles}). The measurement of $\rHI$ is affected by beam smearing
which results in a somewhat larger radius. We correct for this beam-smearing effect to 
obtain the corrected \hone\ radius  $\rHI^{\rm cor}=\sqrt{(\rHI^{\rm obs})^2 - \theta^2}$,
where $\rHI^{\rm obs}$ is the observed radius and $\theta$ the beamsize. The correction is
only 1--4$\arcsec$ (or 1--6\% of $\rHI^{\rm obs}$).

The diameters of the gas disks in our target galaxies are 1--2 times the optical diameter
$D_{25}$, or 3--11 times $\hr$. The average ratio and scatter are
$\rHI/R_{25}=1.35\pm0.22$, somewhat lower than what has been found in earlier studies
\citep[e.g.,][who find the ratio $1.7\pm0.5$]{broeils1997}.
Any correlations with the global properties $M_K$, morphological type, and $B$--$K$ color
are absent \citep[see Fig.~4.6 in][]{martinsson2011} as noted earlier by
\cite{verheyen2001a}. Even though the fainter, bluer galaxies of later morphological type
in our sample are relatively more gas rich, they do not possess relatively larger \hone\
disks with respect to their stellar disks.

With these definitions for the optical and \hone\ diameters, it is clear that the \hone\
disks extend beyond the optically-recognizable stellar disks, which is consistent with
previous work \citep[e.g.,][]{broeils1997,verheyen2001a,swaters2002,noordermeer2005}.
It is noteworthy, however, that the samples studied by \cite{swaters2002} and
\cite{noordermeer2005} were drawn from the
{\it WHISP}\footnote{\url{http://www.astro.rug.nl/~whisp}} survey which comprised a
flux-density-limited parent sample, biased towards more gas-rich galaxies, which tend to
have larger gas disks. Volume-limited samples, such as the Ursa Major sample studied by
\cite{verheyen2001a}, have shown that gas disks are relatively less extended than what has
been concluded from flux-density-limited samples.

% ------------------------------------------------------------------------------
% Figure of HI Mass vs Size

\begin{figure}[t]
\centering
\includegraphics[width=0.5\textwidth]{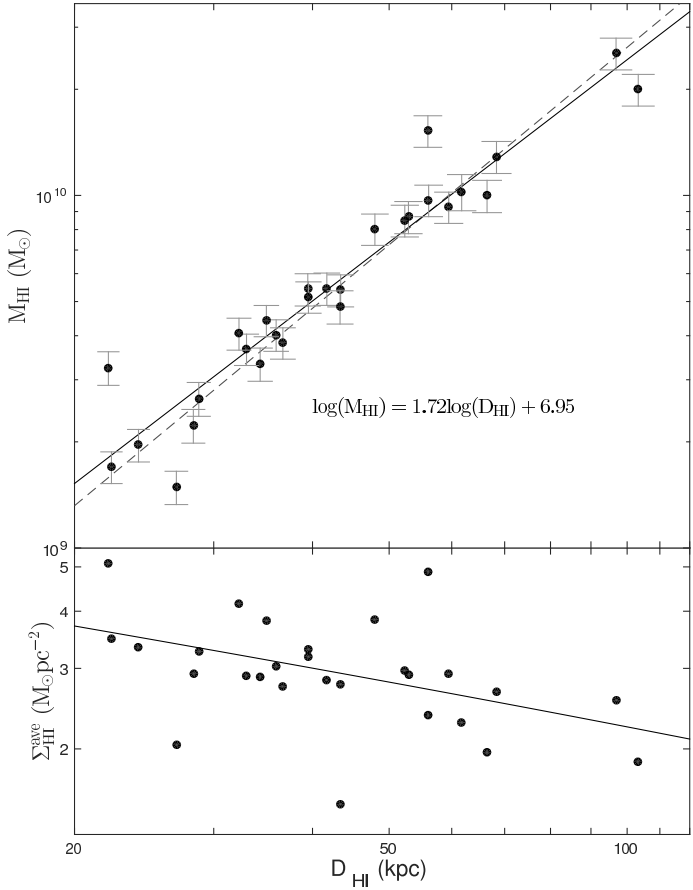}
\caption{{\bf Upper panel:} Correlation between the total \hone\ mass ($\mhi$) and the
beam-corrected diameter of the \hone\ disk ($D_{\rm HI}$=$2\rHI$). The solid line shows
the fit to the data while the dashed line represents the correlation found by
\cite{verheyen2001a}.  {\bf Lower panel:} Average \hone\ mass surface density within
$\rHI$ versus $D_{\rm HI}$. The line indicates the expected slope based on the fit
in the upper panel.
}
\label{fig:Mass_vs_Size}
\end{figure}
% ------------------------------------------------------------------------------

\subsubsection{Size versus mass and stellar kinematics}
Previous studies have revealed a tight correlation between total \hone\ masses and \hone\
diameters of the gas disks \citep{broeils1997,verheyen2001a,swaters2002,noordermeer2005}.
This relation implies an average \hone\ surface density within the \hone\ radius that 
varies only mildly among galaxies. In the upper panel of Fig.~\ref{fig:Mass_vs_Size}, we
plot the total \hone\ masses versus the \hone\ diameters of our target galaxies and we 
recover a similar tight correlation. The solid line illustrates a linear fit to our data 
and is described by

\begin{equation}
\log\left(\mhi\right) = 1.72 \log\left(D_{\rm HI}\right) + 6.95 ,
\label{eq:Mass_vs_Size}
\end{equation}

\noindent
where $\mhi$ is the mass in units of solar masses and $D_{\rm HI}$ the \hone\ diameter
measured in kpc. The dashed line shows the relation found by \cite{verheyen2001a} for
galaxies in Ursa Major and the relations are identical within the errors.

The lower panel of Fig.~\ref{fig:Mass_vs_Size} illustrates that the average \hone\ surface
densities within $\rHI$ of our 28 target galaxies varies from 1.5 to 5 \sdu, with a mean
and rms scatter of 3.0$\pm$0.8~\sdu. The fact that the slope is not quite 2.0 suggests
that the larger \hone\ disks tend to have slightly lower average \hone\ surface densities.

% ------------------------------------------------------------------------------
% kinematic scale length versus HI diameter

\begin{figure}[t]
\centering
\includegraphics[width=0.5\textwidth]{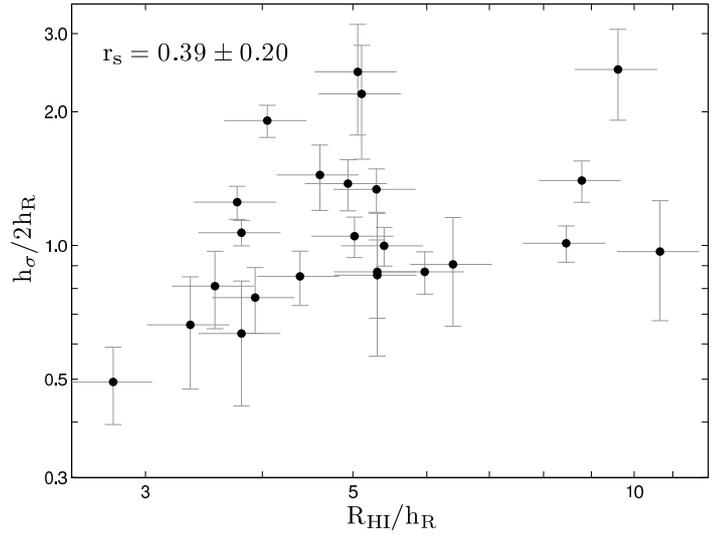}
\caption{Comparison of the relative kinematic scale lengths of the stellar disks and the
relative size of the \hone\ disks. Galaxies in which the stellar velocity dispersion
decreases faster with radius than expected on the basis of the photometric scale length of
the stellar disk ($h_{\sigma}/2\hr<1$), tend to have relatively small \hone\ disks, and
vice versa.
}
\label{fig:hsig_RHI}
\end{figure}
% ------------------------------------------------------------------------------

Although the relative size of the \hone\ disk does not correlate with galaxy luminosity,
color or morphological type, there seems to be a correlation with the ratio of kinematic
to photometric scale lengths, as illustrated in Fig.~\ref{fig:hsig_RHI}. The correlation
is weak ($r_s=0.39\pm0.20$); however, when dividing the sample in two bins, with 12
galaxies in each bin (only including galaxies in the PPak sample for which we have
stellar-kinematic measurements) we find a statistical difference. For galaxies that have
smaller \hone\ disks ($R_{\rm HI} < 5.05\hr$) we find an average ratio
$h_{\sigma}/2\hr = 1.03\pm0.12$, and for galaxies with larger \hone\ disks 
($R_{\rm HI} > 5.05\hr$) we find an average ratio $h_{\sigma}/2\hr = 1.36\pm0.19$.

For purely exponential disks of constant mass-to-light ratio, scale height and stellar
velocity ellipsoid, the stellar velocity dispersion ($\sigma_{*}$) is expected to fall
exponentially with radius, with an e-folding length ($h_{\sigma}$) that is twice the disk
scale length, $h_{\sigma} = 2\hr$ \citepalias{martinsson2013a}.
If $h_{\sigma}/2\hr$$>$1, %$
then the stellar disk is kinematically hotter in the outer
regions (or the central parts are colder) than what would be expected on the basis of the
radial distribution of the stars.
As seen in Fig.~\ref{fig:hsig_RHI}, this seems to be the case in galaxies with
extended \hone\ disks.
A possible, but speculative, explanation for this could be that extended gas disks are the
result of a continuous or discrete accretion of small gas clouds with their associated
stars. Such accretion will likely not be entirely coplanar with the main gas and stellar
disks of the accreting galaxy. The accreted gas may quickly settle into the midplane,
while the accreted stars continue to circle the galaxy in more inclined orbits, thereby
contributing to a higher stellar velocity dispersion in the outer gas-rich disk and a
flaring of the stellar disk.

\subsection{Radial $\Sigma_{\rm HI}$ profiles}
\label{sec:HIprofiles}

% ------------------------------------------------------------------------------
% Mosaic of all Sigma_HI profiles

\begin{figure}[t]
\centering
\includegraphics[width=0.5\textwidth]{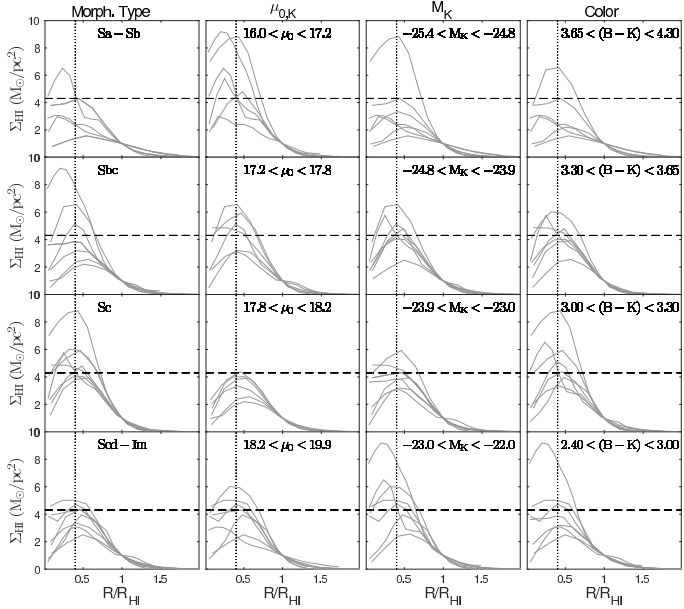}
\caption{Compilation of all 28 \hone\ column density profiles, divided in four
non-equal bins for each of four global optical properties; morphological type, central
disk surface brightness, absolute magnitude, and color. The profiles are scaled in radius
by $\rHI$. The horizontal dashed lines are drawn at the sample median 
$\Sigma_{\rm HI}^{\rm max}$ of 4.3\sdu\ as a reference. The vertical dotted lines are
drawn at the radius where the average profile reaches its peak column density.
}
\label{fig:SigmaRall_ind}
\end{figure}
% ------------------------------------------------------------------------------

To investigate whether the \hone\ column-density profiles (Sect.~\ref{sec:PVsigmaHI})
correlate with global properties of the galaxies, we divide the sample
of 28 galaxies
in 4 groups of $\sim$7 galaxies each along non-uniform intervals of morphological type,
$\mu_{0,K}^{i}$, $M_K$, and color. The results are presented in
Fig.~\ref{fig:SigmaRall_ind}.
Qualitatively, no obvious trends along the columns of Fig.~\ref{fig:SigmaRall_ind}
can be discerned. However, galaxies with the highest column densities tend to be of low
luminosity and high surface brightness, and galaxies with the lowest column densities
tend to be of earlier morphological type, lower surface brightness, higher luminosity, and
redder color.

\begin{figure*}[t]
\centering
\includegraphics[width=0.96\textwidth]{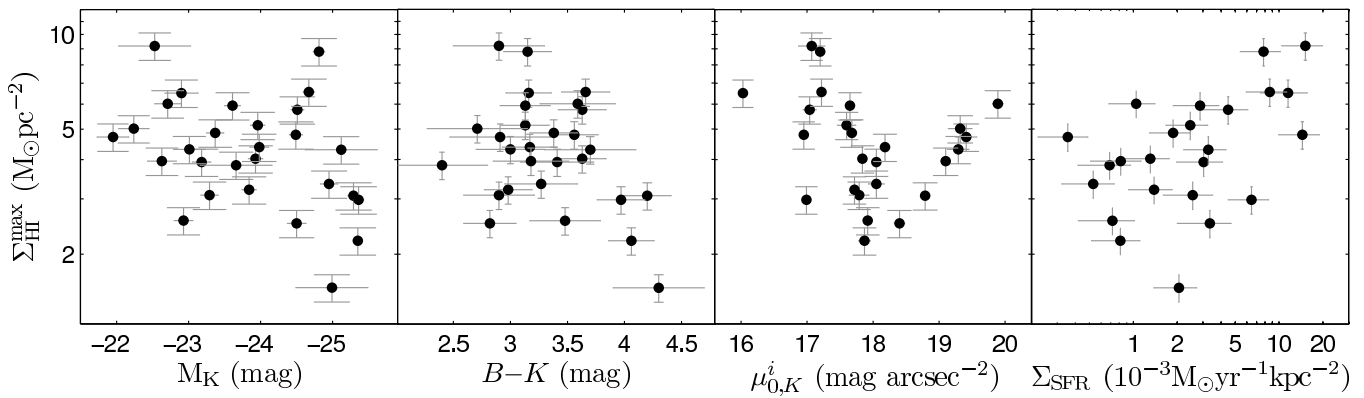}
\caption{Relations between the maximum, azimuthally-averaged \hone\ column densities
($\Sigma_{\rm HI}^{\rm max}$) and global properties of the galaxies. From left to right:
absolute $K$-band magnitude, $B-K$ color, central disk surface brightness, and 
star formation density.
}
\label{fig:SigavSigmax}
\end{figure*}

Figure~\ref{fig:SigavSigmax} shows the observed, azimuthally-averaged, maximum \hone\
column density ($\Sigma_{\rm HI}^{\rm max}$) of every galaxy against its absolute
magnitude, color, surface brightness and star formation rate surface density
($\Sigma_{\rm SFR}$) as derived from the radio-continuum flux (Sect.~\ref{sec:SFR}).
There are only weak correlations in the sense that galaxies with higher peak column
densities tend to be bluer with higher star formation densities. 

The shapes of the radial \hone\ column-density profiles of the galaxies all display the
same general characteristic; rising from a non-zero value in the center of the galaxy to a
maximum value ($\Sigma_{\rm HI}^{\rm max}$) and then declining following a
near-exponential fall-off.
%, crossing the 1~\sdu\ level at the radius $\rHI$ by definition.
This similarity of $\Sigma_{\rm HI}$ profiles of different galaxies has been noted before
\citep[e.g.,][]{cayatte1994,broeils1994}. For each galaxy, the values of 
$\Sigma_{\rm HI}^{\rm max}$ and $D_{\rm HI}$=2$\rHI$ are listed in 
Table~\ref{tab:properties}. It should be noted that the $\Sigma_{\rm HI}$ profiles
considered here have not been corrected for the effects of smearing by the Gaussian beam.
The beam smearing slightly lowers the peak column density, raises the central column
density, and broadens the profile somewhat. The overall shape, however, will not be
altered significantly.

% ------------------------------------------------------------------------------
% B-K and Sigma_SFR vs Sigma_max

% ------------------------------------------------------------------------------

\begin{figure}[t]
\centering
\includegraphics[width=0.5\textwidth]{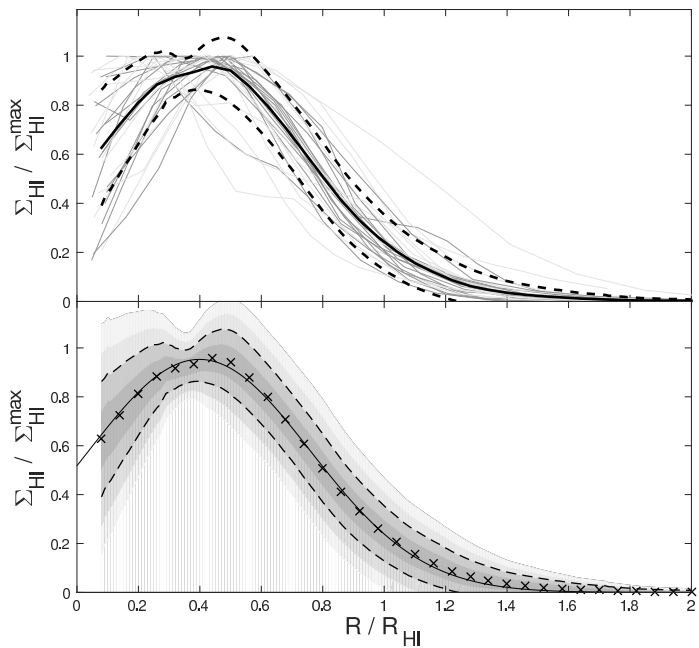}
\caption{Self-similarity of the \hone\ column density profiles.
{\bf Upper panel:} All 28 $\Sigma_{\rm HI}$ profiles scaled in radius and normalized in
amplitude. The thinnest lines correspond to the early-type spirals in our sample. The
thick black line shows the clipped average profile and the dashed lines indicate the rms
scatter at each radius.
{\bf Lower panel:} The crosses follow the thick black line in the upper panel and the
dashed lines are identical to the dashed lines in the upper panel. Gray scales indicate
steps of 0.5 times the rms scatter. The solid black line is the best-fit Gaussian function
following the crosses above ($\Sigma_{\rm HI}$/$\Sigma_{\rm HI}^{\rm max}$)$>$0.2.
}
\label{fig:Sigma_average}
\end{figure}

In the upper panel of Fig.~\ref{fig:Sigma_average}, all 28 radial $\Sigma_{\rm HI}$ 
profiles are collected and drawn after scaling them in radius with $\rHI$ and normalizing
their amplitudes by $\Sigma_{\rm HI}^{\rm max}$. The solid line shows the average profile
shape, calculated after clipping the highest and lowest outlying profiles at every 
resampled radius. The dashed lines indicate the clipped 1$\sigma$ rms scatter at each 
radius. The thinnest lines in the upper panel come from galaxies with earlier
morphological types Sa--Sbc and those tend to deviate most from the average profile shape.
The crosses in the lower panel of Fig.~\ref{fig:Sigma_average} follow the solid line from
the upper panel while the dashed lines are the same and the gray scales indicate intervals
of 0.5$\sigma$.

Next, we try to parameterize the generic shape of the $\Sigma_{\rm HI}$ profiles. Given
the central depression in the \hone\ column densities, it is clear that the \hone\ mass
surface density does not follow the typical exponential radial decline of the stars 
\citep{freeman1970}. For late-type dwarf galaxies, \cite{swaters2002} fitted an
exponential function to the $\Sigma_{\rm HI}$ measurements at the outer radii of the
profiles with fits that follow the observed profile fairly well in many cases.
\cite{wang2014} analyzed the radial distribution of \hone\ gas in a sample of 23
galaxies with unusually high \hone\ content and 19 more ``normal'' galaxies. They found 
that an exponential function with a depression in the central parts of the disk, similar
to what was proposed by \cite{obreschkow2009}, fit the
observed $\Sigma_{\rm HI}$ profiles well for most of these galaxies.

Attempts to fit an exponential to the outer profiles of the galaxies in our sample were
less successful. Instead, we note that the average profile resembles an offset Gaussian
function. The solid line in the lower panel of Fig.~\ref{fig:Sigma_average} is a Gaussian
fit to the data (crosses), including only the data points above
$0.2 \Sigma_{\rm HI}^{\rm max}$. This fit is described by

\begin{equation}
\Sigma_{\rm HI}(R) = \Sigma_{\rm HI}^{\rm max} \cdot 
\rm{e}^{-\frac{\left(R-R_{\rm \Sigma,max}\right)^2}{2\sigma_{\Sigma}^2}},
\label{eq:Gaussian}
\end{equation}

\noindent
where the peak of the profile occurs at $R_{\rm \Sigma,max}=0.40 \rHI$ and the broadening
of the profile is $\sigma_{\Sigma}=0.36 \rHI$. The fitted profile follows the
data surprisingly well, at least out to 1$\rHI$.

From Eq.~\ref{eq:Gaussian}, we derive the radius where the typical \hone\ mass surface
density drops below 50\% of the peak to be $R_{\Sigma,1/2}=0.81 \rHI$. 
We measured directly from the profiles the radius at which the azimuthally averaged column
density drops below 50\% of the peak density. For all 28 galaxies, we found this radius to
span 0.49--1.14 $\rHI$ with an average value of $R_{\Sigma,1/2}=0.80 \rHI$ and a standard
deviation of 0.12$\rHI$. 

We conclude that the $\Sigma_{\rm HI}$ profiles of our target galaxies are remarkably
similar to each other. Possible astrophysical causes for this general shape will be
discussed in a forthcoming paper, taking account of the molecular gas component, disk
stabilities and SFRs.
It is well known that the molecular component closely follows the light distribution with
similar scale lengths \citep{nishiyama2001,regan2001} and the question is whether the
atomic and molecular gas components may conspire to produce an exponential decline of the
total gas disk with the molecular component filling the central depression of the \hone\
distribution, as previously discussed by \cite{bigiel2012}.

% ------------------------------------------------------------------------------
% Figure of averaged Gaussian Sigma_HI profile

% ------------------------------------------------------------------------------

\subsection{Kinematics of the gas disks}
In this section, we will briefly summarize the kinematics of the gas disks by discussing
the shapes of the derived \hone\ rotation curves and the properties of the warps. In
\citetalias{martinsson2013b}, these \hone\ rotation curves were combined with \halp\
rotation curves and their combined properties were discussed in more detail.

\subsubsection{Rotation curve shapes}
The shapes of rotation curves are known to follow some systematic behavior
\cite[e.g.,][]{persic1996}.
The prototypical rotation curve rises monotonically from the center, roughly consistent
with solid-body rotation, then smoothly transitions to a more or less constant rotation
speed that persists beyond the radius of the last measured point
\citep[e.g,][]{AlbadaSancisi1986}. This behavior is typical for Sc-type galaxies without a
strong bulge component and with rotation speeds in excess of $\sim$120~\kms. In such
systems, the baryons and dark matter conspire to produce a flat rotation curve, and most
galaxies in our sample follow this empirical trend.

In galaxies with a significant bulge or very small $\hr$, the rotation curve typically
rises very rapidly, reaches a peak, then declines to where the gravitational potential is
dominated by the dark matter halo. In our sample, UGC~4458 is a galaxy that displays such
a declining rotation curve. Such systems were presented by \cite{casertano1991} as
examples where the disk-halo conspiracy is lifted and the distribution of baryons clearly
leaves a kinematic imprint on the overall shape of the rotation curve.
In dwarf galaxies and galaxies of low surface brightness, the rotation curve often
continues to rise out to the last measured point \citep{swaters2009}. Only in dwarf
galaxies with very extended \hone\ disks do the rotation curves reach out far enough into
the dark matter halo to display a well-defined flat part
\citep[e.g., NGC~3741;][]{begum2005}. These dwarf galaxies with rotation speeds below
$\sim$80~\kms\ are not represented in our sample.

To date, no rotation curve has ever been observed to show a decline in the outer parts
that may signify the boundary of the dark matter halo where the density drops faster than
$\rho \sim R^{-2}$. It is also noteworthy that the detailed shape of the inner rotation
curve can generally be explained by the distribution of the baryons as traced by the
light. An observed rotation curve never rises faster than what would be expected by
assuming a maximal disk. Apart from central, super-massive black holes, there is no
compelling evidence for dark matter within the inner 2$\hr$ if one allows for
unconstrained scaling of the baryonic mass. Hence, the central distribution of the total
dynamical mass that gives rise to the inner rotation curve seems to follow the
distribution of light
\citep[e.g.,][]{kent1986,sancisi2004,swaters2011,fraternali2011,lelli2013}.

% ------------------------------------------------------------------------------
% Figure of RCs

\begin{figure}[t]
\centering
\includegraphics[width=0.5\textwidth]{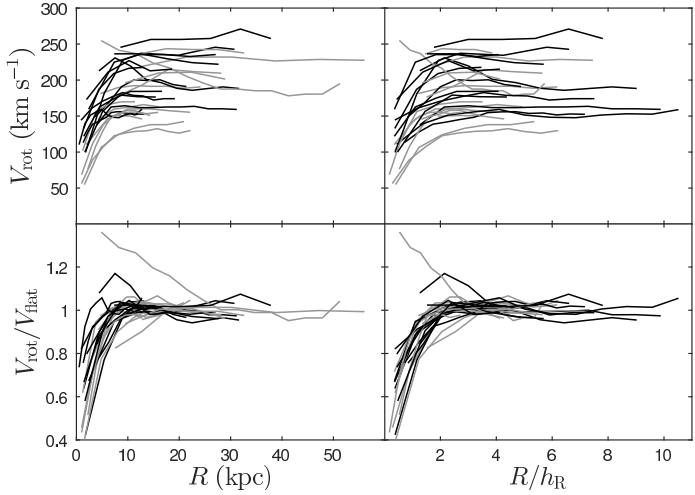}
\caption{Compilation of all 28 derived rotation curves with different scalings
for each panel. The sample of rotation curves is divided in two according to central disk
surface brightness: Gray curves correspond to galaxies of lower surface brightness
($\mu_{0,K}^{i}>17.8$~mag/arcsec$^{\rm 2}$) and black curves correspond to galaxies of 
higher surface brightness ($\mu_{0,K}^{i}<17.8$~mag/arcsec$^{\rm 2}$). In the upper left
panel, no scaling is applied to the rotation curves. In the bottom right panel, the
rotation curves are scaled both in radius and amplitude.
}
\label{fig:RCall}
\end{figure}
% ------------------------------------------------------------------------------

Because of the limited angular resolution of our \hone\ observations with respect to the
sizes of the galaxies, we cannot assess the inner shapes of the rotation curves in detail.
And because of the nearly face-on orientation of the galaxies in our sample, inclination
warps are often impossible to detect and characterize, which hampers the assessment of the
amplitude and slope of the outer \hone\ rotation curves. Hence, we are not in a good 
position to investigate the shapes of the rotation curves in detail. With this in mind, we
plot all 28 beam-corrected \hone\ rotation curves together in Fig.~\ref{fig:RCall}. The
sample is divided in half with black lines indicating galaxies of higher surface
brightness ($\mu_{0,K}^{i} < 17.8$ mag/arcsec$^2$) and gray lines indicating galaxies of
lower surface brightness ($\mu_{0,K}^{i} > 17.8$ mag/arcsec$^2$). The amplitudes of the
rotation curves range from 130 to 270 \kms\ and the maximum radial extent varies from
10--55~kpc. When scaling the rotation curves in radius with $\hr$ and in amplitude by the
rotational velocity of the flat part $V_{\rm flat}$, we see that all rotation curves are
fairly self-similar. Notable exceptions are the declining rotation curves of UGC~3140
(black) and UGC~4458 (gray).

\subsubsection{Warps}
\label{sec:Asymmetries}
Compared to the readily observable stellar disk, the extended \hone\ disks more often show
morphological and kinematic asymmetries. Furthermore, the fact that a warping of the
\hone\ disk is a common feature was already shown by \cite{sancisi1976}, who found that 
four out of five observed edge-on systems were warped. Later observations of less inclined
galaxies showed features in the \hone\ velocity fields that were interpreted as warps; an
(asymmetric) S-shape of the line of nodes or isovelocity contours that open up or close in
on themselves. The method of fitting tilted rings at different radii has been generally
used to quantify warps, revealing that many galaxies are warped
\citep[e.g.,][]{bosma1981b}. We now think that most \hone\ disks are warped and
\cite{garcia2002} even claim that {\it all} galaxies display warped gas disks provided
that they extend well beyond the stellar disk.

The origin and persistence of warps still presents a puzzle. Many attempts have been made
to quantify how often the disks are warped and to explain why they are warped
\citep[e.g.,][]{briggs1990,garcia2002,shen2006,kruit2007}. There are some indications that
the environment may play a role in warping, but the ubiquity of warps suggest that their
origin is not simply the result of interaction with the environment. A constant infall of
\hone\ gas, with an angular momentum vector misaligned to that of the inner disk, might
explain their formation.

When assessing the frequency and properties of the warps detected in our sample of
galaxies, it should be kept in mind that these galaxies were selected to display regular
and symmetric morphologies, as well as regular kinematics of their \halp\ velocity fields
\citepalias{bershady2010a}. Thus, one question to ask is how regular, symmetric, or planar
the \hone\ disks are when the inner \halp\ disks show well-behaved kinematics. Indeed,
if \cite{garcia2002} are correct, then all our galaxies will display warps as their \hone\
disks extend beyond the optical disks (Sect.~\ref{sec:HIsize}).

% ------------------------------------------------------------------------------
% Figure of Warp onset

\begin{figure}
\centering
\includegraphics[width=0.5\textwidth]{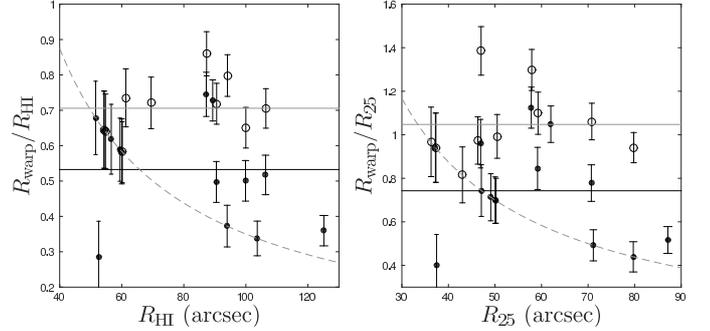}
\caption{Onset radii of the warps detected in the \hone\ disks, scaled by $\rHI$ (left)
and $R_{25}$ (right). Filled symbols indicate the onset radii of $\phi$~warps and open
symbols indicate the onset radii of $i$~warps. The solid black and gray lines show the
straight average of the $i$~warps and $\phi$~warps, respectively. The dashed gray line
indicates $R=35\arcsec$, the edge of the FOV of the PPak IFU.
}
\label{fig:warp_on}
\end{figure}
% ------------------------------------------------------------------------------

A $\phi$~warp is relatively easy to detected in nearly face-on galaxies. We have visually
inspected three data products presented for each galaxy in the Atlas; 1) the observed
\hone\ velocity field, 2) the PV diagram along the kinematic minor axis, and 3) the
results from individual tilted-ring fits at different radii. For many of the galaxies we
detect a warp in only one or two of these three data products because of limited S/N, the
specific geometry of the warp, or the effects of streaming motions.  Nevertheless, 17 out
of 28 galaxies show some indication of a $\phi$~warp. We deem this warp to be significant
in 14 cases and correct accordingly as described in Sect.~\ref{sec:geometry}.

For most of our galaxies the inclination is too unconstrained to be able to detect an
$i$~warp, and we have direct indications for an $i$~warp in only a handful of galaxies.
For the analysis in this paper, we have assumed that the rotation curves are more or less
flat beyond $R$$=$35$\arcsec$, and we enforce a linear $i$~warp in 10 of the galaxies
(Sect.~\ref{sec:geometry}). Even though we do not use the inclinations derived from the
tilted-ring fits, we note that in several cases there is a trend in these measurements
suggesting that there is in fact an $i$ warp somewhat following the enforced warp (see
Atlas).

Table~\ref{tab:Geometry} lists the radii of the onset of the warps for galaxies that have
an adopted $\phi$ and/or $i$~warp and Fig.~\ref{fig:warp_on} shows these onset radii
normalized to the \hone\ radius (left image) and the optical radius  (right image).
To be consistent with our measurements of the stellar and ionized gas kinematic obtained 
with PPak \citepalias{martinsson2013a}, we have initially chosen to not correct for any 
warps inside $R$$=$35$\arcsec$. However, for UGC~3997, which has a clear indications of a
$\phi$~warp starting at $\sim$0.5$R_{25}$ (see Atlas) well inside the PPak FOV,
we make an exception.
\Citet{kruit2007} found that when an \hone~warp is present, it starts at around 1.1 times
the truncation radius of the optical disk (at about 4--5$\hr$).
We find that the warps starts at much smaller radii. It should be noted, however,
that a $\phi$~warp can be mimicked by streaming motions in a strong two-armed spiral disk.

We conclude this section with noting that although we do not detect warps in all galaxies,
our result does not contradict \cite{garcia2002}, since, depending on the S/N of the data
and the orientation of the disk in the sky, we might not be able to detect the warp.
% =========================================================================================

% =========================================================================================
\section{Radio continuum and star formation rates}
\label{sec:SFR}
The radio-infrared relation 
\citep{kruit1971,harwit1975,dickey1984,dejong1985,condon1992,yun2001,boyle2007}
is one of the tightest correlations known in astronomy and offers the possibility of 
deriving SFRs directly from measured radio luminosities, modulo 
contamination from radio emission due to nuclear activity. The correlation relates the 
infrared emission from thermal re-radiation of starlight by dust in star-forming regions
to the nonthermal synchrotron radiation from relativistic particles accelerated by 
supernova explosions \citep{harwit1975,condon1992}.

In this section we describe how the measured 21 cm continuum flux densities
($S_{\rm 21 cm}$) are converted into intrinsic 21 cm luminosities ($L_{\rm 21 cm}$) and
how the SFR is calculated from $L_{\rm 21 cm}$.
We investigate correlations between SFR, specific SFR (sSFR), and SFR density
($\Sigma_{\rm SFR}$) with other galaxy parameters, and find strongest correlations with
the central disk surface brightness of the disk.
For all the analysis in this section, we have excluded the four galaxies with
$< 3 \sigma$ detection in $S_{\rm 21 cm}$ (see~Sect.~\ref{sec:S21}), except for
Table~\ref{tab:properties}, which presents (in italic numbers) the SFR for these
galaxies calculated from $S_{\rm 21 cm}$ as presented in Table~\ref{tab:Fluxes}.

% ------------------------------------------------------------------------------
% Table with measured fluxes

\begin{table}
\caption{
\label{tab:properties}
Derived galaxy properties.}
\centering
\begin{tabular}{|rrrrr|}
\hline
\multicolumn{1}{|c}{UGC}                         &
\multicolumn{1}{c}{SFR}                          &
\multicolumn{1}{c}{$\mhi$}                 &
\multicolumn{1}{c}{$D_{\rm HI}$}                 &
\multicolumn{1}{c|}{$\Sigma_{\rm HI}^{\rm max}$} \\
\multicolumn{1}{|c}{}                        &
\multicolumn{1}{c}{($\msol$~yr$^{\rm -1}$)} &
\multicolumn{1}{c}{($10^9$ $\msol$)}     &
\multicolumn{1}{c}{(arcsec)}                 &
\multicolumn{1}{c|}{($\msol$~pc$^{-2}$)}    \\
\multicolumn{1}{|c}{(1)} &
\multicolumn{1}{c}{(2)}  &
\multicolumn{1}{c}{(3)}  &
\multicolumn{1}{c}{(4)}  &
\multicolumn{1}{c|}{(5)} \\
%
% UGC         SFR                M_HI           R_HI S_HImax
%
\hline
  448 &  1.02 $\pm$ 0.37 &  5.45 $\pm$ 0.57 &  139 &  4.02 $\pm$ 0.48 \\
  463 &  9.38 $\pm$ 3.04 &  2.66 $\pm$ 0.29 &  103 &  4.79 $\pm$ 0.50 \\
 1087 &  1.64 $\pm$ 0.58 &  3.33 $\pm$ 0.36 &  122 &  3.93 $\pm$ 0.45 \\
 1635 &  0.36 $\pm$ 0.15 &  1.49 $\pm$ 0.16 &  122 &  2.56 $\pm$ 0.35 \\
 3140 &  4.36 $\pm$ 1.63 &  9.27 $\pm$ 0.94 &  200 &  5.76 $\pm$ 0.58 \\
 3701 &  0.16 $\pm$ 0.07 &  4.03 $\pm$ 0.41 &  174 &  4.72 $\pm$ 0.56 \\
 3997 &  {\it 0.35 $\pm$ 0.18} &  5.43 $\pm$ 0.57 &  105 &  4.31 $\pm$ 0.46 \\
 4036 &  1.76 $\pm$ 0.58 &  5.39 $\pm$ 0.56 &  179 &  5.14 $\pm$ 0.69 \\
 4107 &  0.71 $\pm$ 0.23 &  1.96 $\pm$ 0.21 &  103 &  4.86 $\pm$ 0.55 \\
 4256 & 14.58 $\pm$ 4.68 & 15.26 $\pm$ 1.57 &  162 &  8.82 $\pm$ 0.88 \\
 4368 &  2.97 $\pm$ 1.13 &  9.70 $\pm$ 0.99 &  207 &  3.08 $\pm$ 0.45 \\
 4380 &  {\it 0.71 $\pm$ 0.31} &  8.70 $\pm$ 0.91 &  109 &  4.38 $\pm$ 0.46 \\
 4458 &  {\it 3.35 $\pm$ 1.57} & 12.88 $\pm$ 1.34 &  214 &  3.07 $\pm$ 0.35 \\
 4555 &  0.93 $\pm$ 0.32 &  3.67 $\pm$ 0.37 &  119 &  3.21 $\pm$ 0.33 \\
 4622 &  2.24 $\pm$ 0.91 & 25.31 $\pm$ 2.62 &  120 &  3.34 $\pm$ 0.35 \\
 6463 &  0.28 $\pm$ 0.11 &  3.82 $\pm$ 0.39 &  190 &  3.84 $\pm$ 0.38 \\
 6869 &  2.83 $\pm$ 0.91 &  3.25 $\pm$ 0.35 &  250 &  9.19 $\pm$ 0.97 \\
 6903 &  0.37 $\pm$ 0.14 &  2.23 $\pm$ 0.24 &  188 &  3.95 $\pm$ 0.42 \\
 6918 &  2.00 $\pm$ 0.73 &  1.70 $\pm$ 0.17 &  213 &  6.51 $\pm$ 0.78 \\
 7244 &  {\it 0.35 $\pm$ 0.16} &  5.16 $\pm$ 0.53 &  129 &  5.02 $\pm$ 0.51 \\
 7416 &  5.34 $\pm$ 1.79 &  8.50 $\pm$ 0.88 &  113 &  4.30 $\pm$ 0.43 \\
 7917 &  2.17 $\pm$ 0.81 &  9.99 $\pm$ 1.05 &  139 &  2.21 $\pm$ 0.29 \\
 8196 & 14.78 $\pm$ 4.76 & 19.98 $\pm$ 2.06 &  181 &  2.98 $\pm$ 0.46 \\
 8230 &  1.62 $\pm$ 0.55 &  4.84 $\pm$ 0.52 &   92 &  1.56 $\pm$ 0.22 \\
 9177 &  7.77 $\pm$ 3.20 & 10.25 $\pm$ 1.21 &  101 &  2.51 $\pm$ 0.31 \\
 9837 &  0.43 $\pm$ 0.15 &  4.06 $\pm$ 0.42 &  157 &  6.02 $\pm$ 0.60 \\
 9965 &  1.44 $\pm$ 0.52 &  4.42 $\pm$ 0.45 &  109 &  5.93 $\pm$ 0.68 \\
11318 & 11.46 $\pm$ 3.69 &  8.03 $\pm$ 0.82 &  120 &  6.56 $\pm$ 0.76 \\
\hline
\end{tabular}
%
%\tablefoot{
%Observed \hone\ and 21-cm continuum fluxes, and calculated \hone\ masses, 
%21-cm luminosities and star-formation rates of the galaxies in the HI~Sample.
%[[NOTE: HI and MHI have no sys error, while the continuum stuff do!!]]
%}
\end{table}

% ------------------------------------------------------------------------------

% ------------------------------------------------------------------------------
% Figure of 21cm luminosity function

\begin{figure}
\centering
\includegraphics[width=0.5\textwidth]{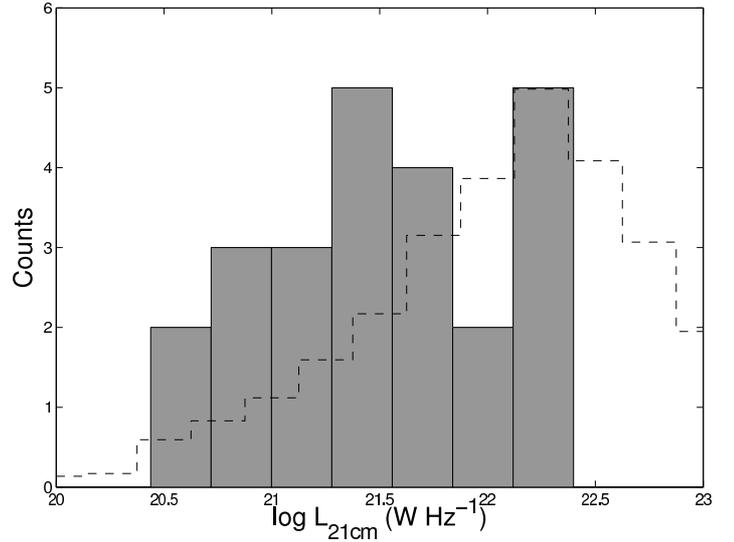}
\caption{Distribution of 21 cm radio continuum luminosities. The solid histogram displays
the distribution of 21 cm radio continuum luminosities of the galaxies in the reduced
\hone\ sample. The dashed histogram shows the distribution for the {\it IRAS} galaxies as
derived by \cite{yun2001}. %, normalized to the same peak counts.
}
\label{fig:L21func}
\end{figure}
% ------------------------------------------------------------------------------

\subsection{21 cm luminosities}
The radio continuum images of our galaxies in the Atlas show that all galaxies
with a significant detection display spatially extended emission, except in the case of
UGC~8196 where only a central point source is detected. This gives confidence that in most
galaxies the 21 cm continuum emission is indeed associated with star formation throughout
their disks.

The intrinsic 21 cm radio luminosities are calculated using the relation from
\cite{yun2001}

\begin{equation}
\log L_{\rm 21 cm} \;[{\rm W Hz^{-1}}] = 20.08 + 2\log D\;[{\rm Mpc}] + 
\log S_{\rm 21 cm} \;[{\rm Jy}],
\label{eq:L21}
\end{equation}

\noindent
where $D$ is the distance, taken from \citetalias{bershady2010a}, and $S_{\rm 21 cm}$ is
the total flux density listed in Table~\ref{tab:Fluxes}. The histogram in 
Fig.~\ref{fig:L21func} shows the distribution of 21 cm luminosities of the galaxies in
our sample, spanning a factor 50 in luminosity and ranging from 5$\times$10$^{20}$ to 
2.5$\times$10$^{\rm 22}$~W~Hz$^{-1}$. The dashed line shows the distribution found by 
\cite{yun2001} for the {\it IRAS}~2Jy sample, with counts normalized to our sample. The 
differences in the distributions come from selection effects, where the {\it IRAS}~2Jy 
sample is biased towards galaxies with higher luminosities, while our sample is biased 
towards galaxies with quiescent disks. This figure illustrates that our galaxies are not
among the most vigorous star forming systems, as expected.

% ------------------------------------------------------------------------------
% Figure of Radio-FIR (60)

\begin{figure}
\centering
\includegraphics[width=0.5\textwidth]{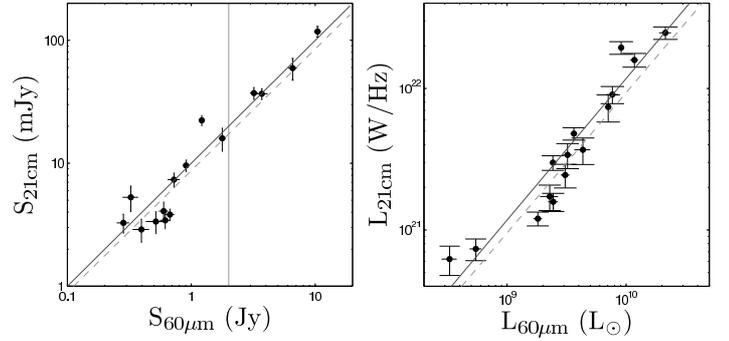}
\caption{Radio-infrared correlation of 15 galaxies in the reduced \hone\ sample
detected by $IRAS$ at $\rm 60\mu m$.
{\bf Left:} Correlation between flux densities. The solid line indicates the simple
relation $S_{\rm 21 cm}$$=$0.01$S_{\rm 60\mu m}$. The dashed line shows the correlation
found by \cite{yun2001} for the $IRAS~{\rm 2Jy}$ sample. The vertical line indicates
their lower flux-density limit of 2~Jy.
{\bf Right:} Correlation between intrinsic radio and infrared luminosities.
}
\label{fig:Radio_FIR_60}
\end{figure}
% ------------------------------------------------------------------------------

% ------------------------------------------------------------------------------
% SFR vs all

\begin{figure*}
\centering
\includegraphics[width=1.0\textwidth]{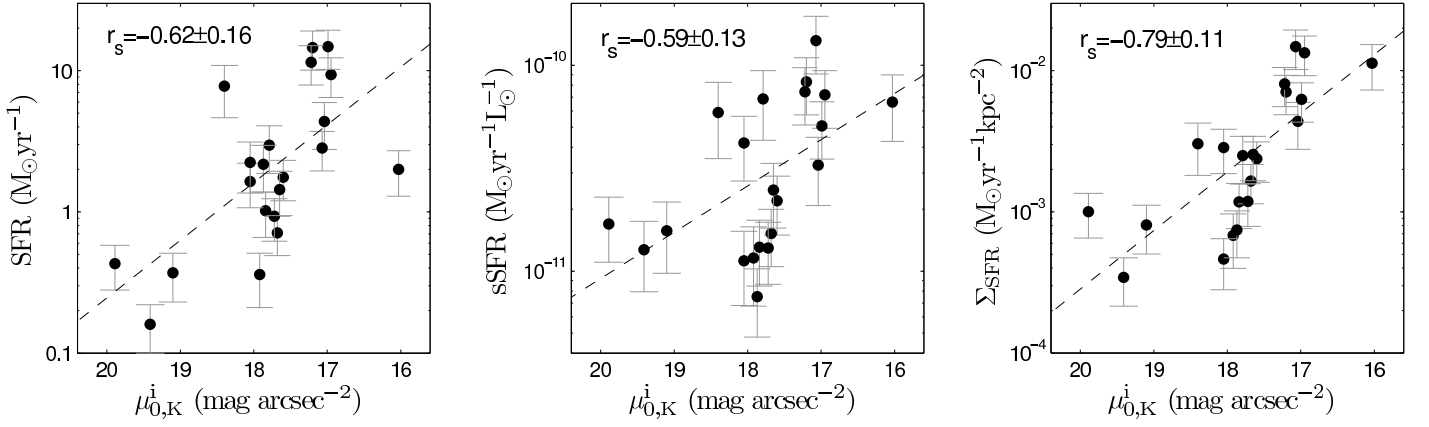}
\caption{Relations between disk surface brightness ($\mu_{0,K}^{i}$) and total SFR (left),
specific SFR (middle), and SFR surface density (right). The dashed lines are linear fits
to the data.
}
\label{fig:SFR_mu}
\end{figure*}
% ------------------------------------------------------------------------------

\subsection{The radio-FIR correlation}
As mentioned above, there is a tight correlation between far-infrared (FIR) and radio 
luminosities. Here we investigate this correlation for the galaxies in our sample, using 
flux densities at $60\mu$m from {\it IRAS} observations \citep{neugebauer1984,moshir1990}.
This is done for the 15 galaxies in our sample with reported detections at $60\mu$m in the
{\it IRAS Faint Source Catalogue}. Figure~\ref{fig:Radio_FIR_60} shows the correlation
between our measured continuum fluxes ($S_{\rm 21cm}$) and the {\it IRAS} fluxes at 
$60\mu$m ($S_{\rm 60\mu m}$), as well as the correlation when converting fluxes to 
luminosities ($L_{\rm 21cm}$ and $L_{\rm 60\mu m}$). The galaxies in our sample follow the
relation described by \cite{yun2001} (dashed gray line), but have the tendency to lie 
slightly above that relation, being somewhat overluminous in the radio continuum flux.
The solid black line shows the simple relation

\begin{equation}
S_{\rm 21 cm} = 0.01 S_{\rm 60\mu m}.
\label{eq:L21rel}
\end{equation}

For lower intrinsic luminosities of $L_{\rm 60 \mu m}$$\leq$$10^9$$\lsol$,
\cite{condon1991} and \cite{yun2001} found a deviation from this linear behavior in the
sense that fainter galaxies become underluminous in their radio emission compared to their
FIR emission. We may see something similar for our galaxies, with a few galaxies around 
$L_{\rm 60\mu m}=3\times10^{9} \lsol$ having $L_{\rm 21cm}$ lower than expected from
the linear relation. However, these are only a handful galaxies, and the two faintest 
galaxies (with large errors) are instead above the relation. Given the small number of 
galaxies in our sample, this offset is not statistically significant.

\subsection{Star formation rates}
\label{sec:SFRs}
To convert the radio luminosities to SFRs, we use the calibration defined by
\cite{yun2001},

\begin{equation}
{\rm SFR}\; \left[{\msol\; \rm yr^{-1}}\right] = \left(5.9 \pm 1.8\right) \times
 10^{-22}\; L_{\rm 21 cm}\; \left[\rm W\; Hz^{-1}\right] .
\label{eq:SFR}
\end{equation}

\noindent
The derived SFRs range from 0.16--15 $\msol$/yr (Table~\ref{tab:properties}). 
Recall that in principle there might be contamination from
active galactic nuclear (AGN) emission in several galaxies, but we do not try to correct
for that here. Based on the width and intensity of the \oiii\ emission line in the central
fiber of the PPak observations \citepalias{martinsson2013a}, UGC~1908, UGC~4036, UGC~6918,
UGC~8196, and UGC~11318 may harbor an AGN. Among the 15 galaxies for which we have $IRAS$
measurements, any strong contribution from AGN-related radio emission would show up as a
radio-excess object in Fig.~\ref{fig:Radio_FIR_60}. There are no galaxies that show a
major excess of radio emission, with the possible exception of UGC~11318 at
$S_{\rm 21 cm}$$=$22~mJy.
The continuum map of this galaxy, however, does not show a
strong central radio point source. Our results are consistent with the radio-FIR
correlation of \cite{yun2001}, who found that only 1.3\% of the galaxies they selected had
an excess of radio emission.

\subsubsection{Correlations with other global properties}
Apart from the total SFR, we also calculate the SFR per unit $K$-band luminosity, or
specific SFR (sSFR=SFR/$\lk$), and the SFR per unit surface area within $R_{25}$, or SFR
surface density ($\Sigma_{\rm SFR}$). We investigate any correlations between these
star formation properties with the global photometric properties $M_K$, $B$--$K$ color,
and $\mu_{0,K}^{i}$.
We find a clear correlation between SFR and $M_K$, in the sense that more luminous
galaxies have a higher SFR. This is not surprising as more massive galaxies will have
higher absolute star formation. There is also a trend that redder galaxies tend to have a
higher SFR, since the most massive galaxies also tend to be redder. However, when looking
at the relative SFR (sSFR or $\Sigma_{\rm SFR}$) the correlations disappear
\citep[see Fig.~4.16 in][]{martinsson2011}. Interestingly though, there are strong
correlations between all star formation properties and the disk surface brightness
(Fig.~\ref{fig:SFR_mu}). Galaxies with high surface brightness have both higher absolute
and relative star formation, where the strongest correlation is between $\Sigma_{\rm SFR}$
and $\mu_{0,K}^{i}$.
The Spearman rank-order coefficients ($r_s$; calculated following
\citetalias{westfall2014}) are presented in the panels of Fig.~\ref{fig:SFR_mu}.
The linear fits to the correlations, indicated by dashed lines, are given by

\begin{equation}
\log (\rm SFR)          =   0.208 - 0.409 \left[\mu_{0,K}^{i} - 18\right] ,
\end{equation}
\begin{equation}
\log (\rm sSFR)         =   -10.6 - 0.225 \left[\mu_{0,K}^{i} - 18\right] ,
\end{equation}
\begin{equation}
\log (\Sigma_{\rm SFR}) =  -2.72  - 0.415 \left[\mu_{0,K}^{i} - 18\right].
%\label{eq:SFR_mu}
\end{equation}

We have also investigated the same star formation properties with three global properties
of the galaxies involving their gas content \citep[see Fig.~4.17 in][]{martinsson2011}:
gas mass per unit luminosity ($\mhi/\lk$), the extent of the gas disk in terms of $K$-band
disk scale lengths ($\rHI/\hr$), and the average \hone\ column density
($\Sigma_{\rm HI}^{\rm ave}$) within $R_{25}$..
We find no clear trends, except that galaxies with higher gas-mass ratios ($\mhi/\lk$)
tend to have lower total SFR. This, again, is because galaxies with higher $\mhi/\lk$ are
less massive (Fig.~\ref{fig:M_rel}) and therefore have less total star formation.
It should be noted that the lack of correlations that we find in this section are possibly
due to large uncertainties in the star formation rates in combination with the limited
range in our observables.

\begin{figure}
\centering
\includegraphics[width=0.5\textwidth]{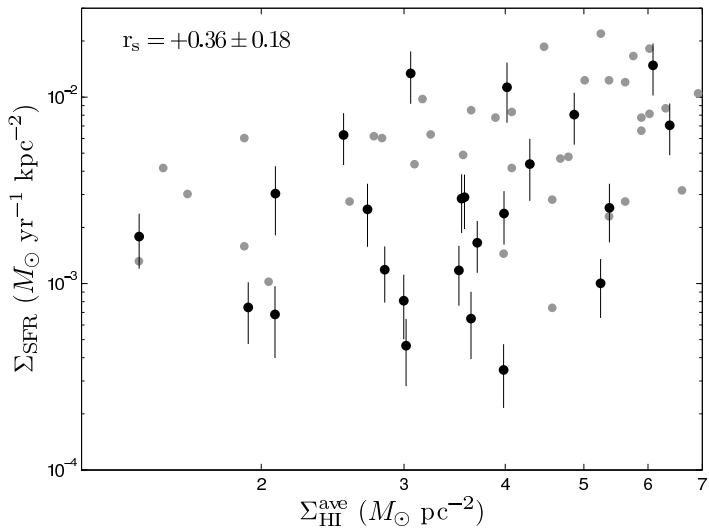}
\caption{Star formation density versus average \hone\ mass surface density for the
galaxies in our sample (black) and the galaxies in \cite{kennicutt1998} (gray).
}
\label{fig:Ken98}
\end{figure}

According to the global Kennicutt-Schmidt law \citep{kennicutt1998}, one would expect a
correlation between the average \hone\ mass surface density in the disk
($\Sigma_{\rm HI}^{\rm ave}$) and $\Sigma_{\rm SFR}$.
However, we do not include the molecular gas in $\Sigma_{\rm HI}^{\rm ave}$, and as
mentioned above the ranges in both parameters are also rather limited.
Figure~\ref{fig:Ken98} shows our data (black symbols) and the data from
\cite{kennicutt1998} (gray symbols, only showing the galaxies in our parameter range)
plotted together.
The Spearman rank-order coefficient of the relation using our data is presented
in the figure, indicating a weak correlation between $\Sigma_{\rm SFR}$ and
$\Sigma_{\rm HI}^{\rm ave}$.
Although similar in distribution, the data from \cite{kennicutt1998}
appear to have a somewhat lower scatter with a stronger indication of a relation.
We can also see that some galaxies in our sample appear to have lower
$\Sigma_{\rm SFR}$ from what one would expect from their $\Sigma_{\rm HI}^{\rm ave}$,
when compared to the results in \cite{kennicutt1998}.
A possible explanation for this could be a selection bias effect. For the DMS, we selected
galaxies to have regular \halp\ kinematics \citepalias{bershady2010a}.
Since one might expect a correlation between gas turbulence and star formation activity,
it is possible that the galaxies in our sample have been selected to have lower
$\Sigma_{\rm SFR}$ than an average galaxy.

% ------------------------------------------------------------------------------
% SFR vs all

% ------------------------------------------------------------------------------
% ========================================================================================

% ========================================================================================
\section{Summary}
\label{sec:Summary_HI}
We have presented the data reduction and a concise analysis of 21 cm spectral-line 
aperture-synthesis observations of 28 spiral galaxies from the DiskMass~Survey. The data
were collected with three different arrays (WSRT, GMRT and VLA), but the final data
products for the various galaxies have similar angular resolutions of $\sim$15$\arcsec$, 
similar velocity resolutions of 8--13~\kms, and similar rms noise levels in the channel
maps of 0.50--0.75 mJy/beam. We have derived a homogeneous suite of data products for all
galaxies, including radio-continuum and total \hone\ maps, global \hone\ profiles, \hone\
velocity fields, rotation curves and warp geometries based on tilted-ring fits,
PV diagrams, and radial \hone\ density profiles. The data products of prime
interest are the extended \hone\ rotation curves and the radial \hone\ column-density
profiles that were used in \citetalias{martinsson2013b} for the purpose of rotation curve 
mass decompositions. The various data products are presented in the accompanying Atlas. 

We find good agreements with reported literature values for both the 21 cm continuum
fluxes and the \hone\ line widths that we measure. We find somewhat lower integrated 
\hone\ fluxes compared to literature values and we ascribe this to the single-dish fluxes
being overcorrected for beam attenuation. We also find good correspondence between the
systemic velocities of the \hone\ disks compared to those from the stellar and \oiii\
measurements from PPak IFU observations \citepalias{martinsson2013a}, with a marginally
significant offset of 2.5~\kms. There is excellent agreement between the position angles
of the kinematic major axes of the \hone\ disks and the stellar and \oiii\ disks.

All galaxies in our sample have \hone\ disks that are more extended than the stellar disks
with an average ratio $\rHI/R_{25}=1.35\pm0.22$, somewhat lower than found earlier in the
literature. The galaxies follow a well-defined \hone\ mass-diameter relation, implying
that the average \hone\ column density within $\rHI$ varies only mildly among the galaxies
in our sample, ranging between 1.5--5 \sdu. However, larger \hone\ disks tend to have
slightly lower column densities on average.
Galaxies with the largest \hone\ disks compared to their stellar disks appear to have
kinematically hotter outer stellar disks, possibly due to dynamical influence on the stars
by the gas disk.

The radial \hone\ mass surface density profiles show very similar behavior among the
galaxies in our sample. After normalizing the amplitudes of the individual profiles and
scaling them radially with $\rHI$, the average profile is surprisingly well fit by an
offset Gaussian. The radius where the peak of the profile occurs is
$R_{\rm \Sigma,max}$$=$0.40$\rHI$, %$
and the Full Width at Half Maximum of the profile is
FWHM=0.85$\rHI$. Combining this result with the tight \hone\ mass-size relation
(Eq.~\ref{eq:Mass_vs_Size}) makes it possible to estimate the radial 
$\Sigma_{\rm HI}$ profile from a measurement of the total \hone\ flux only. This was used
in \citetalias{martinsson2013b} for the galaxies lacking reduced \hone\ data, where we
instead derived the radial $\Sigma_{\rm HI}$ profiles from single-dish measurements
reported in the literature.

From inspections of the observed \hone\ velocity fields, the minor-axis PV diagrams, and
the results from tilted-ring fits at different radii, we find that 17 out of
28 galaxies show an indication for a warp in position angle. Based on the presumed
flatness of the outer rotation curves, we obtain an indirect indication for an inclination
warp in 10 galaxies. The warp often starts at a radius smaller than the size of the optical
disk, which is somewhat in disagreement with earlier studies. %\citep{kruit2007}.
It should be noted, however, that in nearly face-on galaxies, a minor intrinsic warp may
manifest itself by a major twist of the line of nodes. 

Rotation curves were derived from tilted-ring fits to the \hone\ velocity fields, 
following the geometry of a possible warp, and corrected for beam smearing on the
basis of the major-axis PV diagrams. Rotational velocities range from 
130--270 \kms\ and reach out to 10--55 kpc from the galaxy centers, well into the regime
where the galaxy potential is dominated by the dark matter halo. After normalizing the
rotation curves in amplitude and scaling them in radius by $\hr$, all rotation curves have
similar shapes. A notable exception is the rotation curve of UGC~4458, which declines from
the galaxy center out to $4\hr$ beyond which it remains more or less flat.

The 15 galaxies in our sample with $IRAS$ 60~$\mu$m measurements adhere to the 
radio-infrared correlation. Star formation rates were calculated on the basis of the
21 cm continuum fluxes and vary from 0.16 to 15 $\msol$/yr.
%These SFRs were used in \cite{westfall2014}.
The strongest correlation exists between the average star formation surface density within
$R_{25}$ ($\Sigma_{\rm SFR}$) and the central $K$-band disk surface brightness.
% ========================================================================================

% ========================================================================================
% ----------------Acknowledgements--------------------
\acknowledgement{
We give a special thank to Sambit Roychowdhary, then at NCRA in Pune, for helping us carry
out some of the GMRT observations.
We are grateful for the tips we got from Jayaram Chengalur at NCRA and
K.\ S.\ Dwarakanath at RRI in Bangalore regarding the reduction of the GMRT data.
T.P.K.M.\ acknowledges financial support from the Spanish Ministry of Economy and
Competitiveness (MINECO) under grant number AYA2013-41243-P.
He is grateful to the Kapteyn Astronomical Institute for their hospitality as a long-term
guest during the time much of this paper was being written. Part of this work was done at
Leiden Observatory.
M.A.W.V.\ and T.P.K.M.\ acknowledge financial support provided by NOVA, the Netherlands
Research School for Astronomy, and travel support from the Leids Kerkhoven-Bosscha Fonds.
M.A.W.V.\ acknowledges support from the FP7 Marie Curie Actions of the European
Commission, via the Initial Training Network DAGAL under REA grant agreement 289313. 
M.A.B.\ acknowledges support from NSF/AST-1009471.
K.B.W.\ acknowledges grants OISE-754437 (NSF) and 614.000.807 (NWO).
We finally thank the referee for constructive comments and suggestions.
}
% ========================================================================================

% ========================================================================================
% ---------------------References----------------------
\bibliography{./Martinsson-DMS-X}
\bibliographystyle{./aa_DMS_X}
\setlength{\bibsep}{1.3pt}
% -------------------------------------------------------------
% ========================================================================================

% ========================================================================================
\newpage

\appendix
{\small

\section{Notes on individual galaxies}
\label{app:gal}
Here we present a few notes on individual galaxies.
See \citetalias{martinsson2013a} for more detailed notes on the data products
of individual galaxies from PPak, for comments on \oiii\ emission, stellar and
\oiii\ kinematics, kinematic flaring, known supernovae within the galaxies, and for
notes on nearby field stars.
\begin{list}{}
  {
  \settowidth{\labelwidth}{\bf UGC 00000:}
  \setlength{\labelsep}{1em}
  \setlength{\itemsep}{0.2em}
  \setlength{\parskip}{0.2em}
  \setlength{\leftmargin}{\labelwidth}
  \setlength{\rightmargin}{0pt}
  }
  \item[{\bf UGC   448:}] IC 43. Has a significant bulge with the second highest
   bulge-to-disk ratio in the sample ($B/D$$=$$0.32$; \citetalias{martinsson2013a}).
   The rotation curve rises quite sharply, and already reaches the flat part on the
   first measurement. The projected rotation curve indicates an $i$~warp. Large
   elongated beam along minor axis, also seen in the PV diagram. This galaxy has a
   close companion (UGC~449)
   3.5\arcmin\ directly to the north with $\vsys \sim 400$~\kms\ higher than UGC~448.
  \item[{\bf UGC   463:}] NGC 234. PPak and SparsePak data studied in detail in
  \citetalias{westfall2011b}.
  \item[{\bf UGC  1087:}] The projected rotation curve indicates an $i$~warp.
  \item[{\bf UGC  1635:}] IC 208. Has a close companion (UGC~1636) 4\arcmin\ south-east,
  with a $\vsys$ similar to UGC~1635. Rather gas poor, also in \oiii\ gas
  (\citetalias{martinsson2013a}).
  \item[{\bf UGC  3140:}] NGC 1642. Very close to face on with $\itf$$=$$14\arcdeg$.
  The projected rotation curve indicates an $i$~warp.
  There is a small offset between the \hone\ and \halp\ rotation curves
  (\citetalias{martinsson2013b}), possibly due to \hone\ asymmetries.
  \item[{\bf UGC  3701:}] The projected rotation curve indicates an $i$~warp. The least
  luminous galaxy in the reduced \hone\ sample ($M_K=-22.0$). Rotation curve rises slowly.
  \item[{\bf UGC  3997:}] Classified as Im by RC3. Has a close companion (UGC~3990)
  3\arcmin\ south-west, with a $\vsys$ similar to UGC~3997. It has a $\phi$~warp starting
  already at $15\arcsec$. This warp is not seen in the ionized gas kinematics from the
  optical IFU data. There is a small offset between the \hone\ and \halp\ rotation curves
  (\citetalias{martinsson2013b}), maybe due to the warp in position angle, corrected for
  in the \hone\ rotation curve but not in the \halp\ rotation curve.  
  \item[{\bf UGC  4036:}] NGC 2441. Has a close companion (LEDA~21981) 4.5\arcmin\
  south-west, with a $\vsys$ similar to UGC~4036. Its spiral arms can clearly be seen in
  the 21 cm continuum map.
  \item[{\bf UGC  4107:}] Rather typical galaxy in the sample. Has high-quality data for 
  both the 21 cm and optical IFU data (\citetalias{martinsson2013a}).
  \item[{\bf UGC  4256:}] NGC 2532. This galaxy has two close companions
  (SDSS~J081025.21+340015.8 \& SDSS~J081021.17+340158.7) about 4\arcmin\ to the north of
  UGC~4256, with a bridge in the \hone\ gas between the three galaxies. We exclude the
  companion galaxies from any analysis. There is a small offset between the \hone\ and
  \halp\ rotation curves (\citetalias{martinsson2013b}), maybe due to its kinematic
  lopsidedness.
  \item[{\bf UGC  4368:}] NGC 2575. Inclination measurements from our tilted-ring fitting
  indicate that this galaxy has an $i$~warp. However, except for UGC~6869, this is the
  most inclined galaxy in the sample ($\itf=45\arcdeg$), and a correction for this warp
  would change the rotation curve only marginally.
  \item[{\bf UGC  4380:}]  Low-inclination galaxy ($\itf=14\arcdeg$). The projected
  rotation curve indicates an $i$~warp.
  \item[{\bf UGC  4458:}]  NGC 2599, Mrk 389.  Earliest morphological type in our sample
  (Sa), with the largest bulge-to-disk ratio ($B/D$$=$0.72; \citetalias{martinsson2013a}).
  Rotation curve declining from 300~\kms\ to 200~\kms.
  \hone\ data taken from the {\it WHISP} survey.
  \item[{\bf UGC  4555:}] NGC 2649. Large elongated beam aligned with the minor axis,
   affecting the observed inclination. This is seen clearly in the residual map between
   the observed and modeled (using $\itf$) velocity fields.
  \item[{\bf UGC  4622:}] The projected rotation curve indicates an $i$~warp.
  The most distant galaxy in the sample ($\vsys$$=$$12830$ \kms; $D$$=$$178$~Mpc).
  \item[{\bf UGC  6463:}] NGC 3687. Not included in the PPak Sample, but has high quality
  SparsePak data.
  \item[{\bf UGC  6869:}] NGC~3949. High-surface-brightness member of the Ursa Major
  cluster.  It has the largest $\Sigma_{\rm HI}^{\rm max}$ of all galaxies and is the most
  nearby galaxy in the sample at $\vsys=800$~\kms. It is kinematically lopsided.
  This galaxy is more inclined than the rest of the sample and is therefore
  not included in the DMS (or PPak) sample.
  \item[{\bf UGC  6903:}] Barred galaxy with rather low surface brightness. Poorest
  quality of stellar-kinematic data in our sample (one hour observation with PPak;
  \citetalias{martinsson2013a}).
  \item[{\bf UGC  6918:}] NGC 3982. The projected rotation curve indicates an $i$~warp.
  High-surface-brightness member of the Ursa Major cluster. Very high quality kinematic
  data. Classified as a Seyfert~1.9 \citep{veron2006}. Warped and lopsided extension to
  the \hone\ gas; PPak kinematics are regular \citepalias{martinsson2013a}.
  Included in DMS pilot sample as presented in early publications
  \citep{verheyen2004,bershady2005,westfall2009}.
  \item[{\bf UGC  7244:}] NGC 4195. Barred galaxy.
  The receding part of the \hone\ rotation curve rises steeper than the approaching side.
  This is not seen in the \halp\ rotation curve, and results in
  some offsets between the \hone\ and \halp\ rotation curves \citepalias{martinsson2013b}.
  In optical images \citepalias{bershady2010a}, this galaxy looks rather peculiar, with an
  offset bar and bent spiral arms.
  \item[{\bf UGC  7416:}] Barred galaxy. Strong spiral-arm pattern seen in the total
  \hone\ map with higher $\Sigma_{\rm HI}$ in the arm regions, even though this galaxy
  is observed with a relatively large beam. Not included in the PPak Sample.
  \item[{\bf UGC  7917:}] NGC 4662. Has a large bar. Gas-poor galaxy, also in \oiii\ 
  \citepalias{martinsson2013a}.
  \item[{\bf UGC  8196:}]  NGC 4977. Early-type spiral (SAb). Third highest bulge-to-disk
  ratio in the sample ($B/D$$=$0.24; \citetalias{martinsson2013a}). %$
  The projected rotation curve indicates an $i$~warp. Rather poor data, with
  strange behavior in the geometry measurements. Gas-poor galaxy, also in \oiii\ 
  \citepalias{martinsson2013a}. Seems to have a very extended low-surface-brightness
  stellar disk.
  \item[{\bf UGC  8230:}] IC 853. The second earliest morphological type in the sample (Sab).
  Gas-poor galaxy with the lowest $\Sigma_{\rm HI}^{\rm max}$.
  Not included in the PPak sample.
  \item[{\bf UGC  9177:}] One of the most inclined galaxies in the sample
  ($\itf=40\arcdeg$).
  \item[{\bf UGC  9837:}] Very regular kinematics, nicely modeled with low residuals
  between data and model.
  \item[{\bf UGC  9965:}] IC 1132. The projected rotation curve indicates an $i$~warp.
  Close to face on ($\itf=12\arcdeg$). Shows an extreme $\phi$~warp.
  \item[{\bf UGC 11318:}] NGC 6691.  Barred galaxy. Lowest inclination in the sample
  ($\itf=6\arcdeg$), yielding a very low-amplitude projected rotation curve.
\end{list}
% ---------------------------------------------------------------------------------------
% ========================================================================================

% ========================================================================================
\section{Atlas of 21 cm radio synthesis observations}
\label{sec:Atlas}
The Atlas page of every galaxy presents a variety of data products including
two-dimensional maps of the sky, PV diagrams, the global \hone\ line profile,
the \hone\ column density profile, panels characterizing the \hone\ kinematics and 
rotation curve, and a table containing some results and contour levels for the various
maps.

\subsection{The maps}
There are six maps for every galaxy and all maps for a particular galaxy are on the same
scale, showing the same area of the sky. The angular dimension of the maps, however,
varies between different galaxies depending on the size of their \hone\ disk. The upper
row of three maps shows an optical image taken from the blue POSS-II plates (left), the
21 cm radio continuum map (middle), and a velocity-integrated total \hone\ map (right).
The lower row of three maps shows the observed \hone\ velocity field (left), the modeled
velocity field (middle), and the residual map as the difference between the observed and
model velocity fields (right). In all maps, we have marked the adopted morphological
center with a small white cross. The FWHM dimension of the synthesized beam is indicated
in the lower left corners of the continuum, total \hone, velocity field and residual maps.

In the optical image, we indicate the isophotal radius at a blue surface brightness level
of 25 mag/arcsec$^2$ ($R_{25}$) as reported by NED, projected to an ellipse using our
adopted position angle in the inner part of the galaxy ($\pa$) and inclination from the
inverse Tully-Fisher relation ($\itf$). The morphological type is shown in the
upper left corner. All optical images have the same gray-scale stretch to emphasize the
different surface brightness levels of the galaxies.

The continuum map shows the distribution of the 21 cm continuum flux with the same
gray-scale stretch for all galaxies, but with contour levels dependent on the rms noise in
the image ($\sigma_{\rm cont}$). The contour levels are drawn at 2, 4, 8, 16, 32, 64,
128$\times \sigma_{\rm cont}$, where $\sigma_{\rm cont}$ is reported in the Table in the
bottom right corner of every Atlas page.

All \hone\ column-density maps have been reproduced with the same gray-scale stretch for
all galaxies, with contour levels showing 1, 3, 5, 7, 10, 15, 20, 25, 30, 35 
$\msol \rm pc^{-2}$. The thicker outer contour indicates 1~$\msol$~pc$^{-2}$.
The bar in the upper left corner of the \hone\ maps indicates 10 kpc based on the distance
tabulated in \citetalias{bershady2010a} (except for UGC~6869 for which the distance was
taken from NED).

The observed and model velocity fields of a galaxy have identical gray scales with a
stretch that depends on the width of the \hone\ line ($W_{20}$). White and black
isovelocity contours show approaching and receding velocities, respectively. The thick,
black line corresponds to the systemic velocity of the galaxy ($\vsys$). The spacing
between the isovelocity contours is reported in the Atlas Table.

The residual map shows the difference between the observed and model velocity fields
with white contours showing negative residuals and black contours showing positive
residuals. The zero-residual contour is omitted. The velocity intervals between the
residual contours are reported in the Table.

\subsection{Panels describing the HI kinematics}
The three panels to the right of the maps show the measured and adopted systemic 
velocity and orientation of the \hone\ gas disk as a function of radius, based on the
outcome from our tilted-ring modeling of the observed \hone\ velocity fields
(Sect.~\ref{sec:geometry}). The errorbars are the formal uncertainties as reported by
the tilted-ring fitting routine.

The upper panel shows the best-fitting $\vsys$ of the gas kinematics in every ring. The
velocity range of 30 \kms\ centered on the average $\vsys$ is the same for every
galaxy. The solid, horizontal line indicates the weighted average of the measured
$\vsys$ values, which was adopted as the $\vsys$ of the galaxy and forced to be constant
with radius in subsequent fits. In some galaxies, a significant trend of $\vsys$ with
radius seems to be present, but this could be the results of kinematic lopsidedness or an
asymmetric warp and we have ignored such trends.

The middle panel shows the best-fitting position angle of the kinematic major axis on the
receding side of the galaxy. The solid line indicates the adopted position angle which,
after being constant at $\pa$ in the center, often has a constant slope at larger radii to
account for a $\phi$~warp. In three cases (UGC~3140, UGC~6918 and UGC~8196) we change the
slope at larger radii to follow the trend of measured position angles.

The lower panel shows formal measurements of the best-fitting inclinations ($i$) of the
rings as a function of radius. The large errorbars reflect the fact that in most cases it
is impossible to obtain reliable measurements. We therefore adopted inclinations based on
the Tully-Fisher relation. For ten of our galaxies we have introduced an $i$~warp. The 
adopted $i$ as a function of radius is indicated by the solid line. The two dashed lines
show the estimated error on the $\itf$ inclination. In the cases of an $i$~warp, we
use the same absolute error on the inclination at all radii. 

The square panel below the geometry panels shows the derived rotation curve. The small
crosses indicate the best-fitting circular velocity of the \hone\ gas in every tilted ring
as a function of radius, following the geometry of the observed \hone\ velocity field as
indicated by the panels described above. The filled and open circles in this panel show
the beam-corrected rotation curve on the receding and approaching sides of the galaxy as
motivated by the PV diagrams. The solid curve going through the midpoints
of the solid and open circles shows the beam-corrected rotation curve 
(Sect.~\ref{sec:RCbeamcor}) that we have adopted. The dashed horizontal line indicates
the expected $\vmax$ based on the corrected width of the global \hone\ line 
($W_{20}^{\rm cor}/2$). The solid horizontal line shows the expected flat rotation curve
based on the absolute $K$-band magnitudes derived from the 2MASS images and the 
$M_K$-$V_{\rm flat}$ TF-relation (\citetalias{martinsson2013a}).

\subsection{Position-velocity diagrams}
The two PV diagrams show slices along the kinematic minor (left) and major (right) axes of
the galaxy, following the constant position angle $\pa$ as defined by the weighted average
of the inner points inside the radius $R=35\arcsec$. The contours show 2, 4, 6, 9, 12, 15,
20, 25 times the noise in the PV diagrams ($\sigma_{\rm pvd}$), where $\sigma_{\rm pvd}$
can be found in the Atlas Table. The dashed horizontal and vertical lines indicate $\vsys$
and the adopted dynamical center of the galaxy. The width of the \hone\ profile is
indicated with the two horizontal arrows at $\vsys \pm (W_{20}/2)$. In these diagrams we
plot with small crosses the projected, tilted-ring based rotation curve, while the solid
and open circles indicate the projected rotation curves corrected by eye for the effects
of beam smearing. The cross in the lower right corners indicates the velocity resolution
and beam size at the position angle of the slice.

\subsection{Radial HI column density profile and global HI profile}
The radial \hone\ mass surface density profile shows the average \hone\ mass surface
density ($\Sigma_{\rm HI}$) in each 10$\arcsec$-wide tilted ring, following their 
orientations as described above, corrected to face-on. The solid line indicates the 
average surface density in the entire ring while the filled and open circles show the 
average density in the receding and approaching halves of the ring, respectively. The 
optical and \hone\ radii of the galaxies ($R_{25}$ and $\rHI$) are indicated by vertical
arrows for an easy comparison; $\rHI$ is always at a larger radius than $R_{25}$.

The global \hone\ profile shows the flux density as a function of heliocentric velocity.
$\vsys$ is indicated with a vertical dashed line, while $W_{20}$ is indicated with a
horizontal dotted line.

\subsection{Atlas table}
The table in the lower right lists the coordinates of the adopted morphological center 
(RA \& Dec), systemic velocity ($\vsys$), position angle in the inner region ($\pa$), and
 the adopted inclination ($\itf$). It presents the measured total continuum flux 
density ($S_{\rm 21cm}$), the \hone\ peak flux density in the global profile 
($S_{\rm HI,max}$), the integrated \hone\ flux ($\int S_{\rm HI} {\rm d}V$), and the peak
of the \hone\ mass surface density profile ($\Sigma_{\rm HI,max}$). It also lists the rms
noise levels in the continuum map and PV diagrams ($\sigma_{\rm cont}$
\& $\sigma_{\rm pvd}$), as well as the isovelocity contour levels in the velocity fields.
Finally, it includes the observed width of the \hone\ line ($W_{20}$), the size of the
\hone\ disk ($R_{\rm HI}$), the beam size, and the velocity resolution of the observation.
}

% ------------------------ Atlas -----------------------
% HI ATLAS

\onecolumn

\subsection{The Atlas}

 \begin{figure}[!b]
 \centering
 \includegraphics[width=1.0\textwidth]{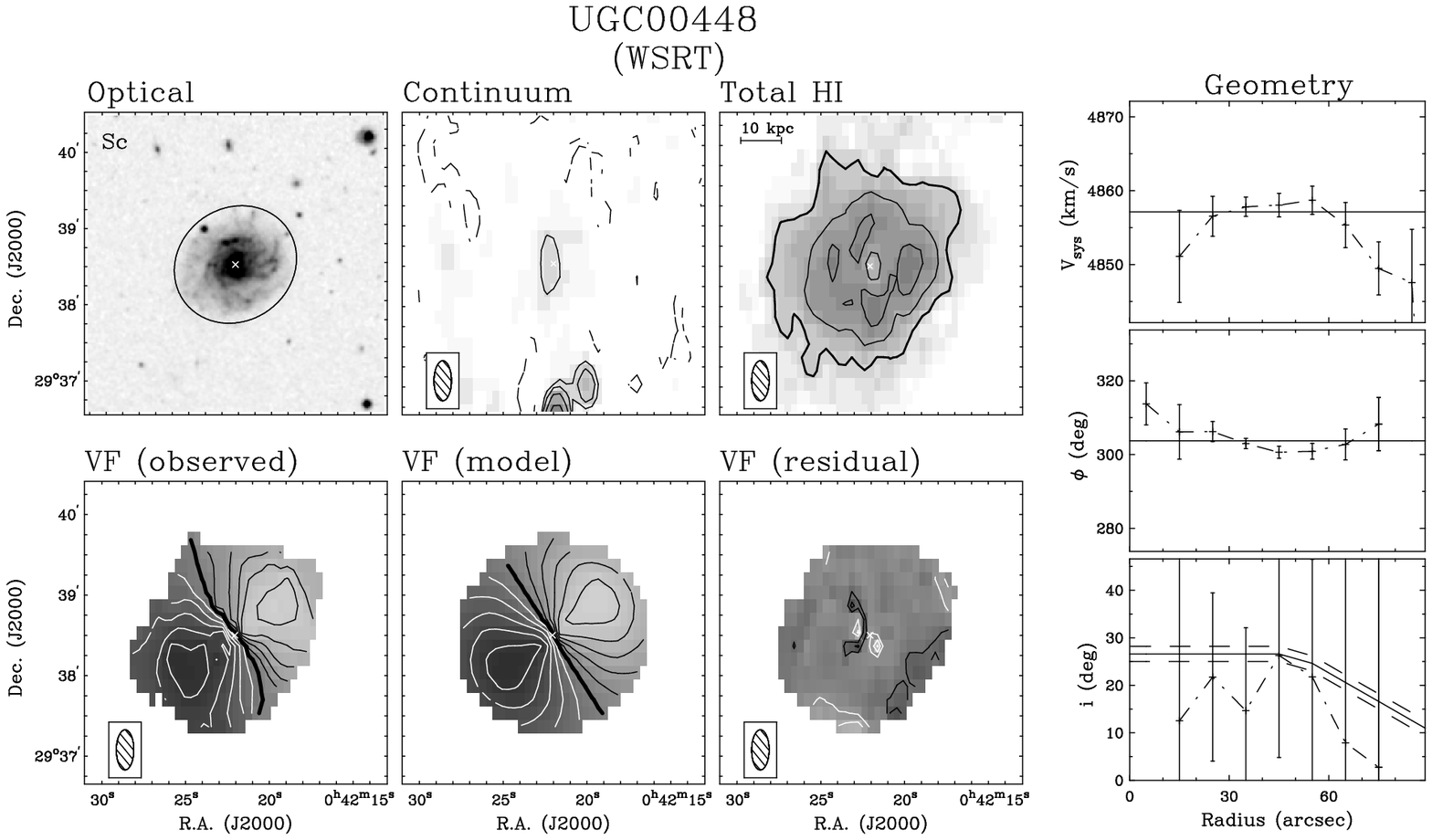}
 \end{figure}

 \begin{figure}[!b]
 \centering
 \includegraphics[width=1.0\textwidth]{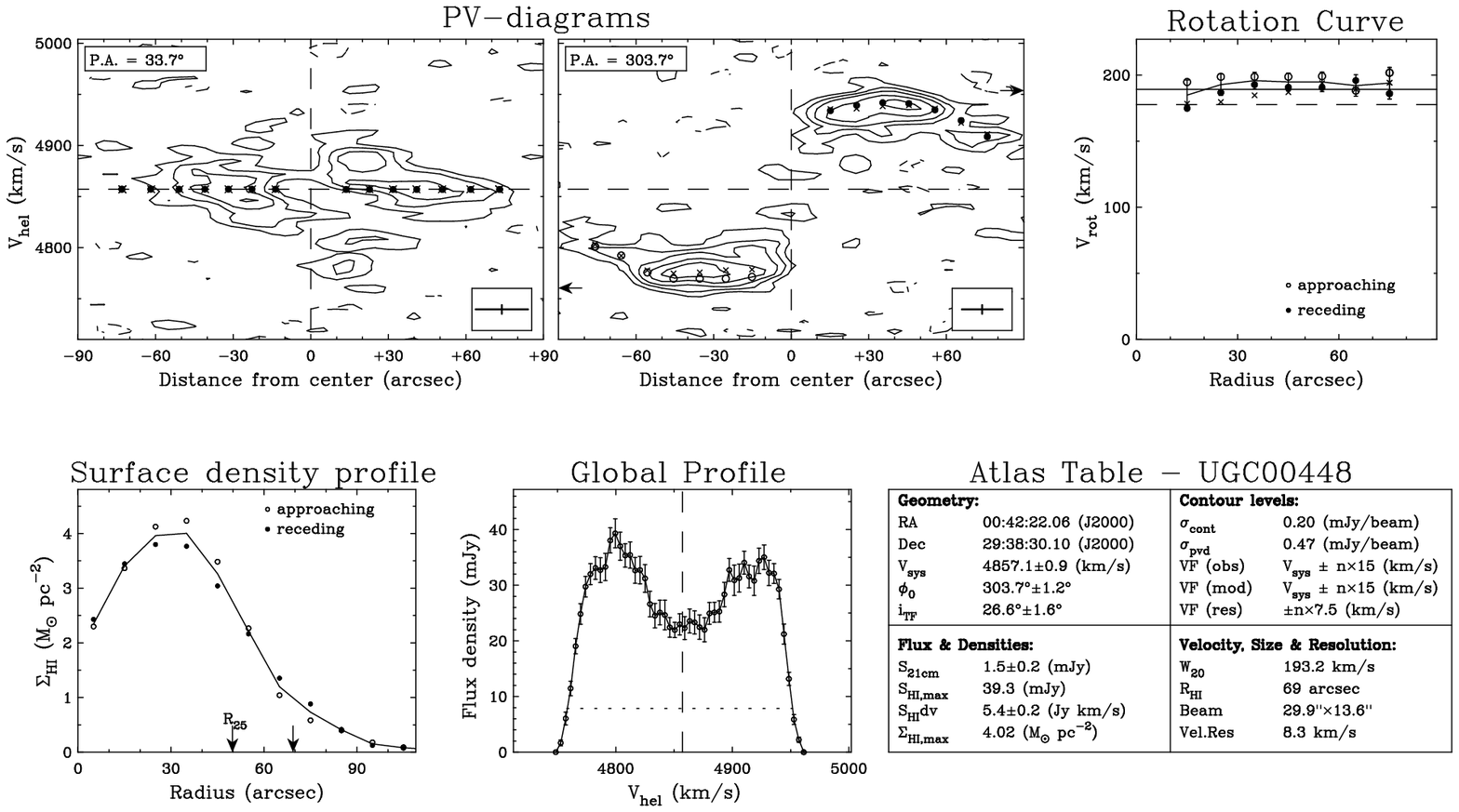}
 \end{figure}

 \begin{figure}
 \centering
 \includegraphics[width=1.0\textwidth]{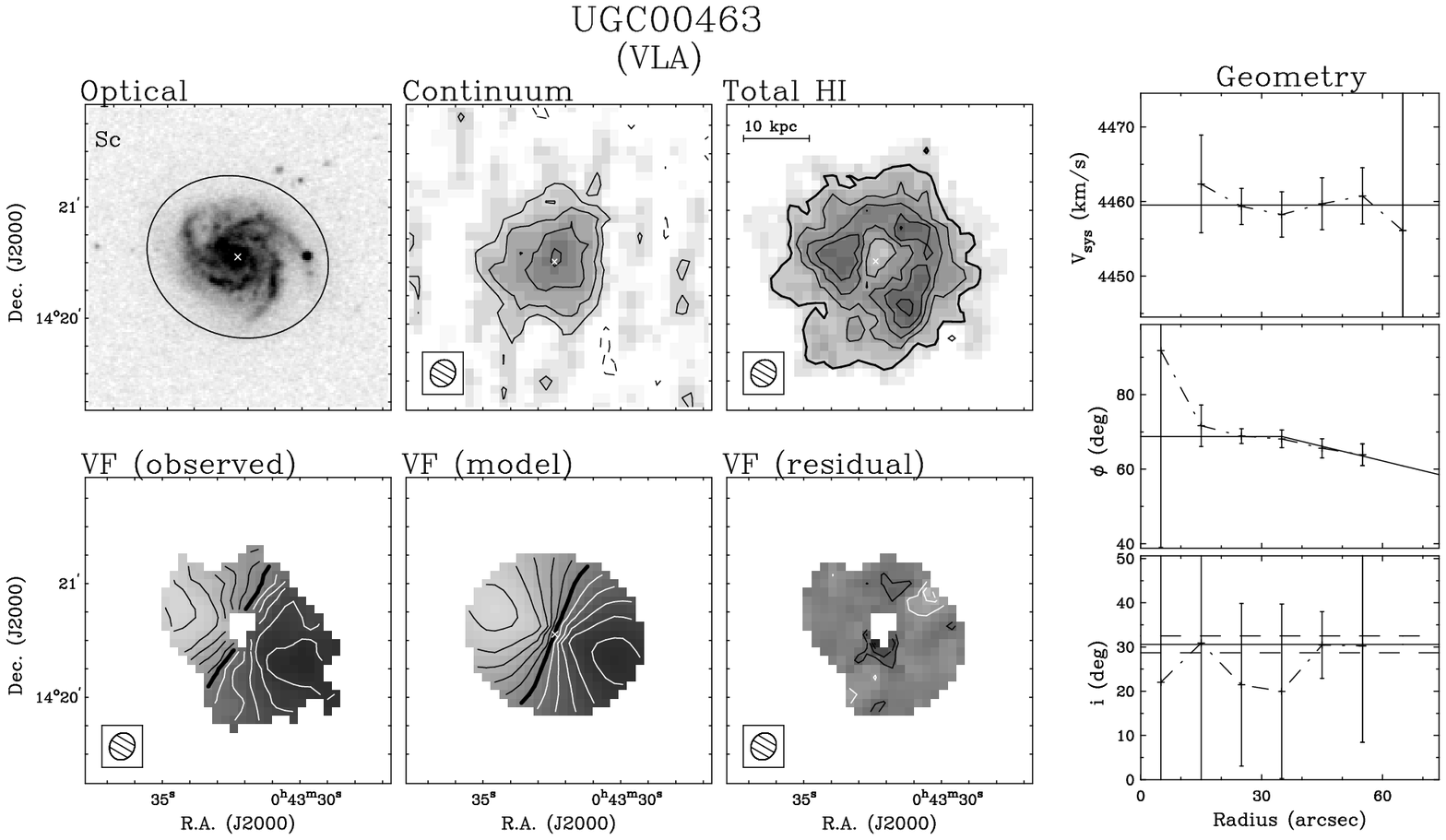}
 \end{figure}

 \begin{figure}
 \centering
 \includegraphics[width=1.0\textwidth]{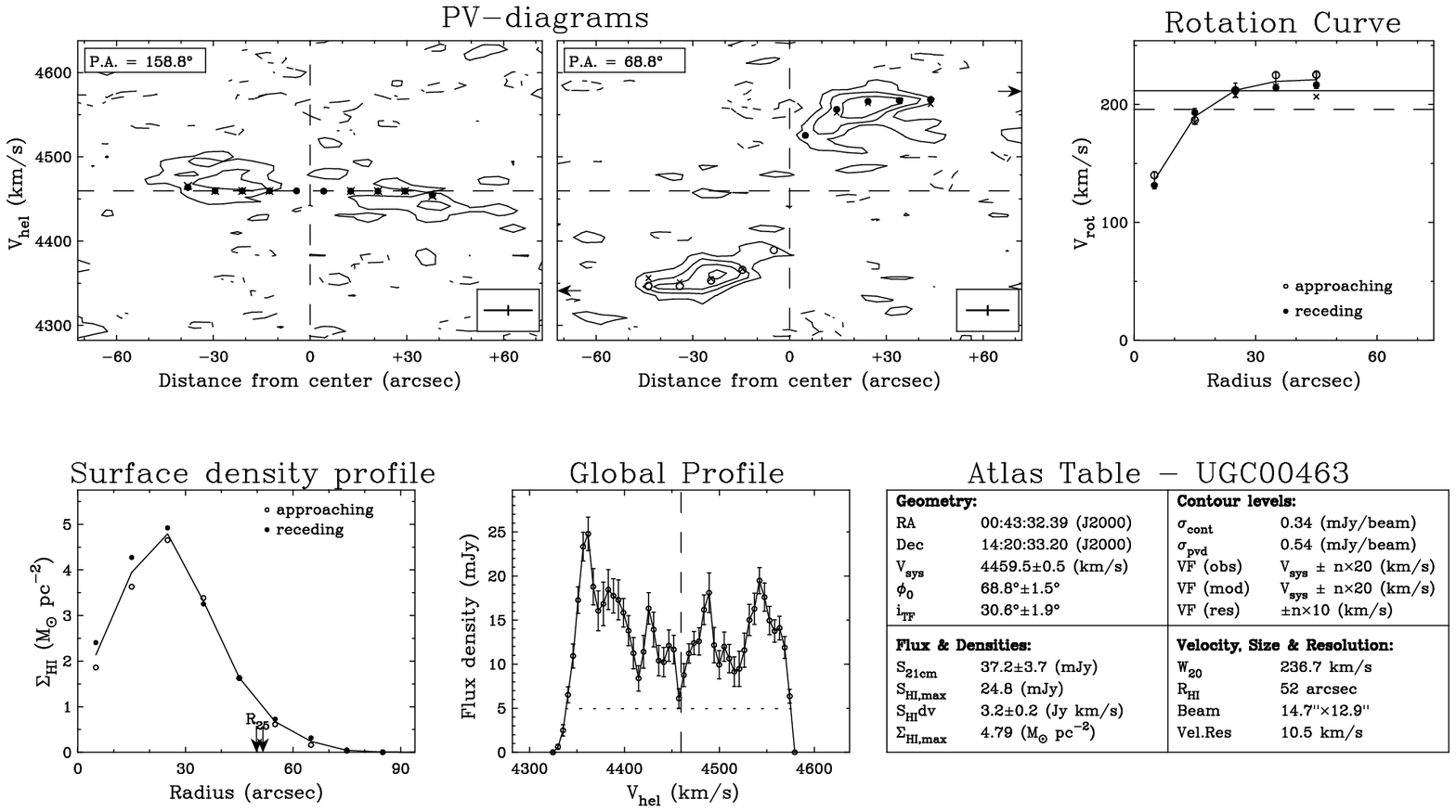}
 \end{figure}

 %\begin{figure}
 %\centering
 %\includegraphics[width=1.0\textwidth]{HIAtlas/UGC01081}
 %\caption{UGC01081}
 %\label{fig:ppak}
 %\end{figure}

 \begin{figure}
 \centering
 \includegraphics[width=1.0\textwidth]{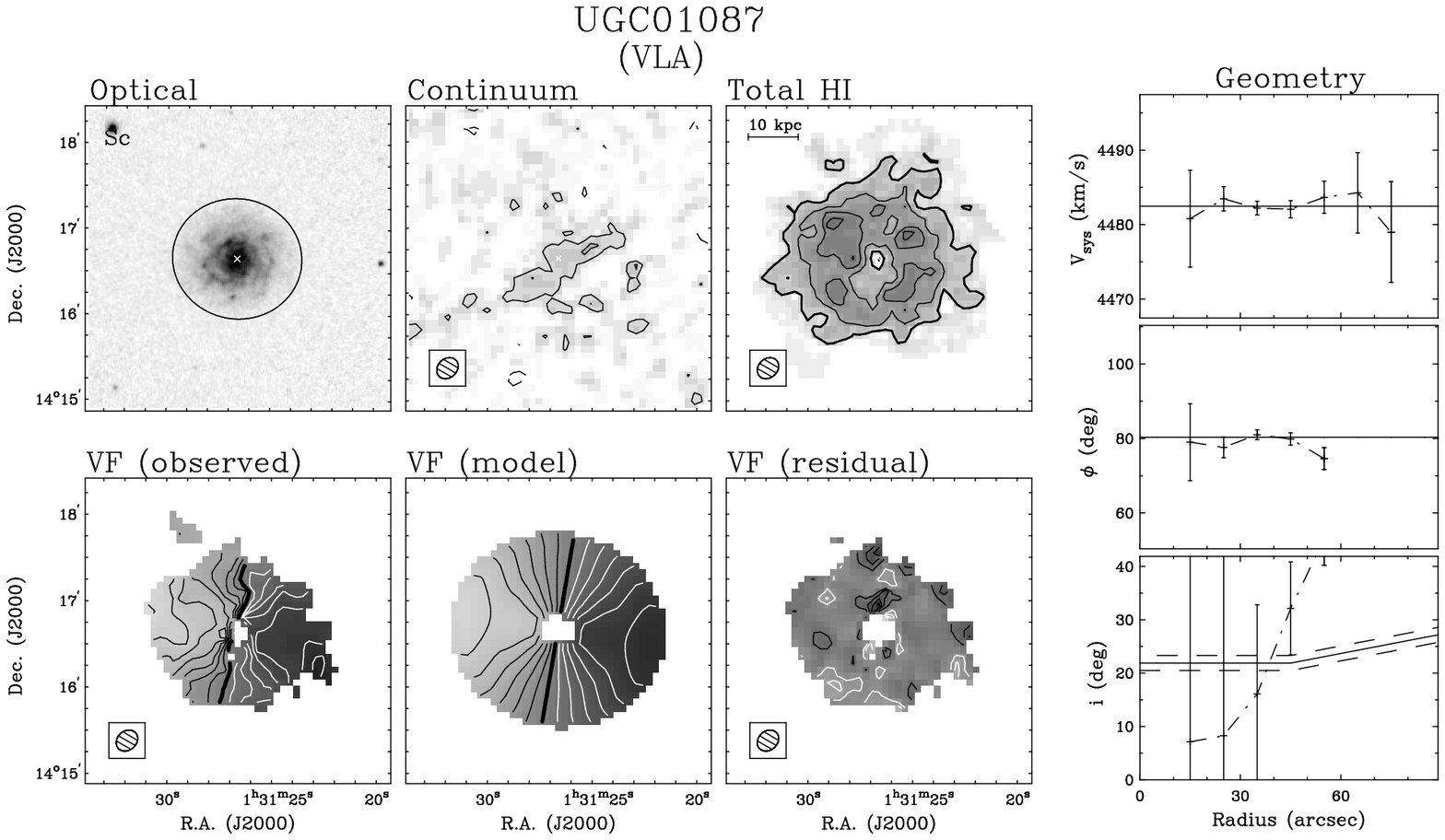}
 \end{figure}

 \begin{figure}
 \centering
 \includegraphics[width=1.0\textwidth]{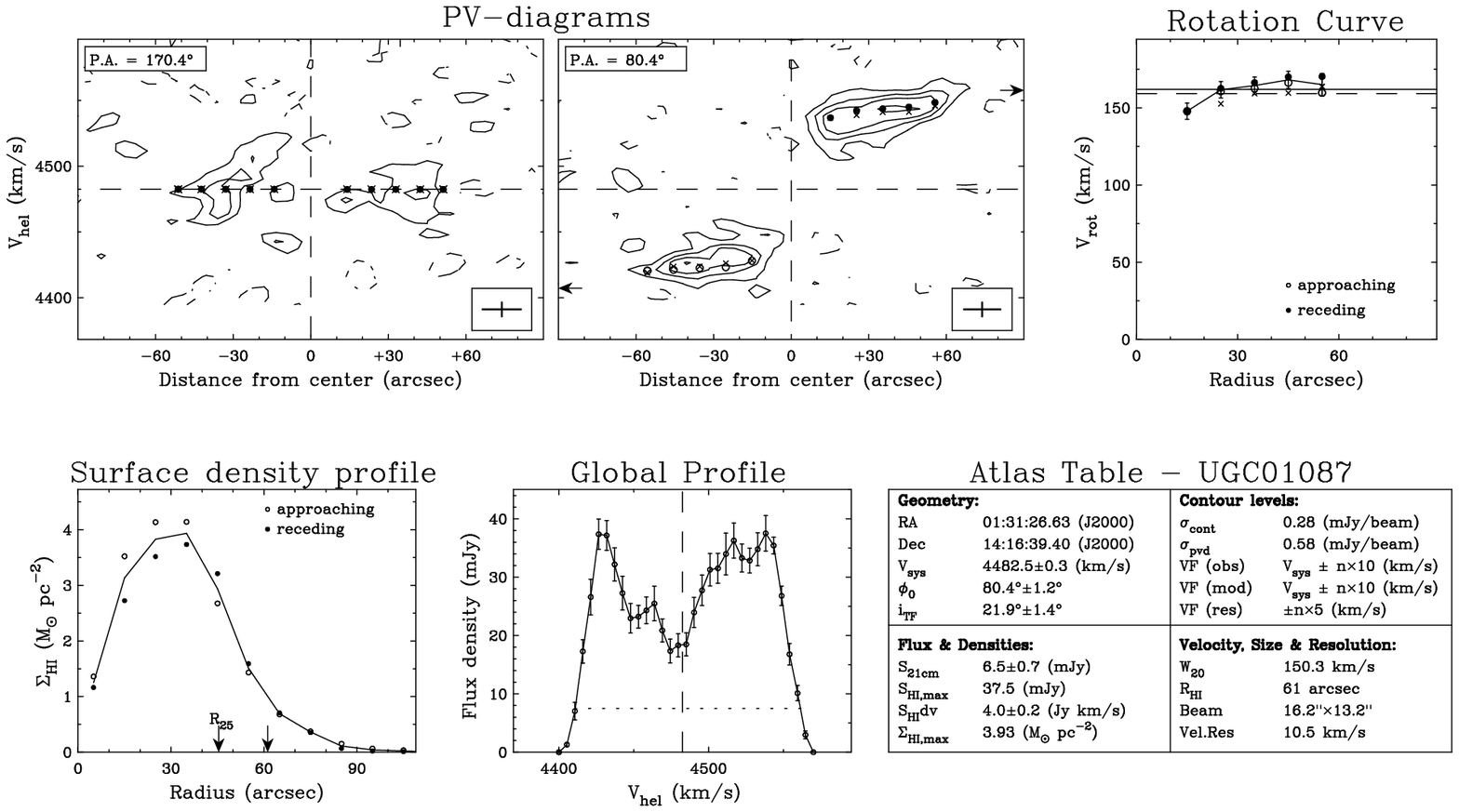}
 \end{figure}

 %\begin{figure}
 %\centering
 %\includegraphics[width=1.0\textwidth]{HIAtlas/UGC01529}
 %\caption{UGC01529}
 %\label{fig:ppak}
 %\end{figure}

 \begin{figure}
 \centering
 \includegraphics[width=1.0\textwidth]{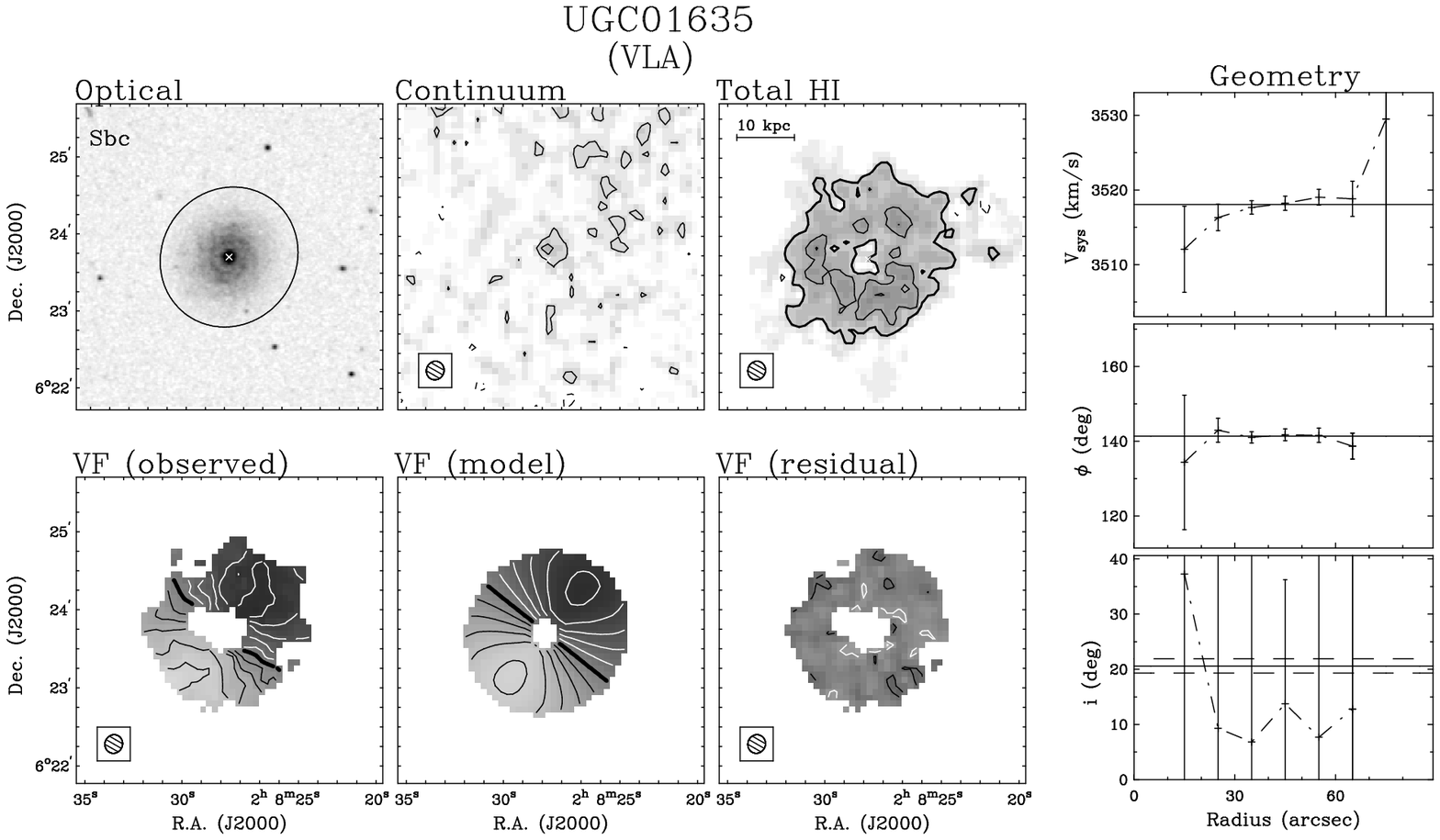}
 \end{figure}

 \begin{figure}
 \centering
 \includegraphics[width=1.0\textwidth]{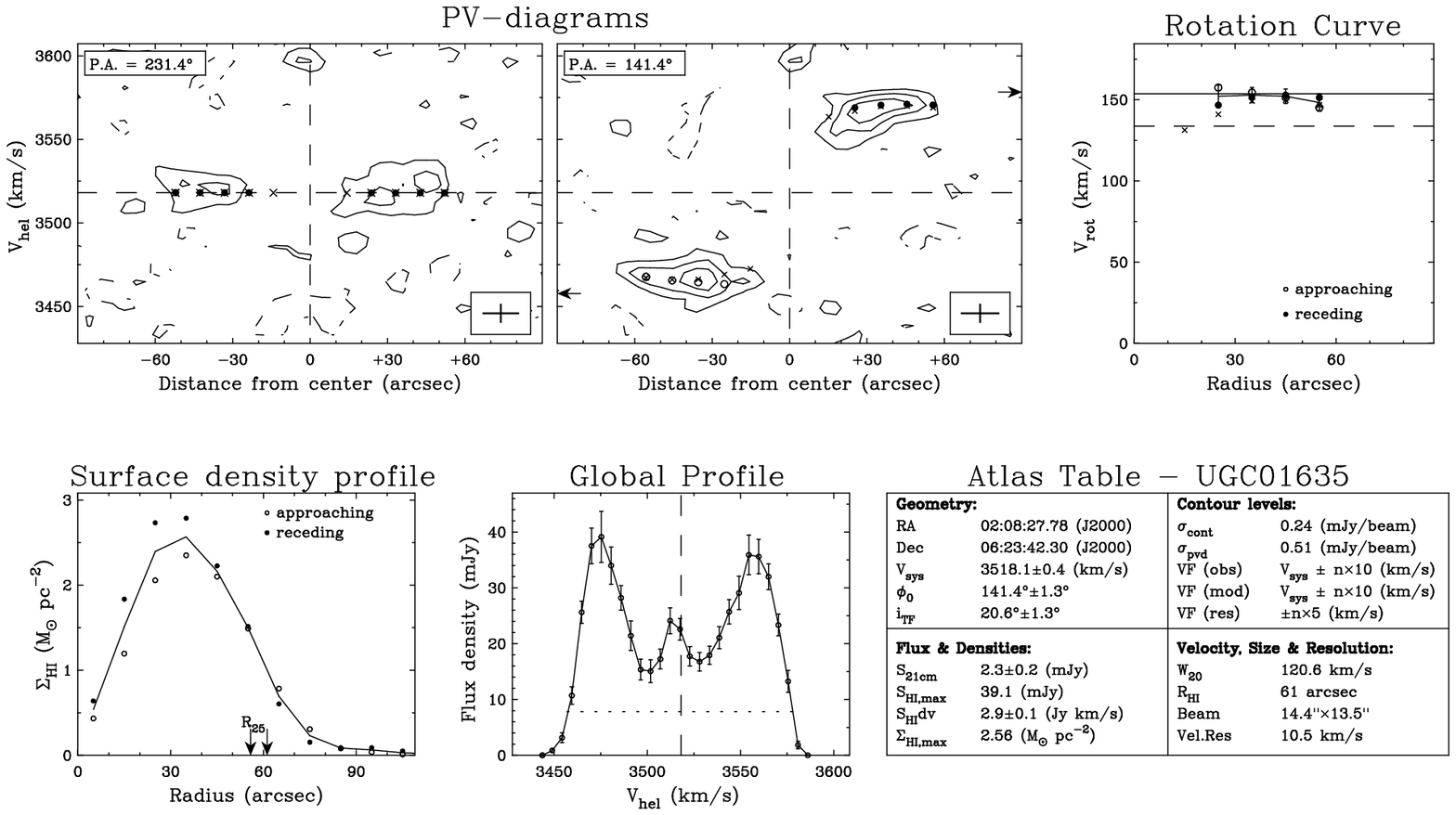}
 \end{figure}

 %\begin{figure}
 %\centering
 %\includegraphics[width=1.0\textwidth]{HIAtlas/UGC01862}
 %\caption{UGC01862}
 %\label{fig:ppak}
 %\end{figure}

 %\begin{figure}
 %\centering
 %\includegraphics[width=1.0\textwidth]{HIAtlas/UGC01908}
 %\caption{UGC01908}
 %\label{fig:ppak}
 %\end{figure}

 %\begin{figure}
 %\centering
 %\includegraphics[width=1.0\textwidth]{HIAtlas/UGC03091}
 %\caption{UGC03091}
 %\label{fig:ppak}
 %\end{figure}

 \begin{figure}
 \centering
 \includegraphics[width=1.0\textwidth]{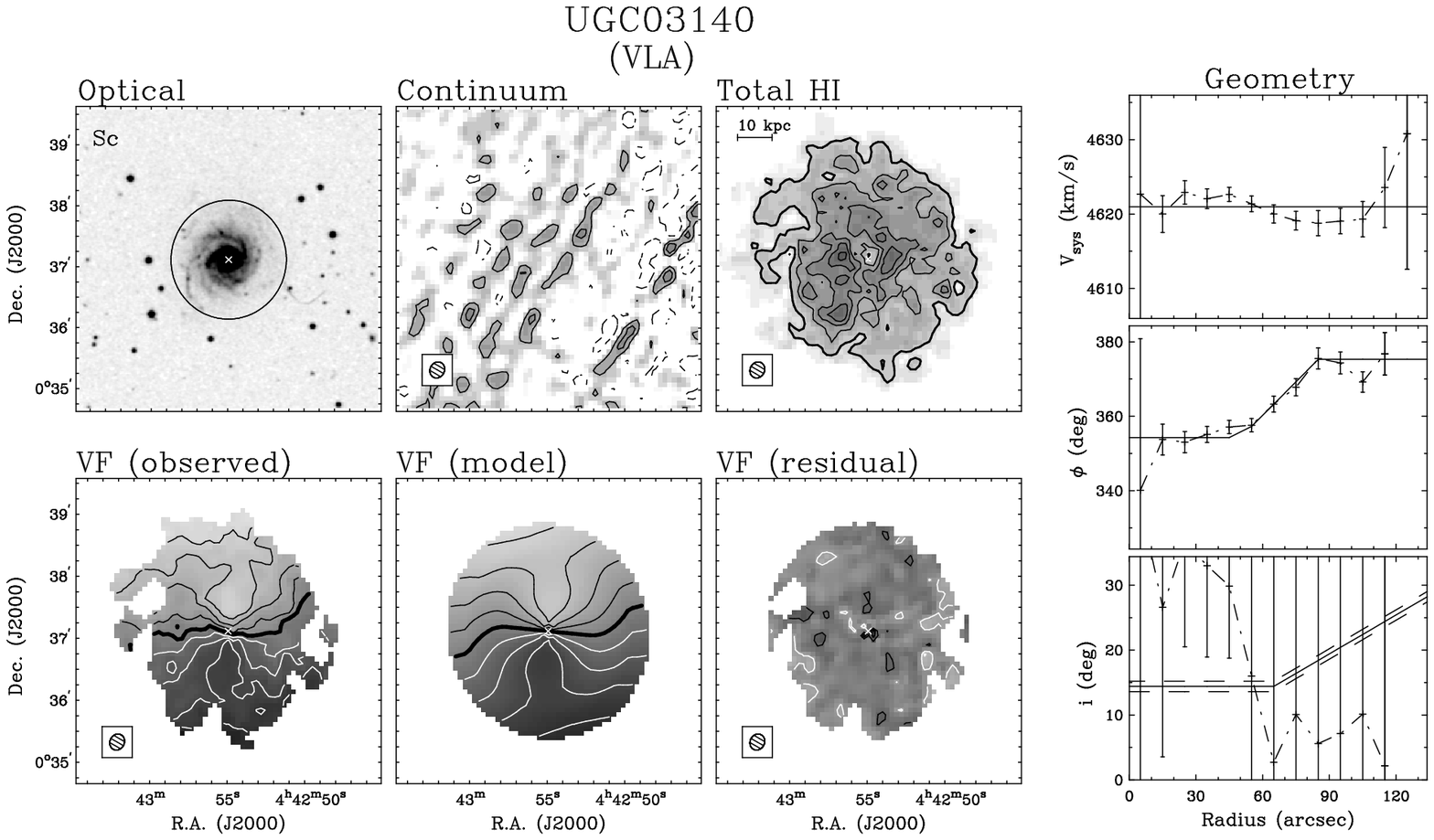}
 \end{figure}

 \begin{figure}
 \centering
 \includegraphics[width=1.0\textwidth]{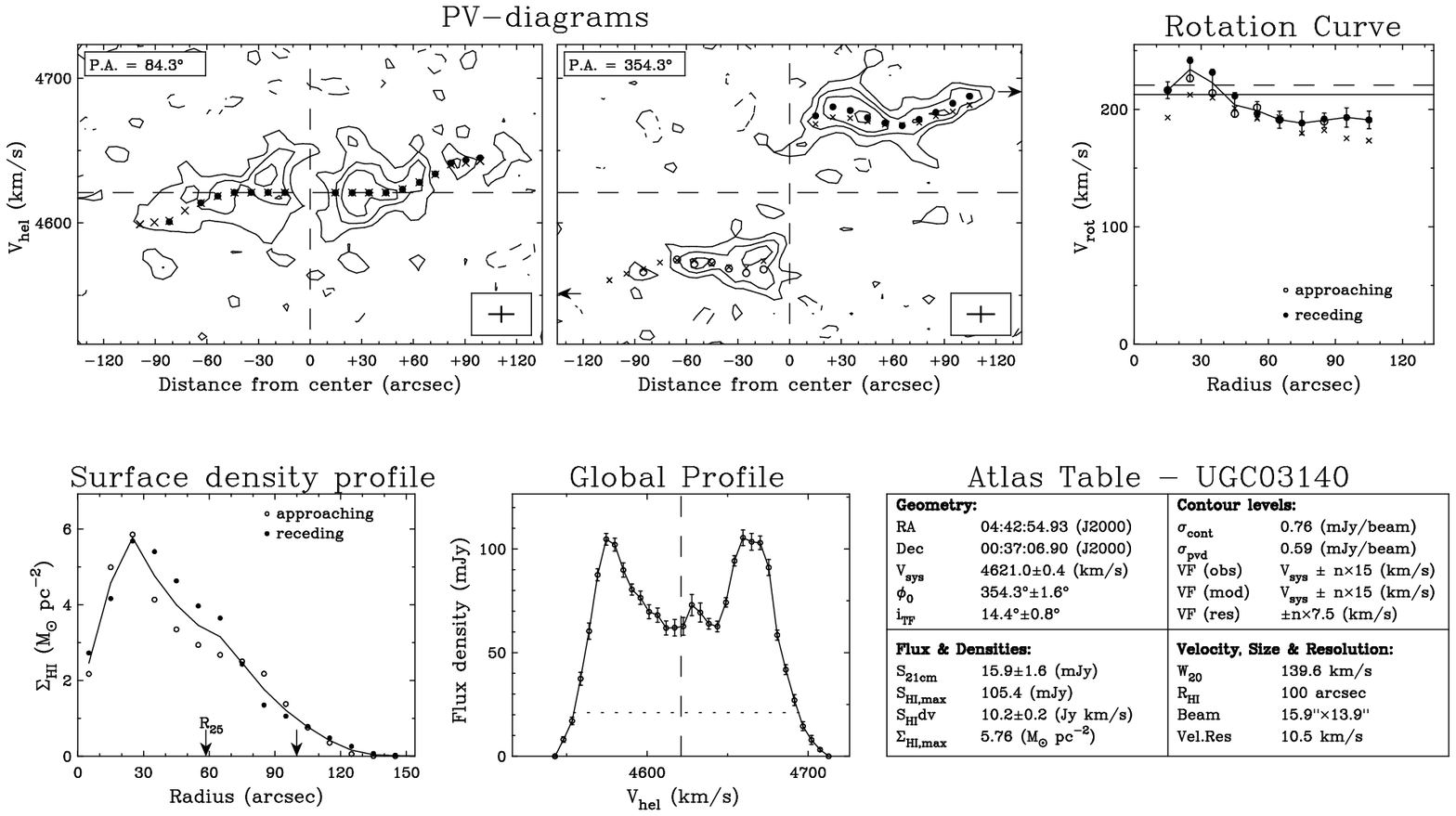}
 \end{figure}

 \begin{figure}
 \centering
 \includegraphics[width=1.0\textwidth]{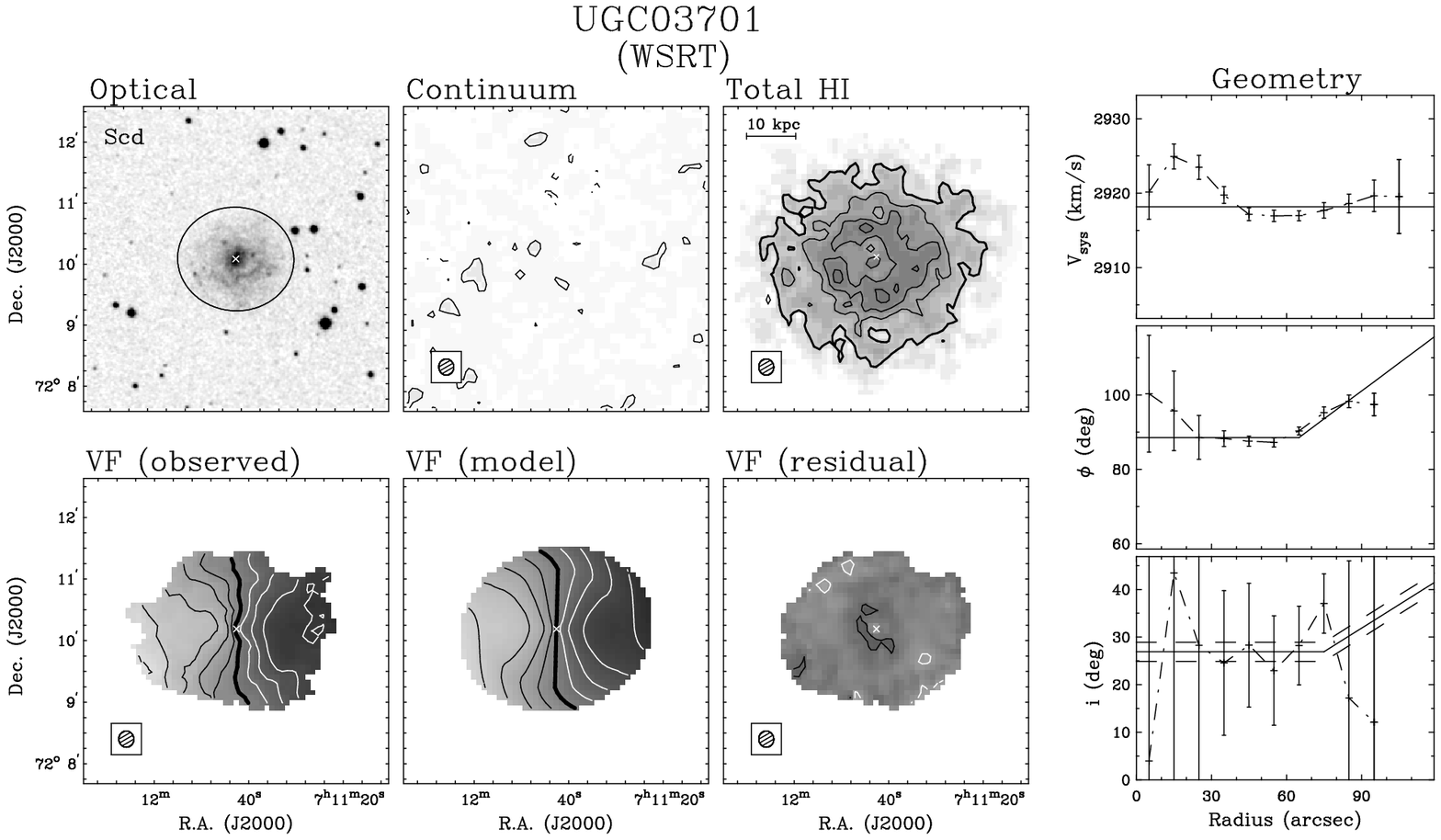}
 \end{figure}

 \begin{figure}
 \centering
 \includegraphics[width=1.0\textwidth]{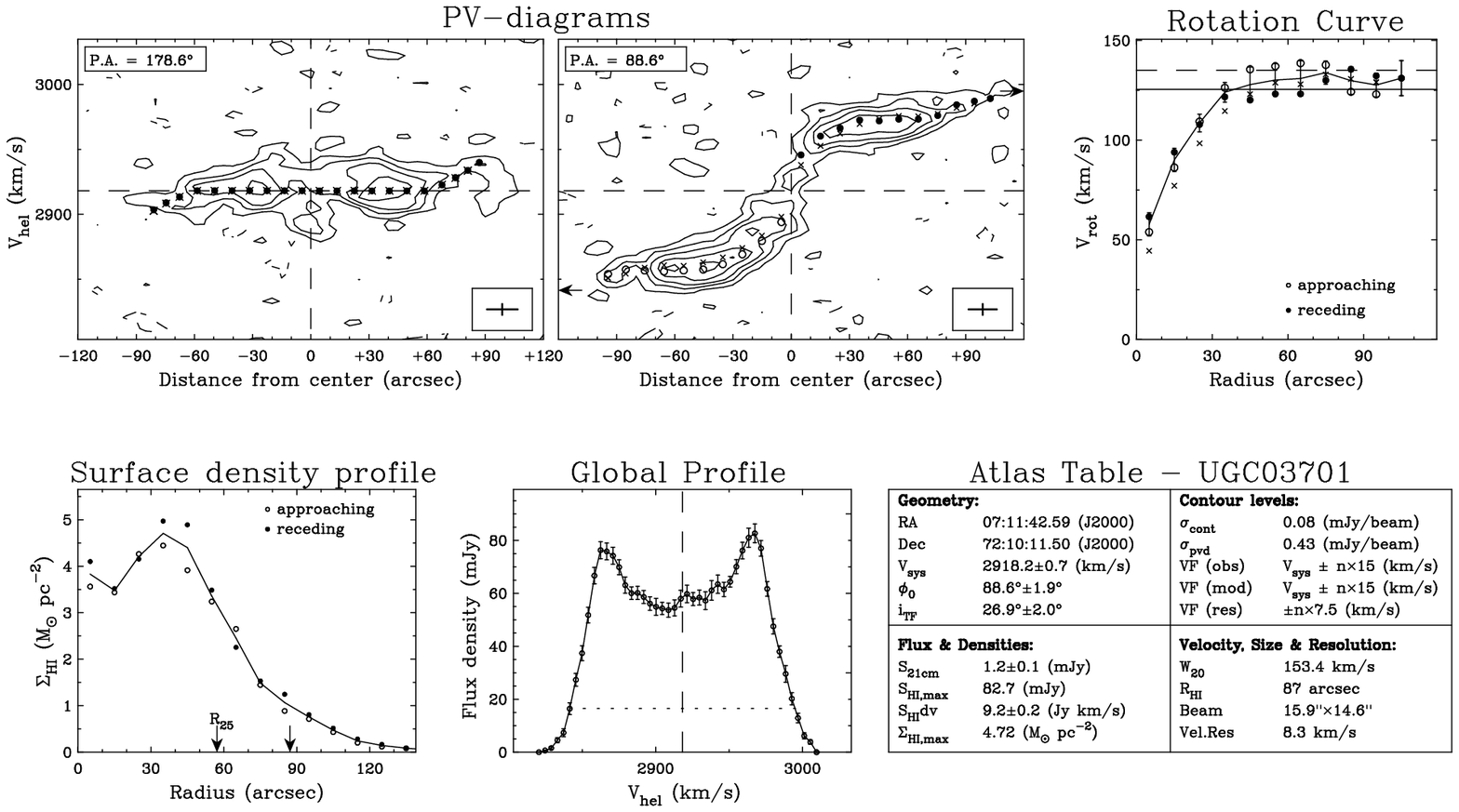}
 \end{figure}

 %\begin{figure}
 %\centering
 %\includegraphics[width=1.0\textwidth]{HIAtlas/UGC03701.3.1.ps}
 %\end{figure}

 %\begin{figure}
 %\centering
 %\includegraphics[width=1.0\textwidth]{HIAtlas/UGC03701.3.2.ps}
 %\end{figure}

 \begin{figure}
 \centering
 \includegraphics[width=1.0\textwidth]{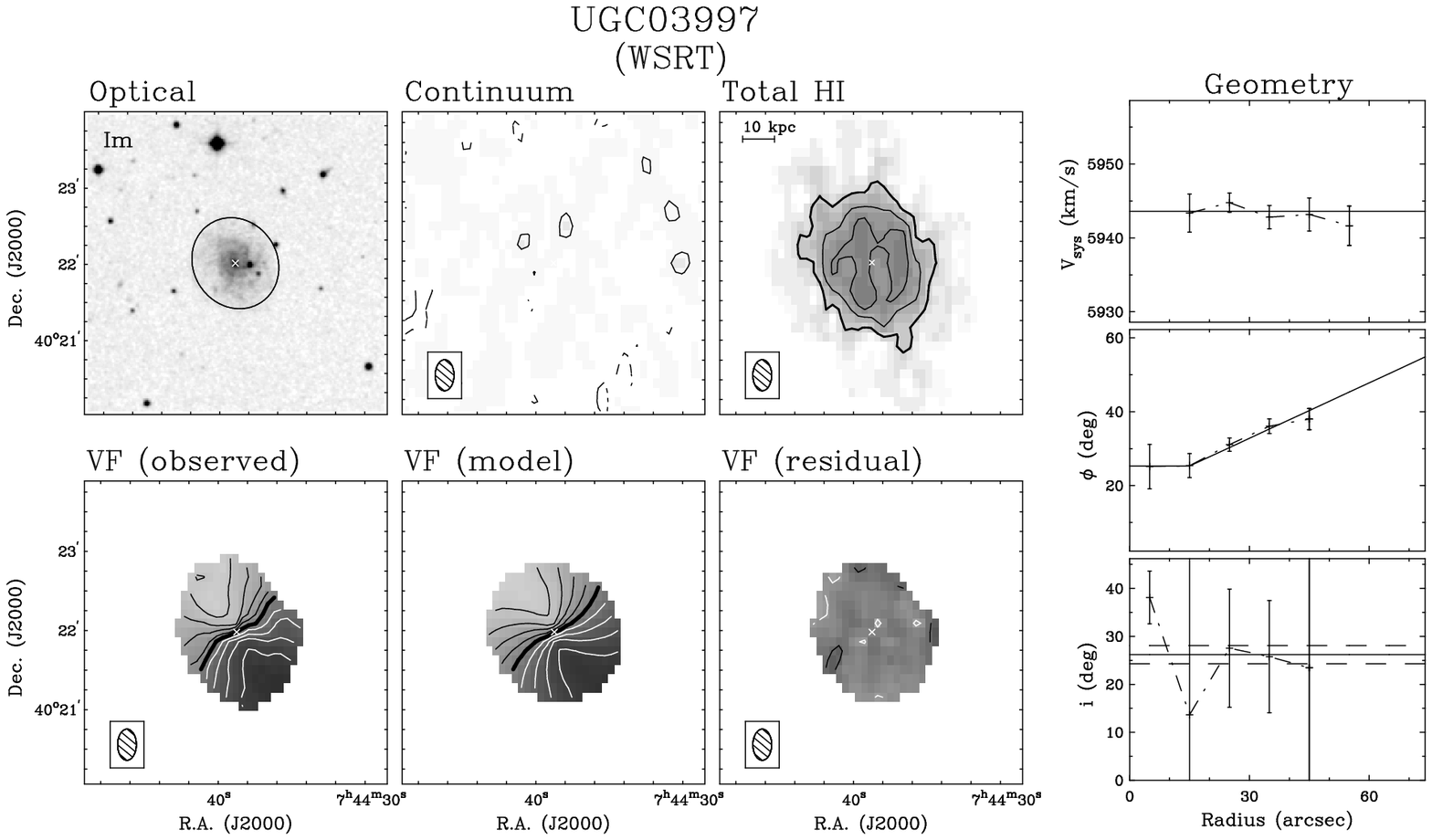}
 \end{figure}

 \begin{figure}
 \centering
 \includegraphics[width=1.0\textwidth]{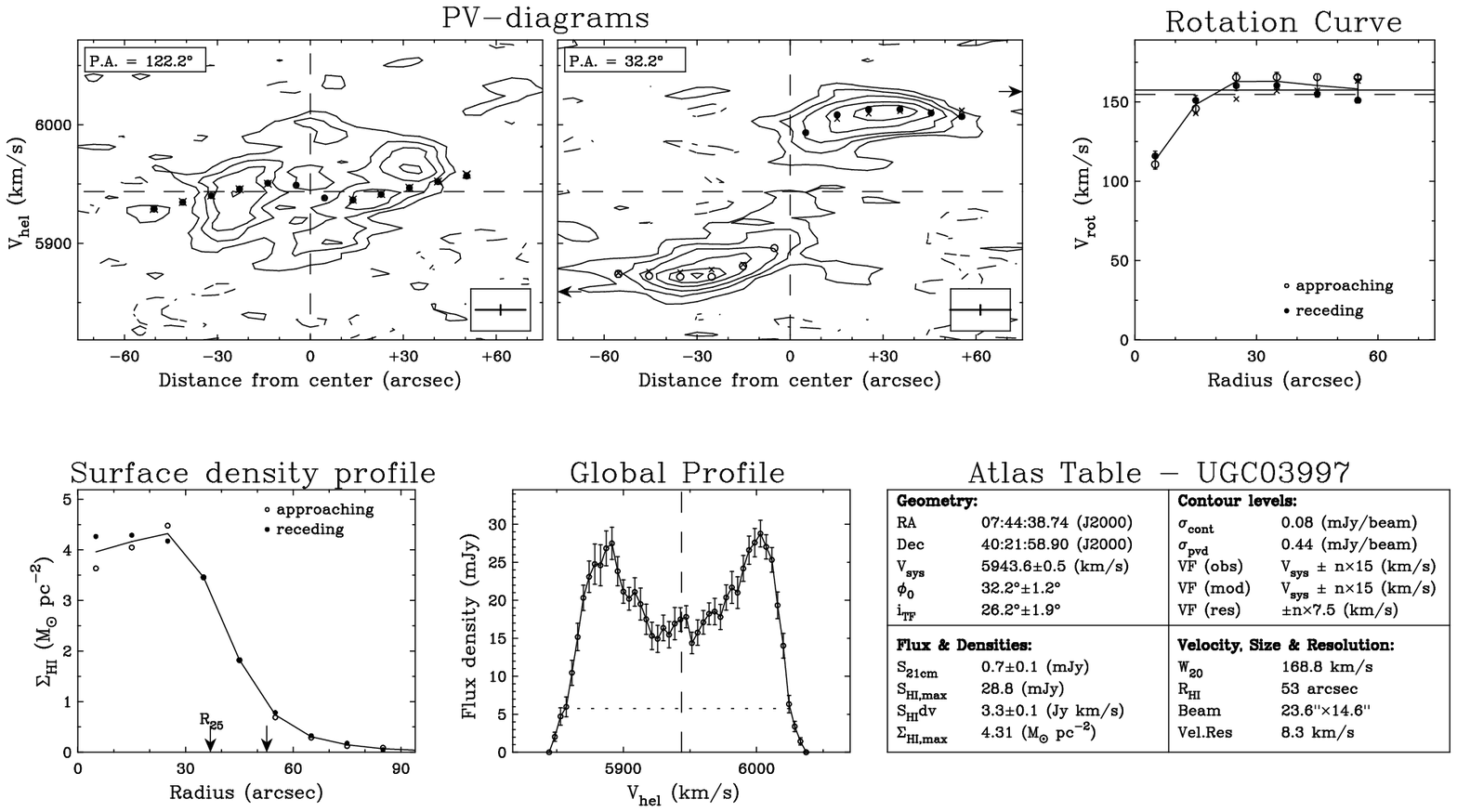}
 \end{figure}

\clearpage

 \begin{figure}
 \centering
 \includegraphics[width=1.0\textwidth]{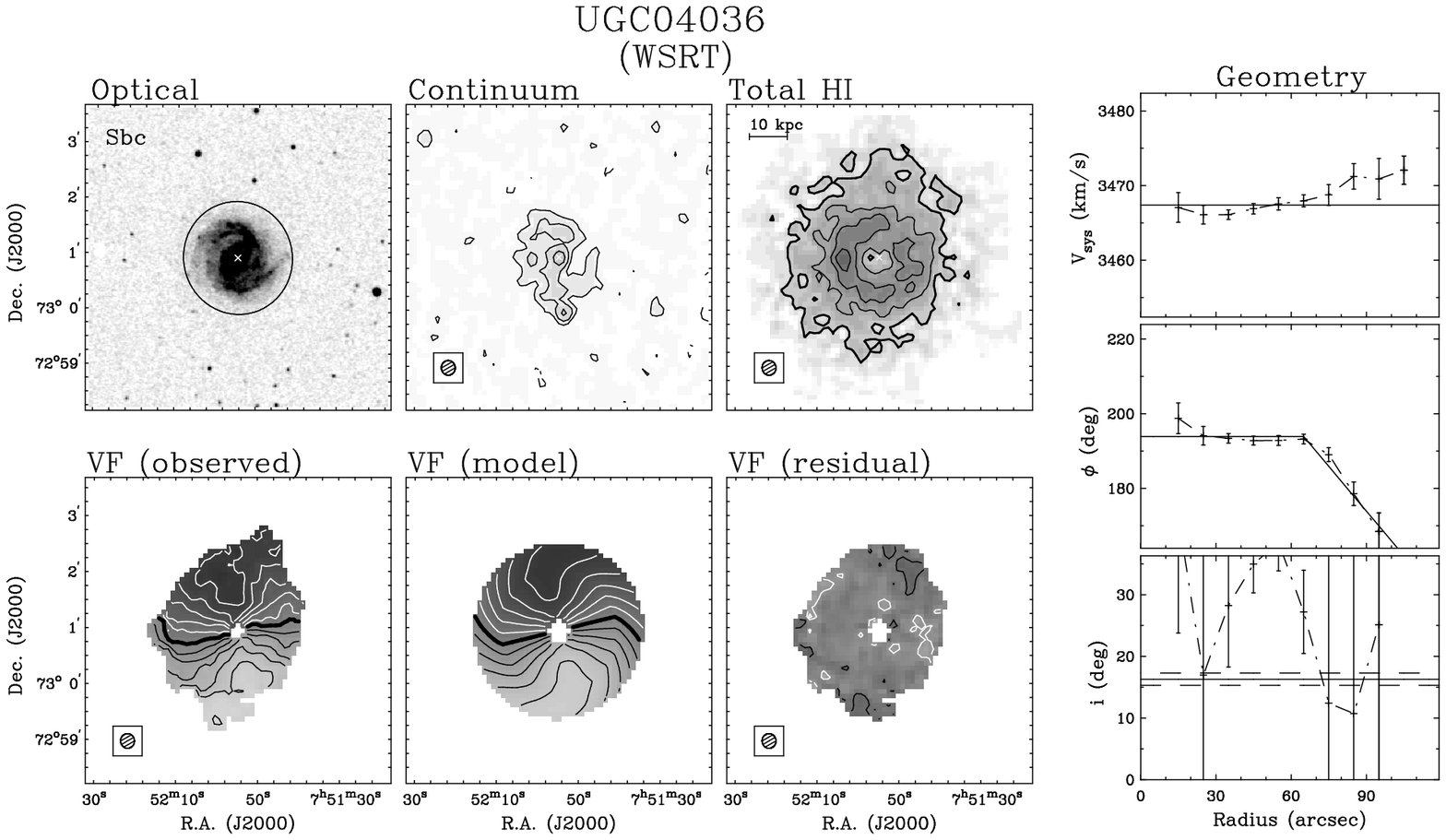}
 \end{figure}

 \begin{figure}
 \centering
 \includegraphics[width=1.0\textwidth]{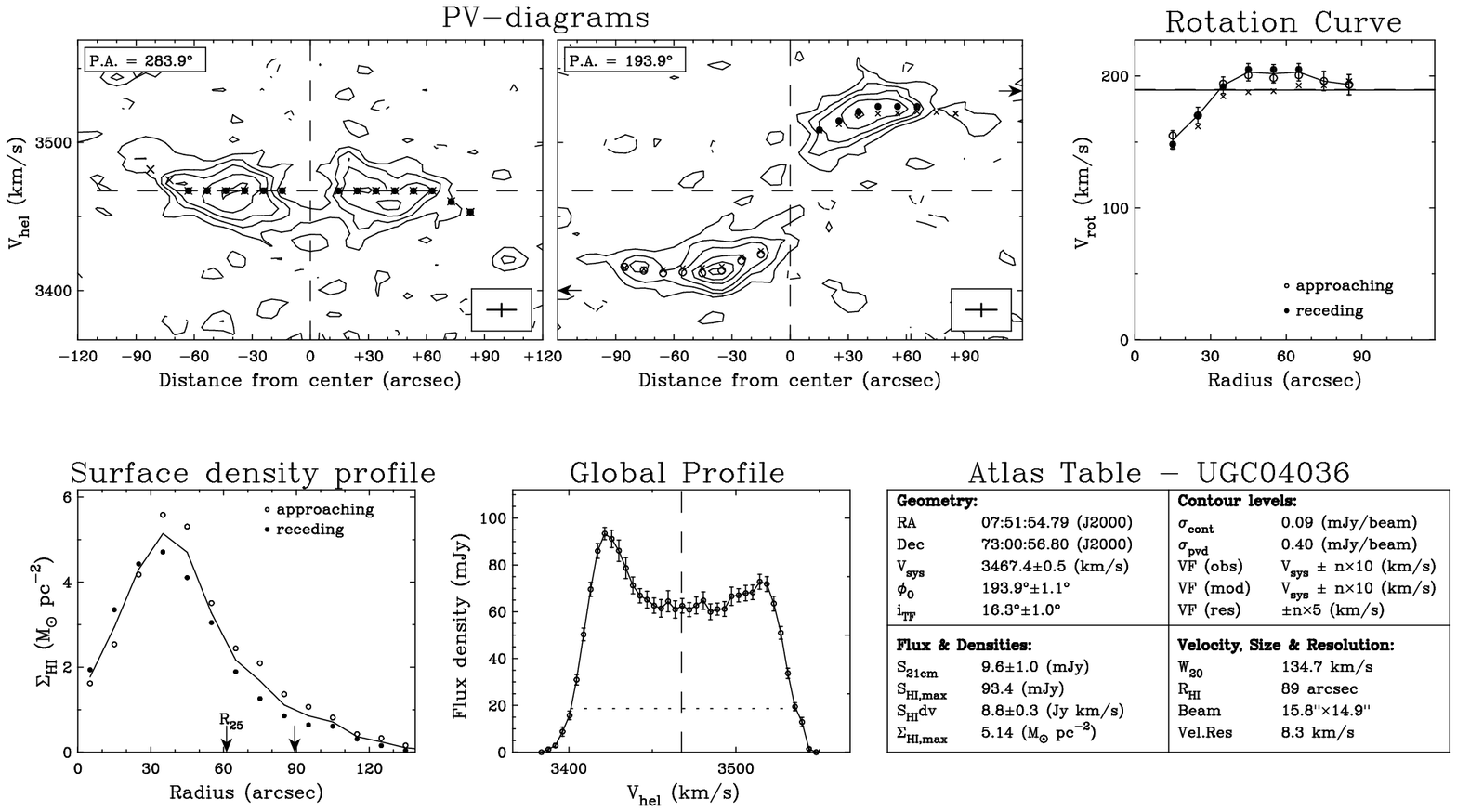}
 \end{figure}

 \begin{figure}
 \centering
 \includegraphics[width=1.0\textwidth]{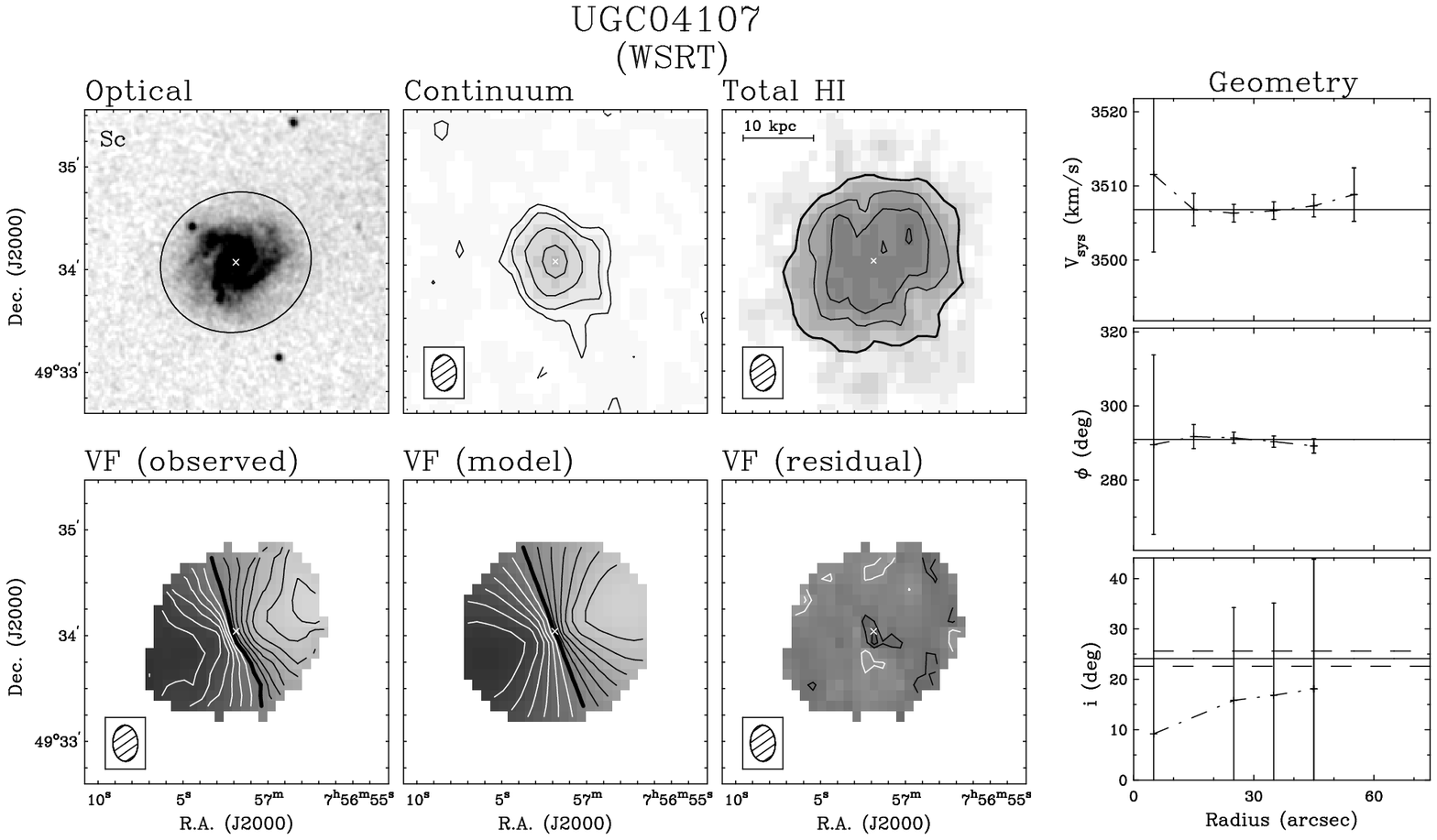}
 \end{figure}

 \begin{figure}
 \centering
 \includegraphics[width=1.0\textwidth]{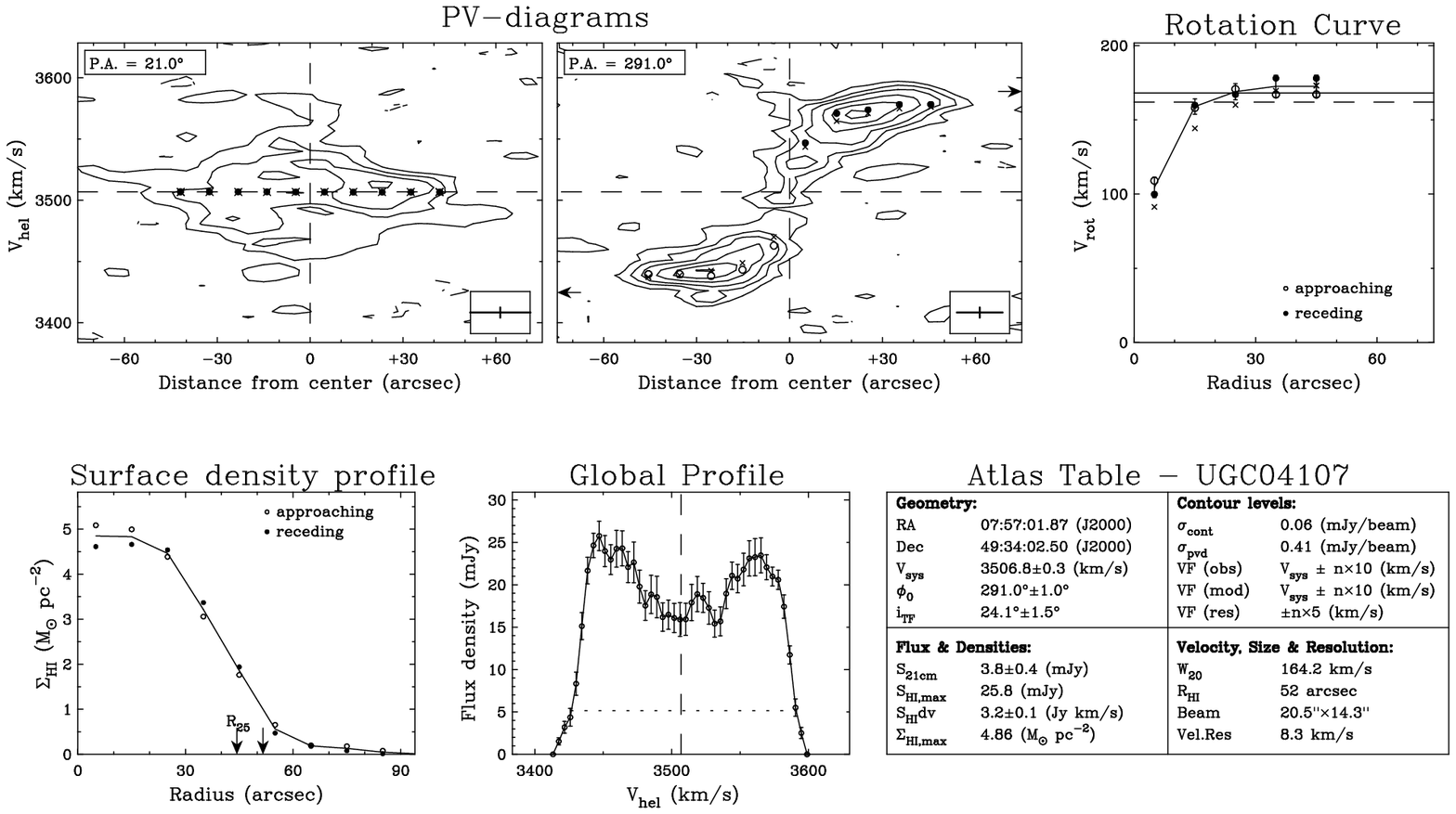}
 \end{figure}

 \begin{figure}
 \centering
 \includegraphics[width=1.0\textwidth]{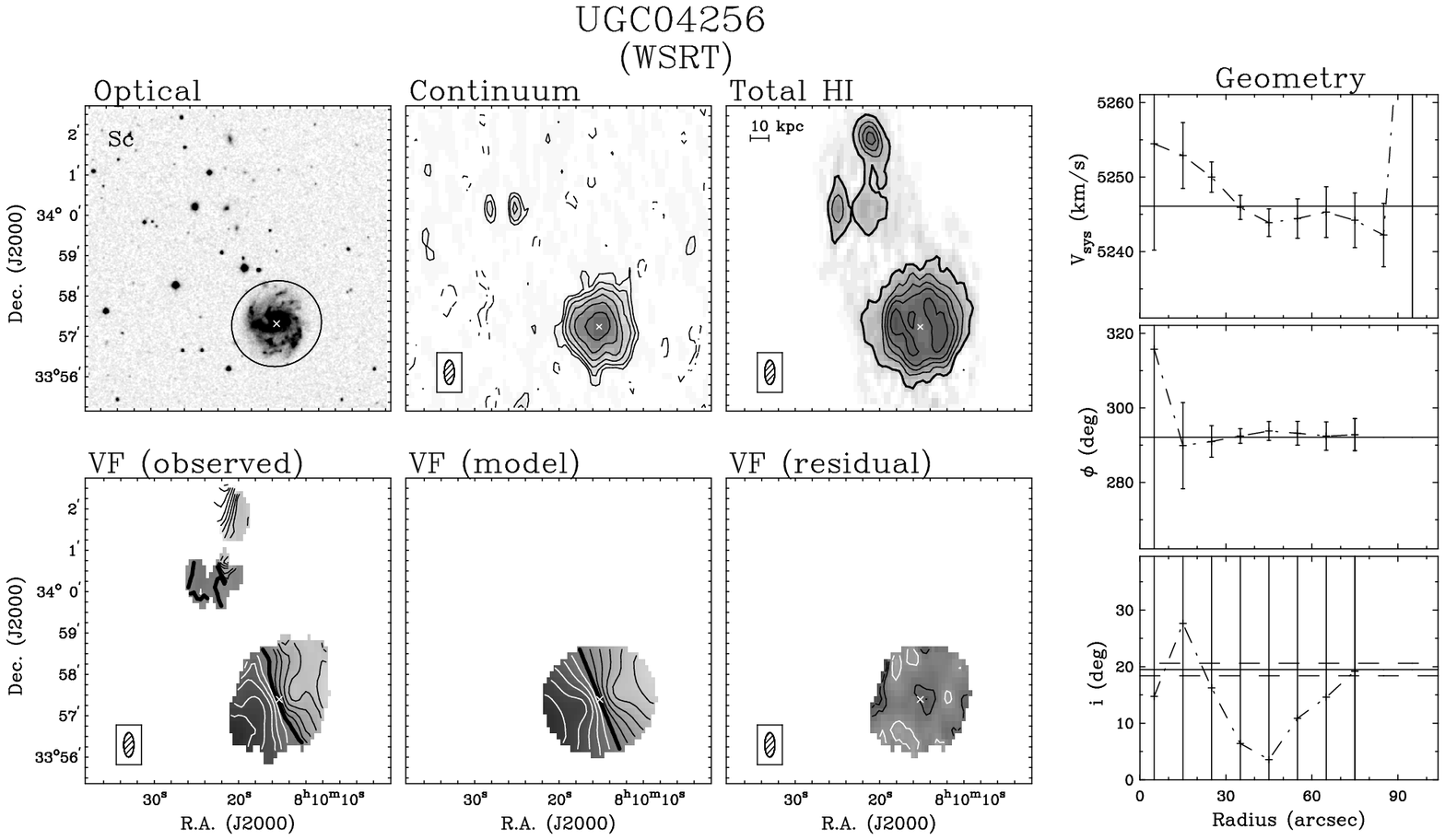}
 \end{figure}

 \begin{figure}
 \centering
 \includegraphics[width=1.0\textwidth]{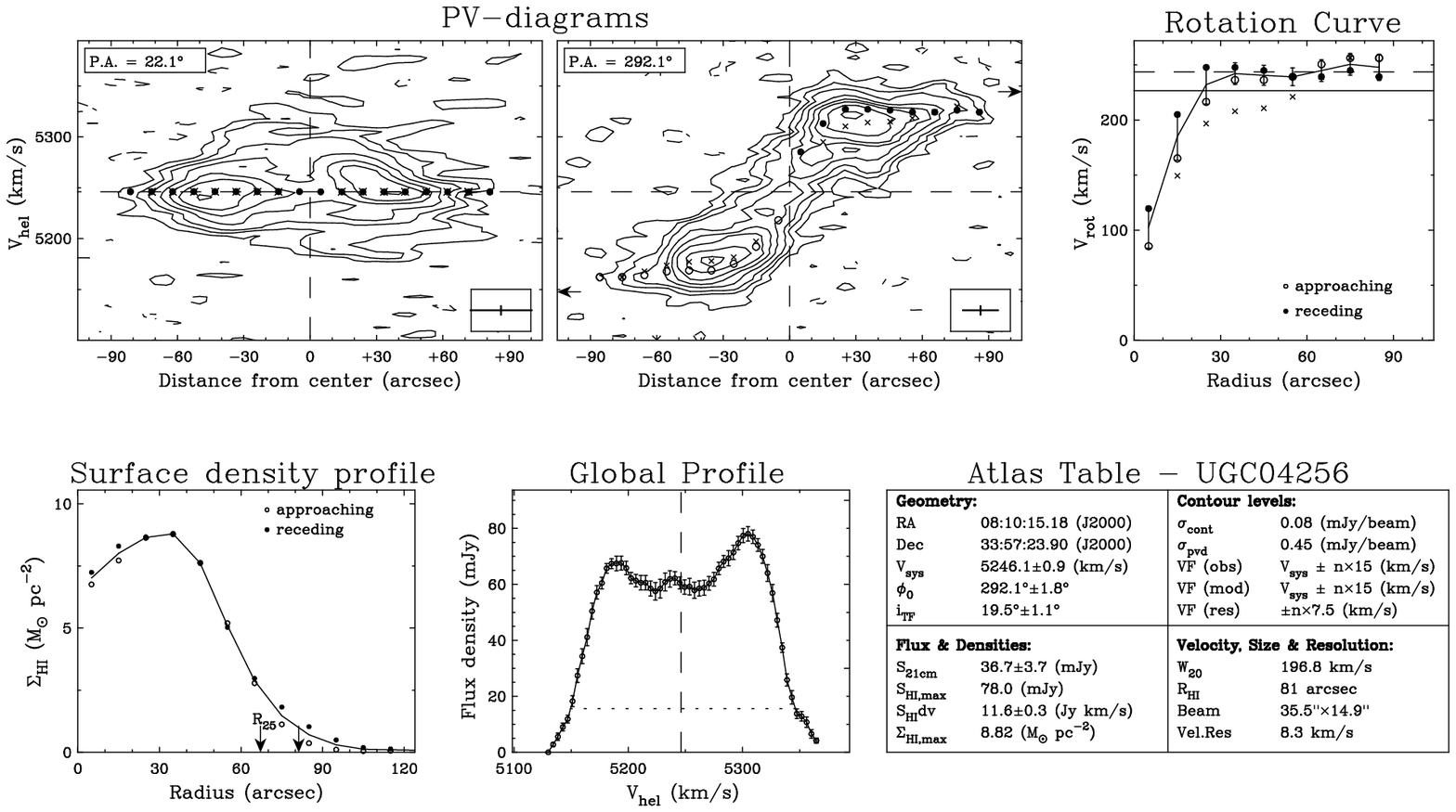}
 \end{figure}

 \begin{figure}
 \centering
 \includegraphics[width=1.0\textwidth]{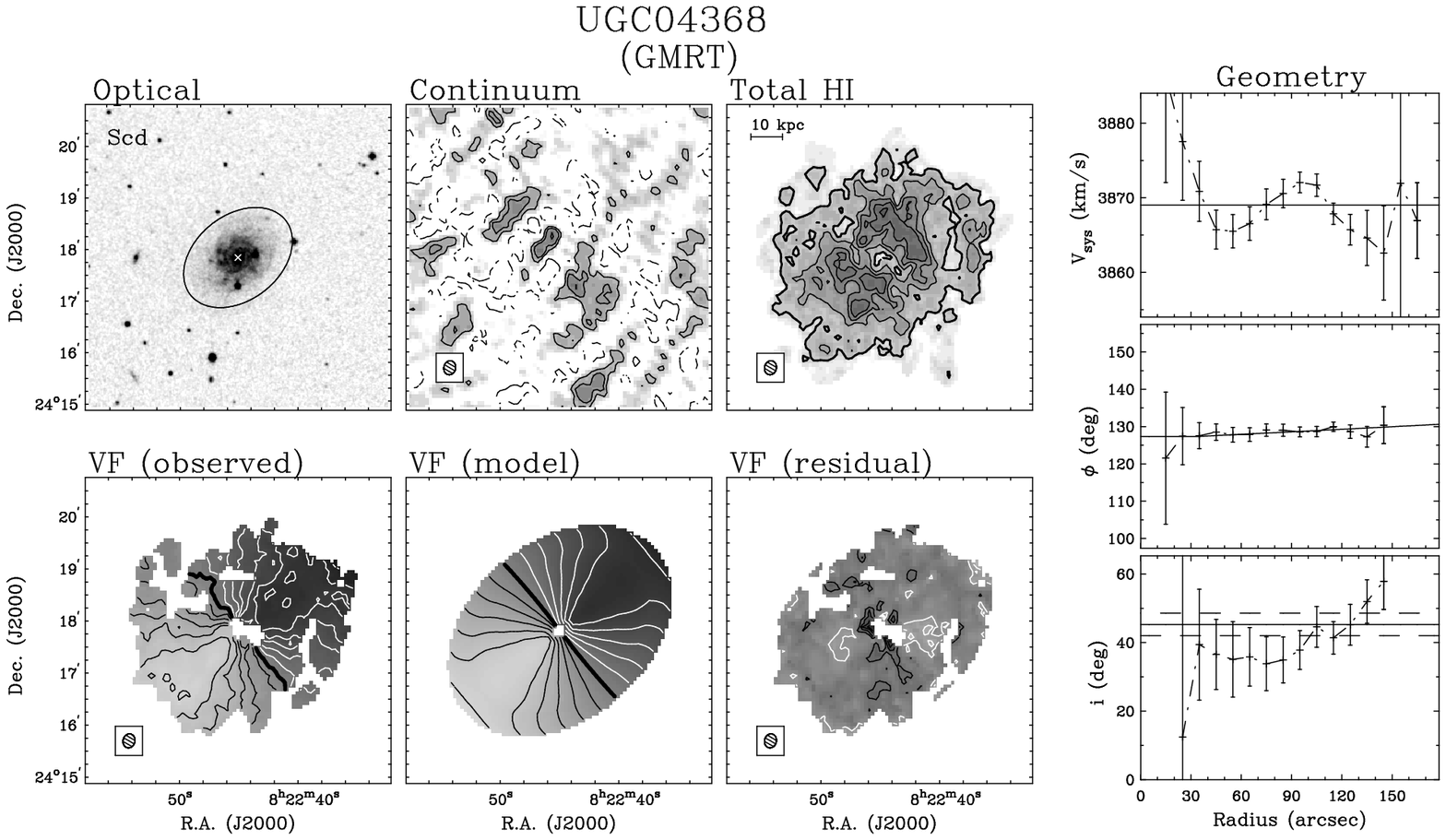}
 \end{figure}

 \begin{figure}
 \centering
 \includegraphics[width=1.0\textwidth]{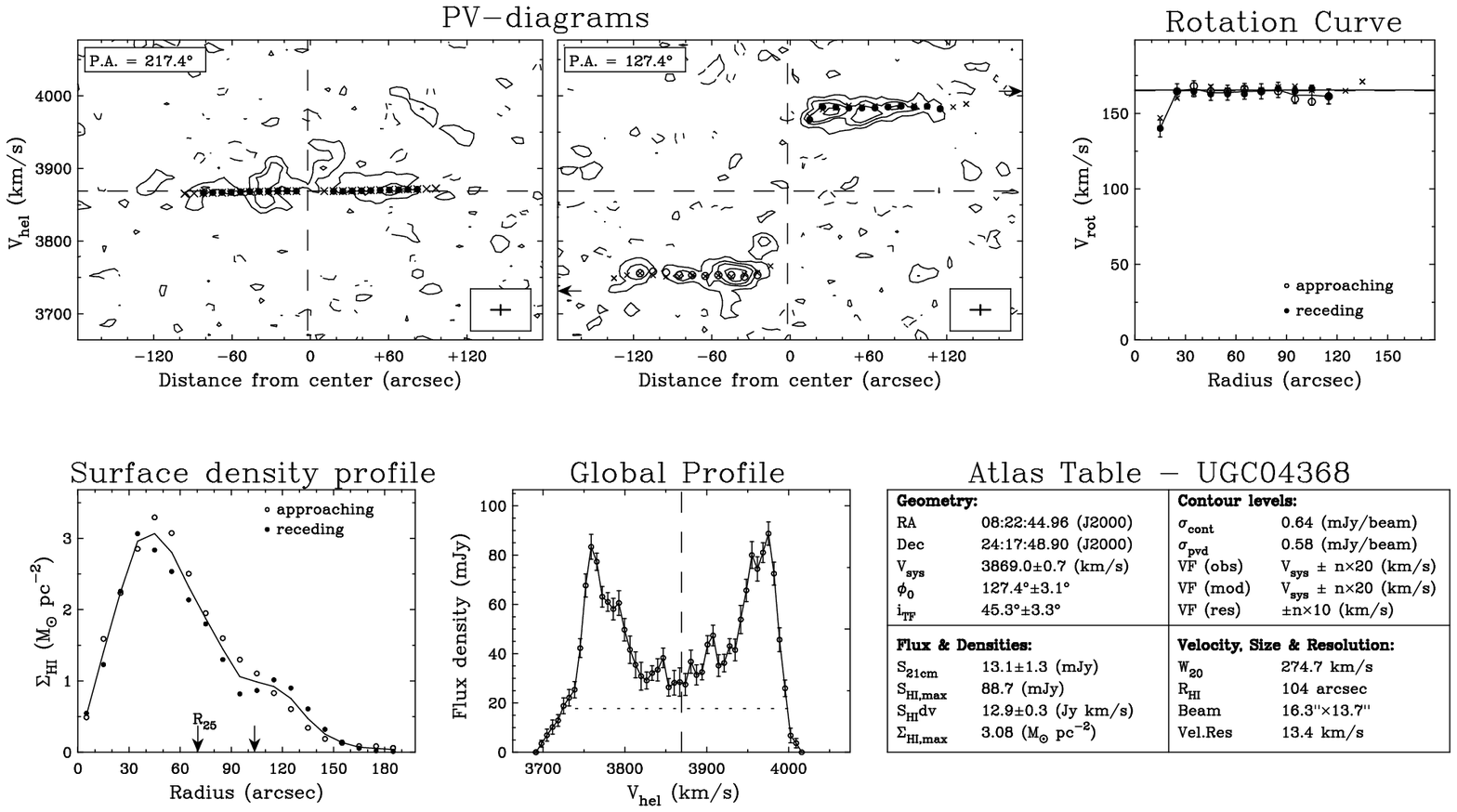}
 \end{figure}

 \begin{figure}
 \centering
 \includegraphics[width=1.0\textwidth]{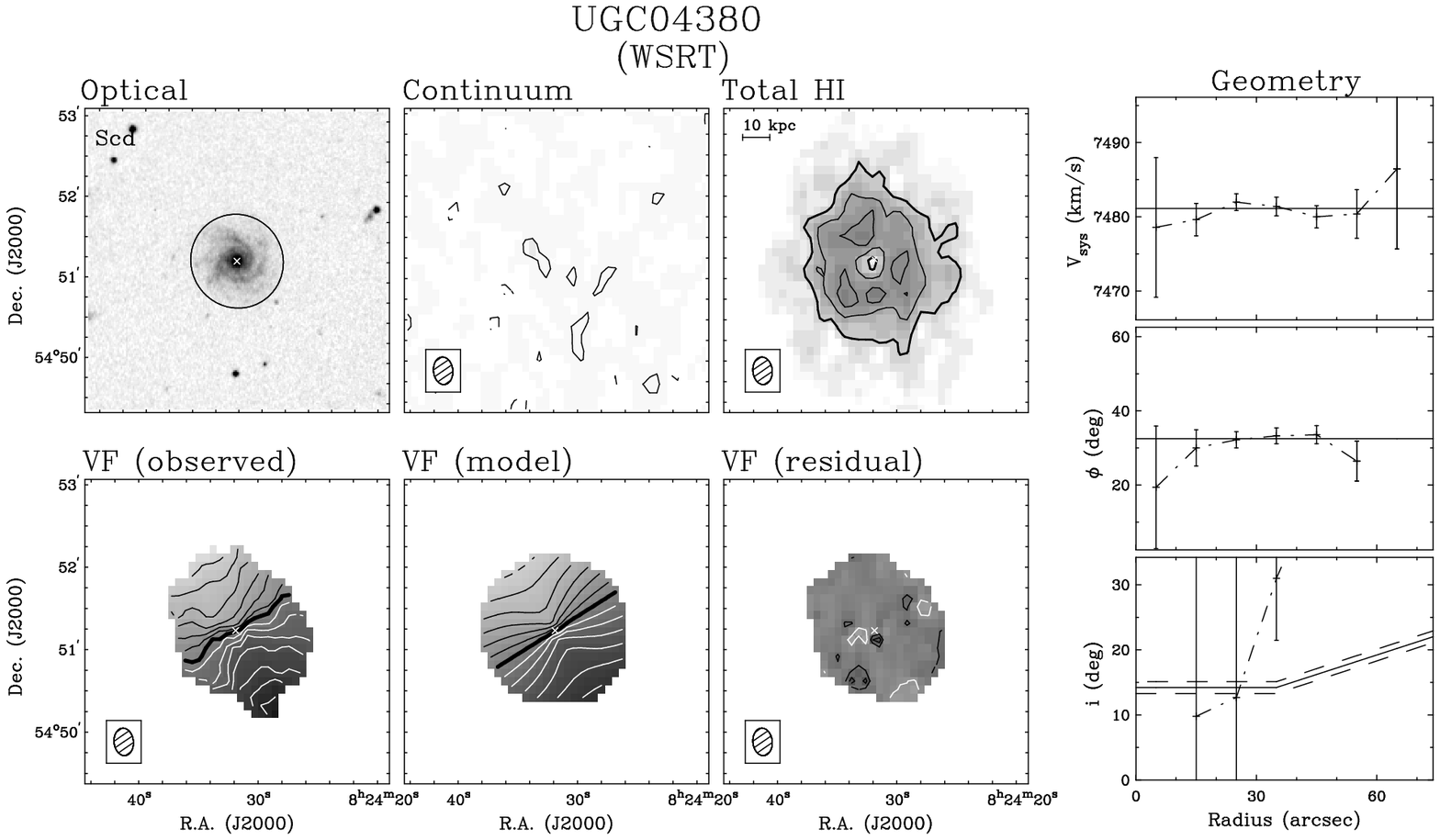}
 \end{figure}

 \begin{figure}
 \centering
 \includegraphics[width=1.0\textwidth]{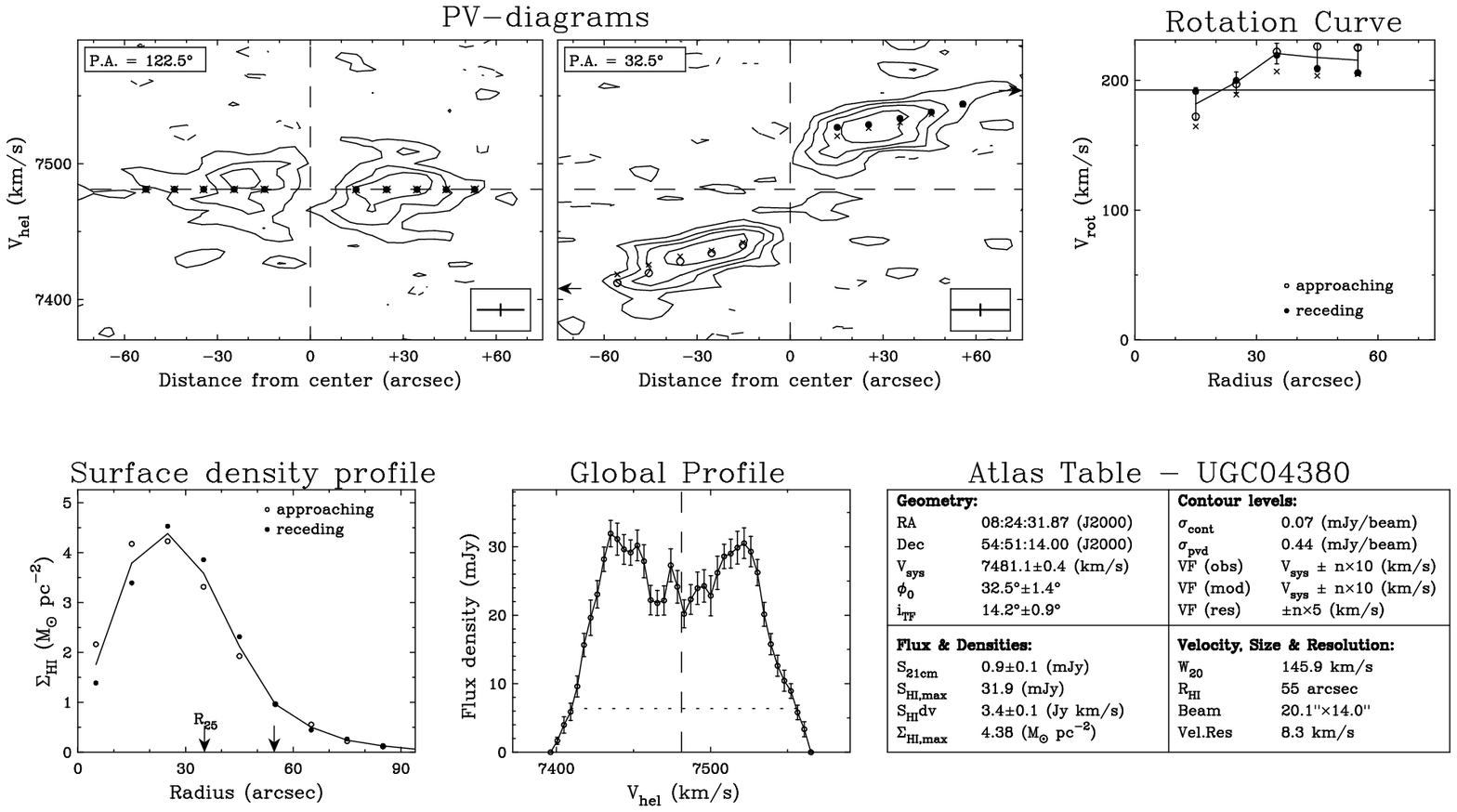}
 \end{figure}

 \begin{figure}
 \centering
 \includegraphics[width=1.0\textwidth]{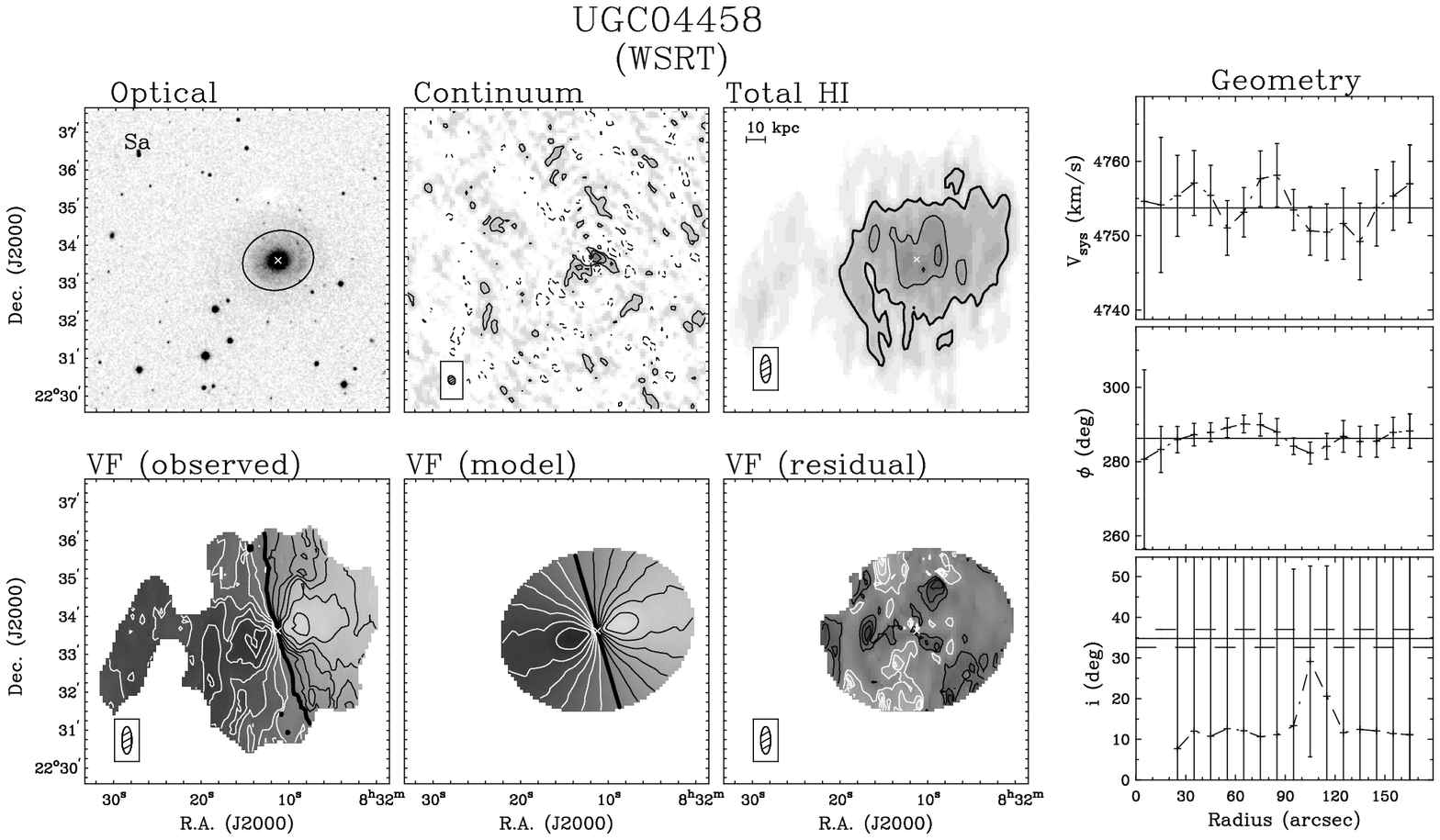}
 \end{figure}

 \begin{figure}
 \centering
 \includegraphics[width=1.0\textwidth]{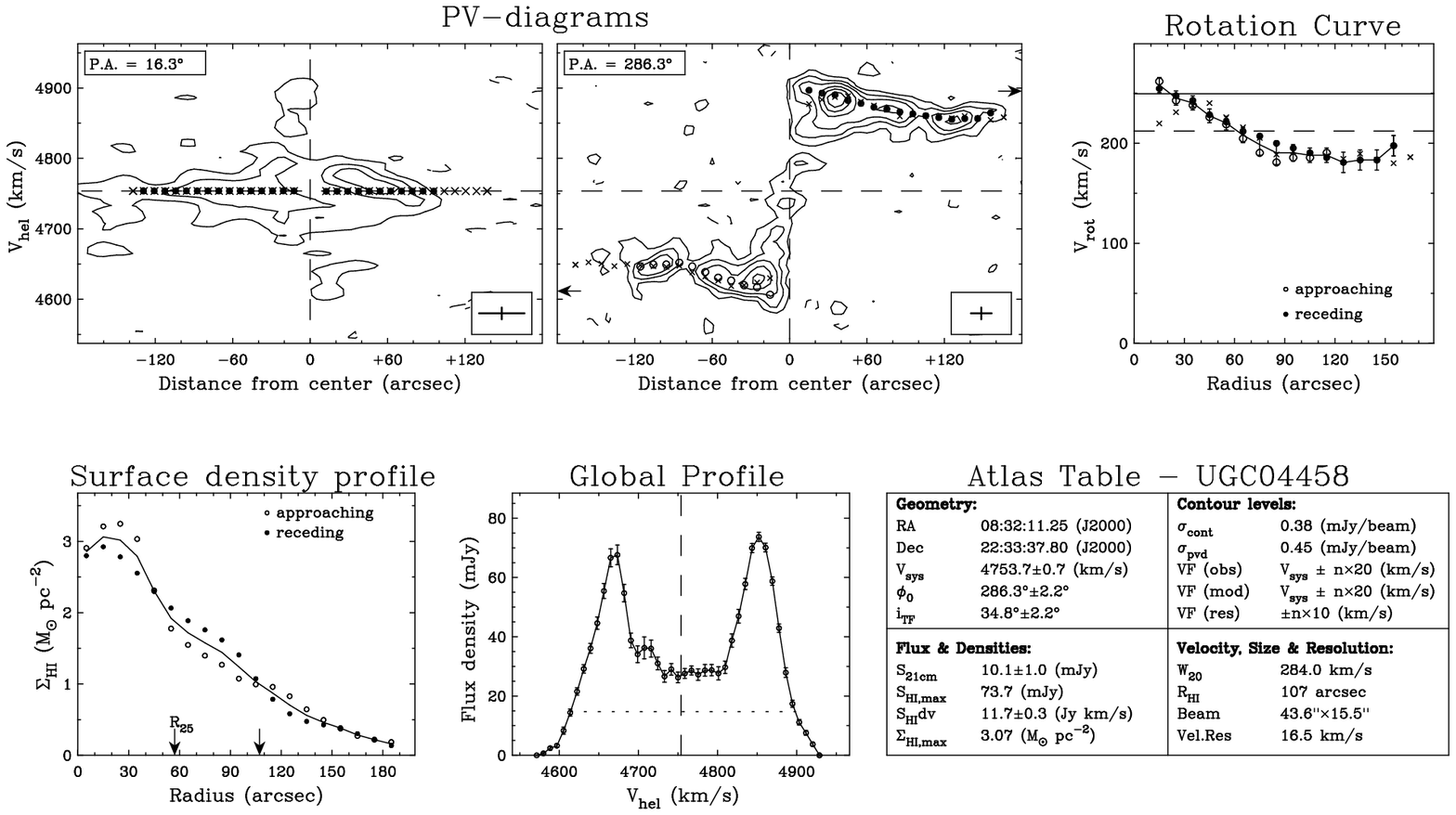}
 \end{figure}

\clearpage

 \begin{figure}
 \centering
 \includegraphics[width=1.0\textwidth]{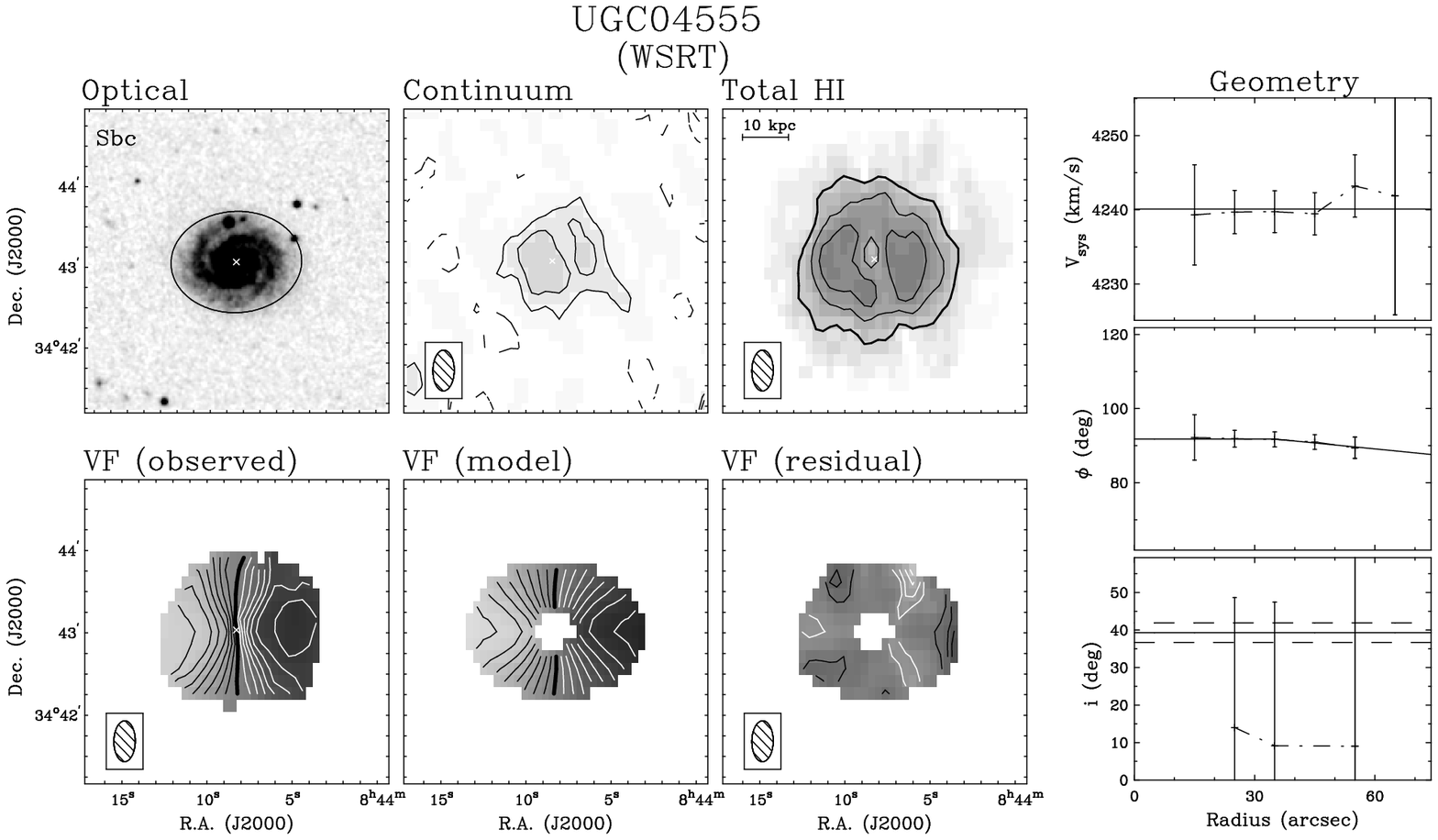}
 \end{figure}

 \begin{figure}
 \centering
 \includegraphics[width=1.0\textwidth]{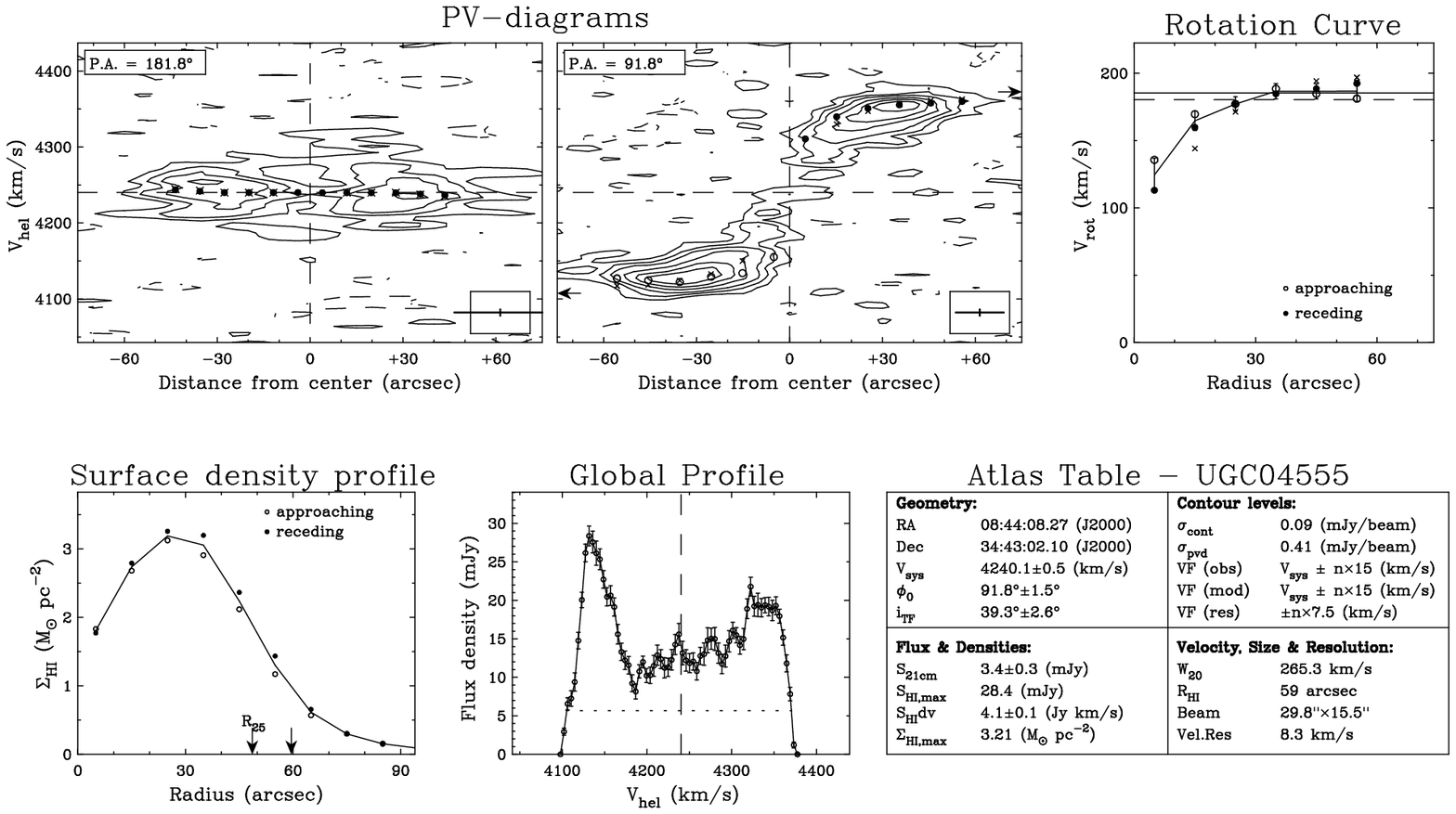}
 \end{figure}

 \begin{figure}
 \centering
 \includegraphics[width=1.0\textwidth]{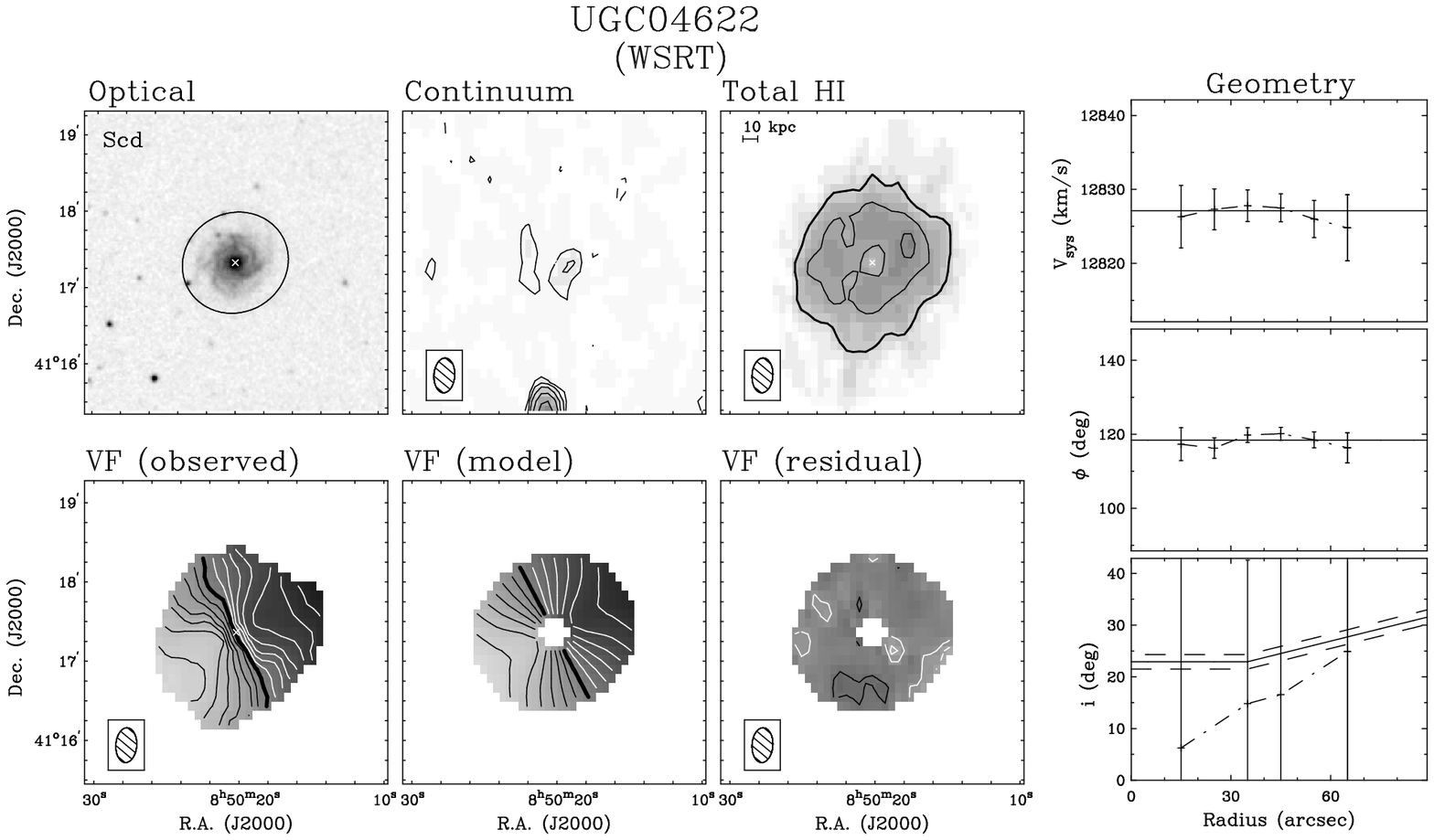}
 \end{figure}

 \begin{figure}
 \centering
 \includegraphics[width=1.0\textwidth]{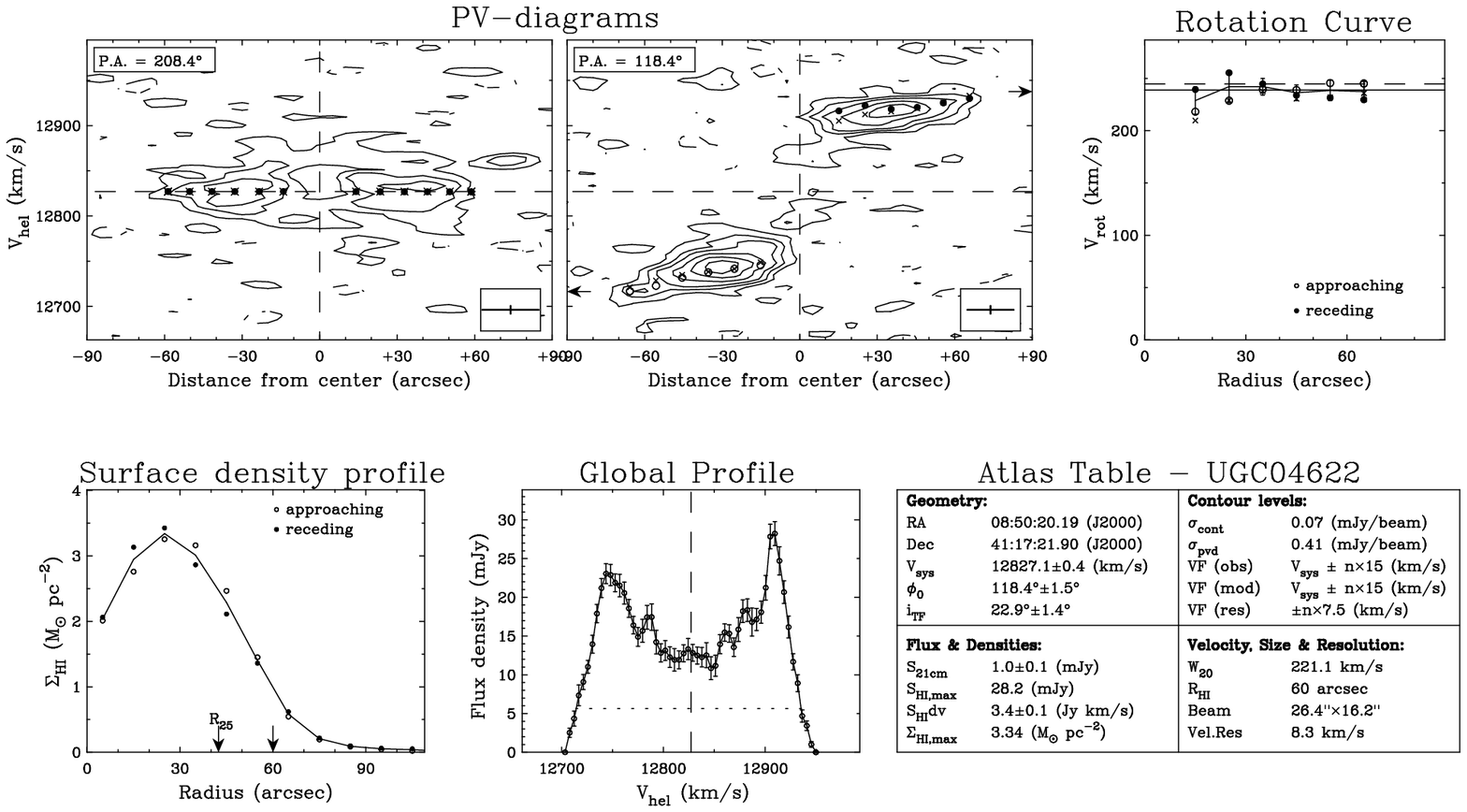}
 \end{figure}

 %\begin{figure}
 %\centering
 %\includegraphics[width=1.0\textwidth]{HIAtlas/UGC04622.3.1.ps}
 %\end{figure}

 %\begin{figure}
 %\centering
 %\includegraphics[width=1.0\textwidth]{HIAtlas/UGC04622.3.2.ps}
 %\end{figure}

 \begin{figure}
 \centering
 \includegraphics[width=1.0\textwidth]{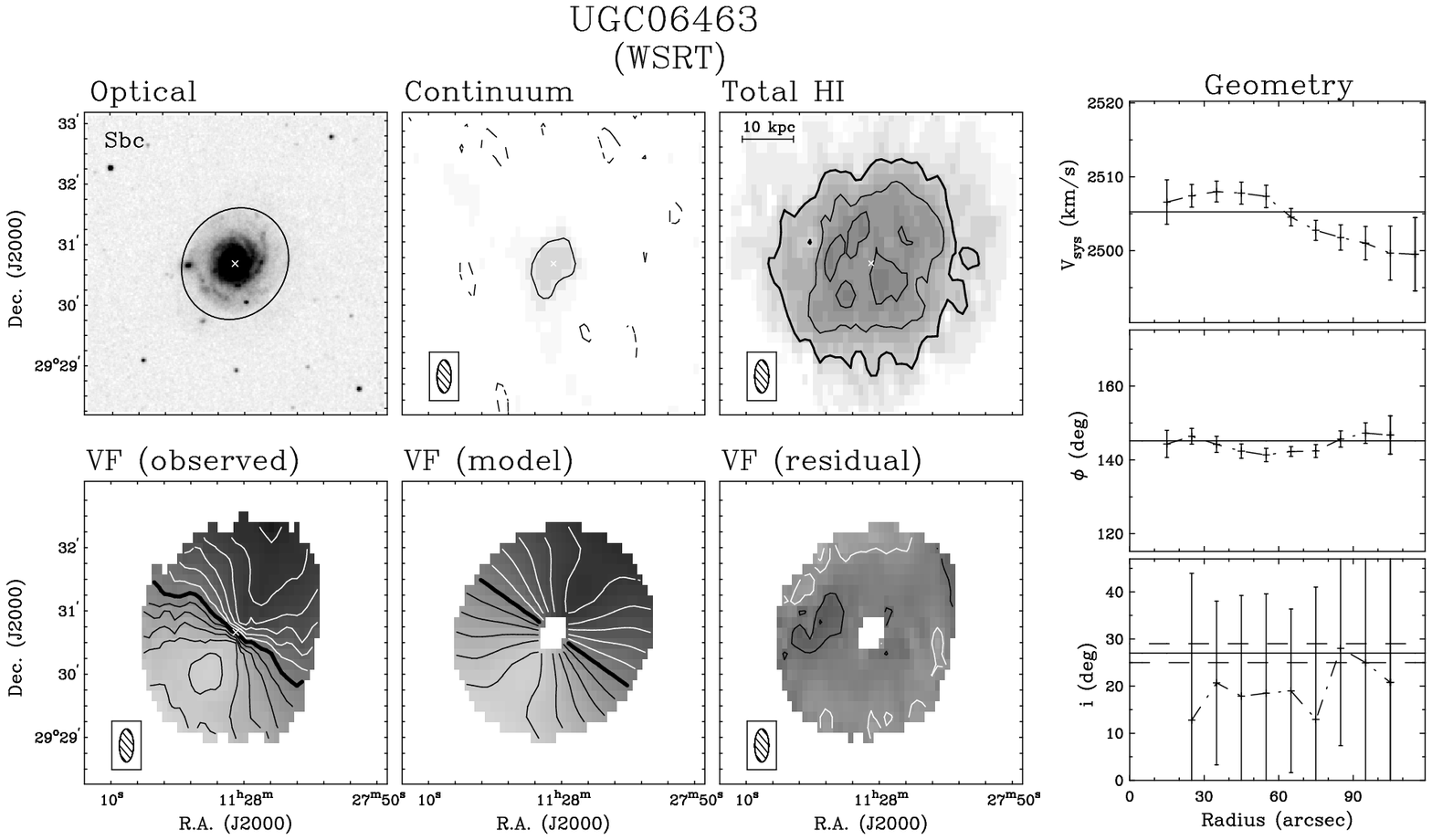}
 \end{figure}

 \begin{figure}
 \centering
 \includegraphics[width=1.0\textwidth]{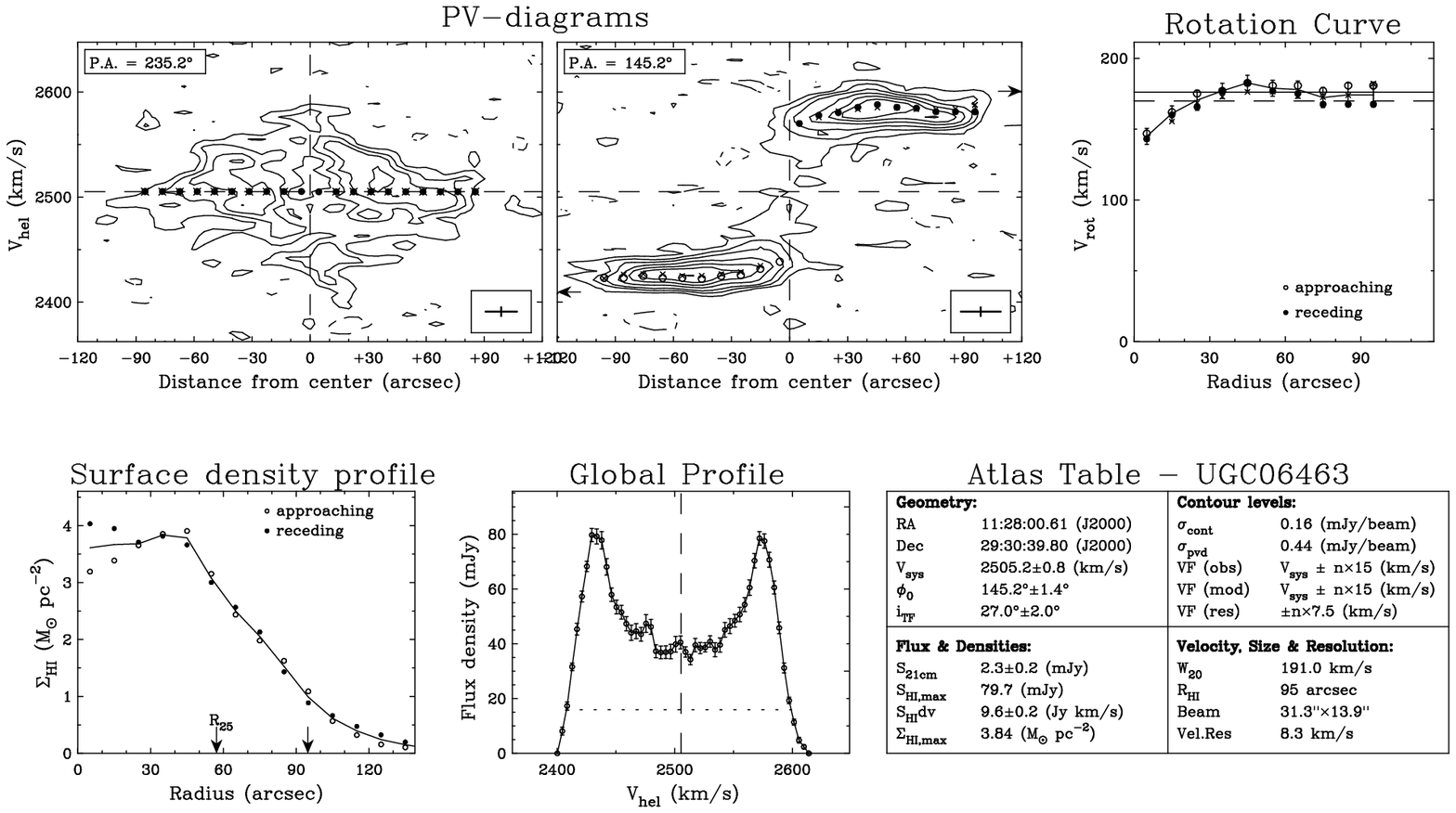}
 \end{figure}

 \begin{figure}
 \centering
 \includegraphics[width=1.0\textwidth]{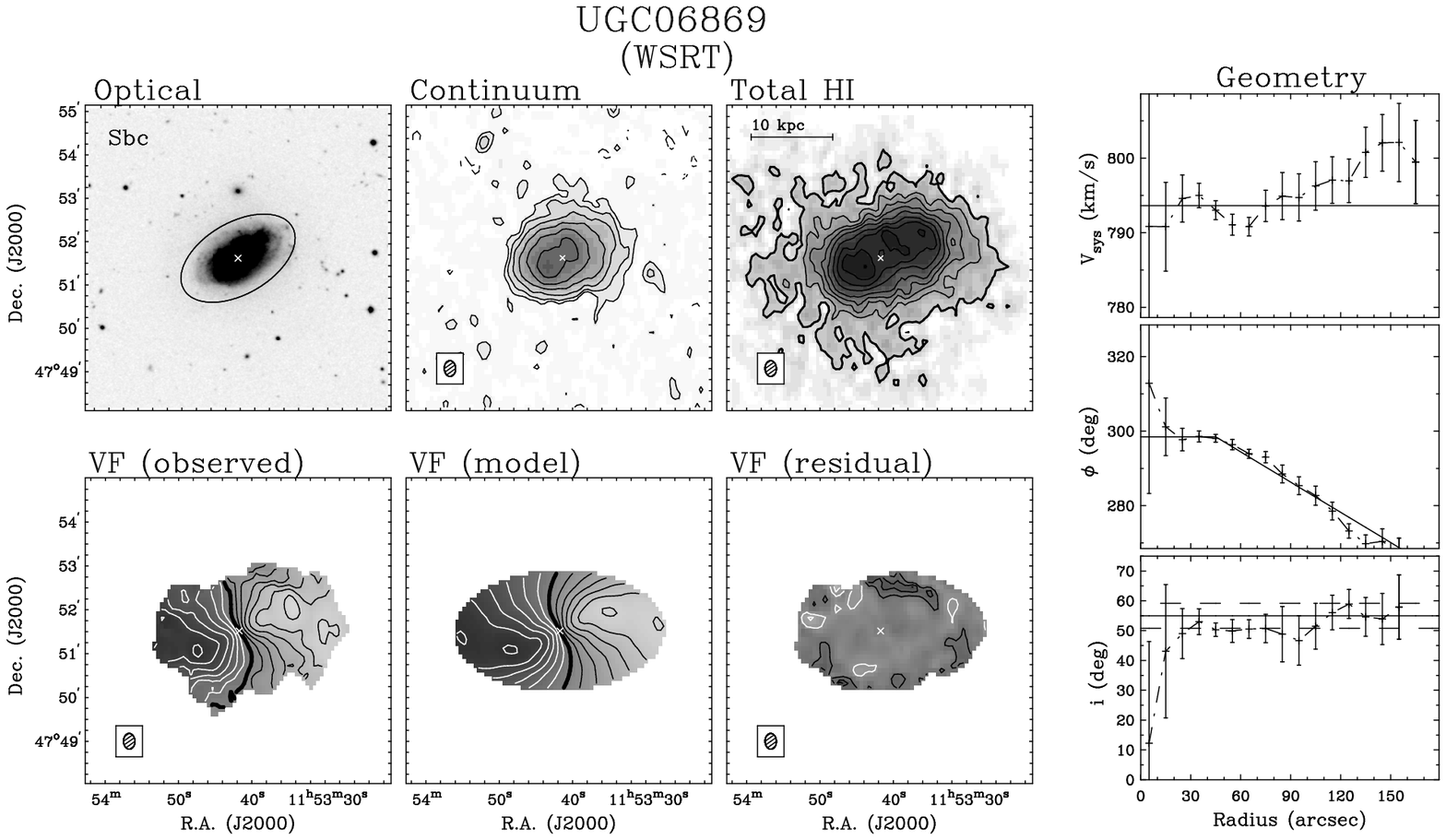}
 \end{figure}

 \begin{figure}
 \centering
 \includegraphics[width=1.0\textwidth]{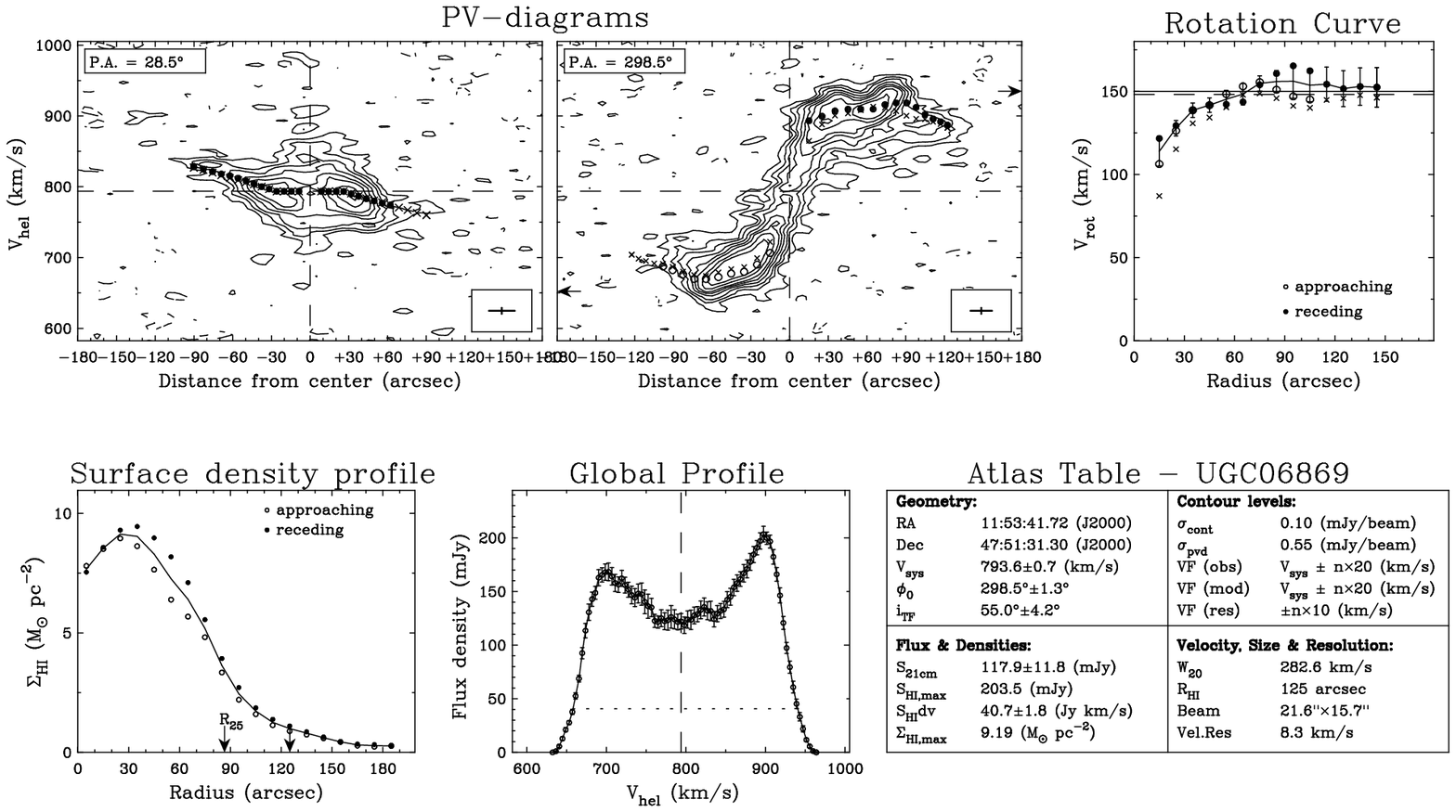}
 \end{figure}

 \begin{figure}
 \centering
 \includegraphics[width=1.0\textwidth]{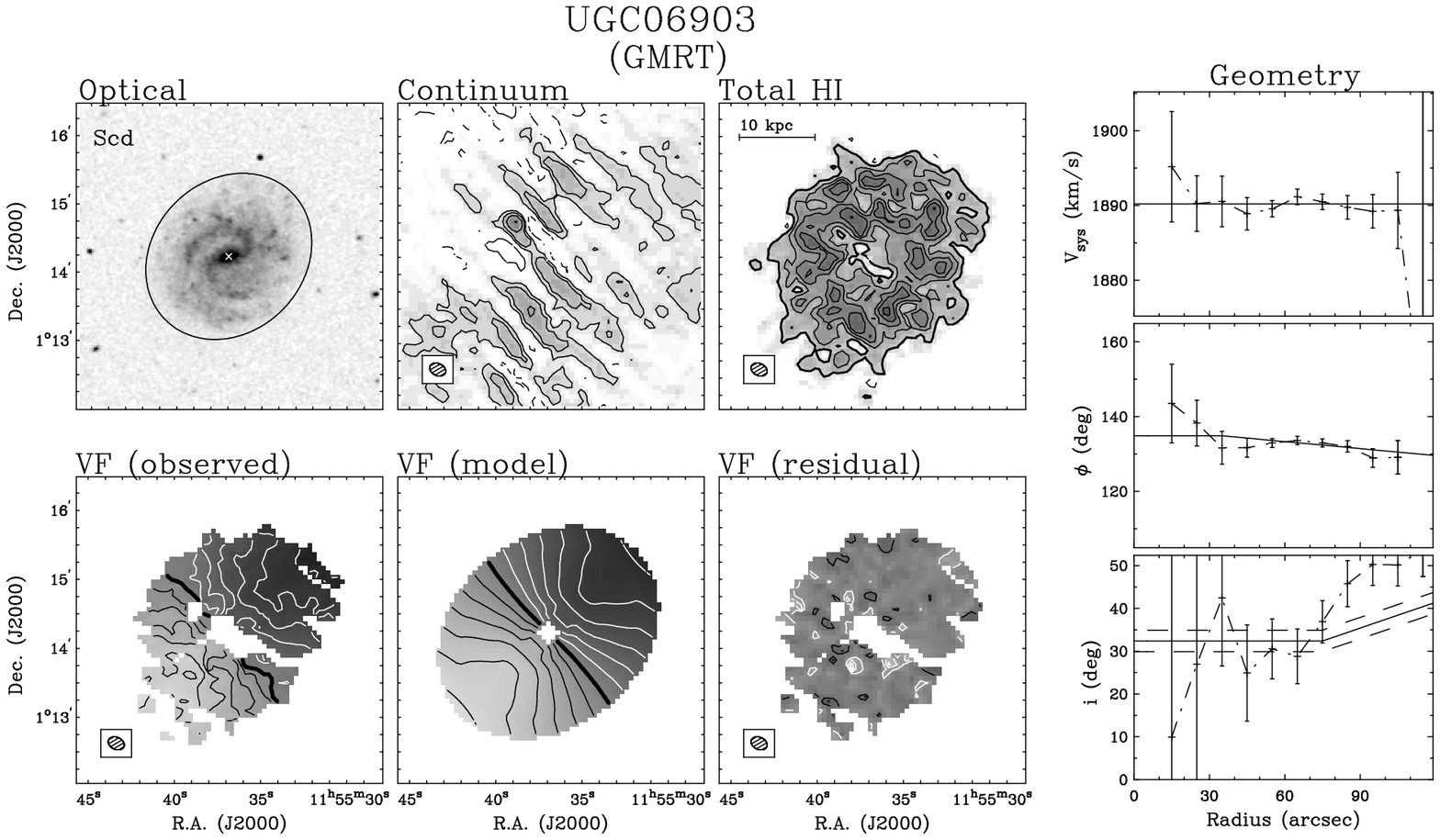}
 \end{figure}

 \begin{figure}
 \centering
 \includegraphics[width=1.0\textwidth]{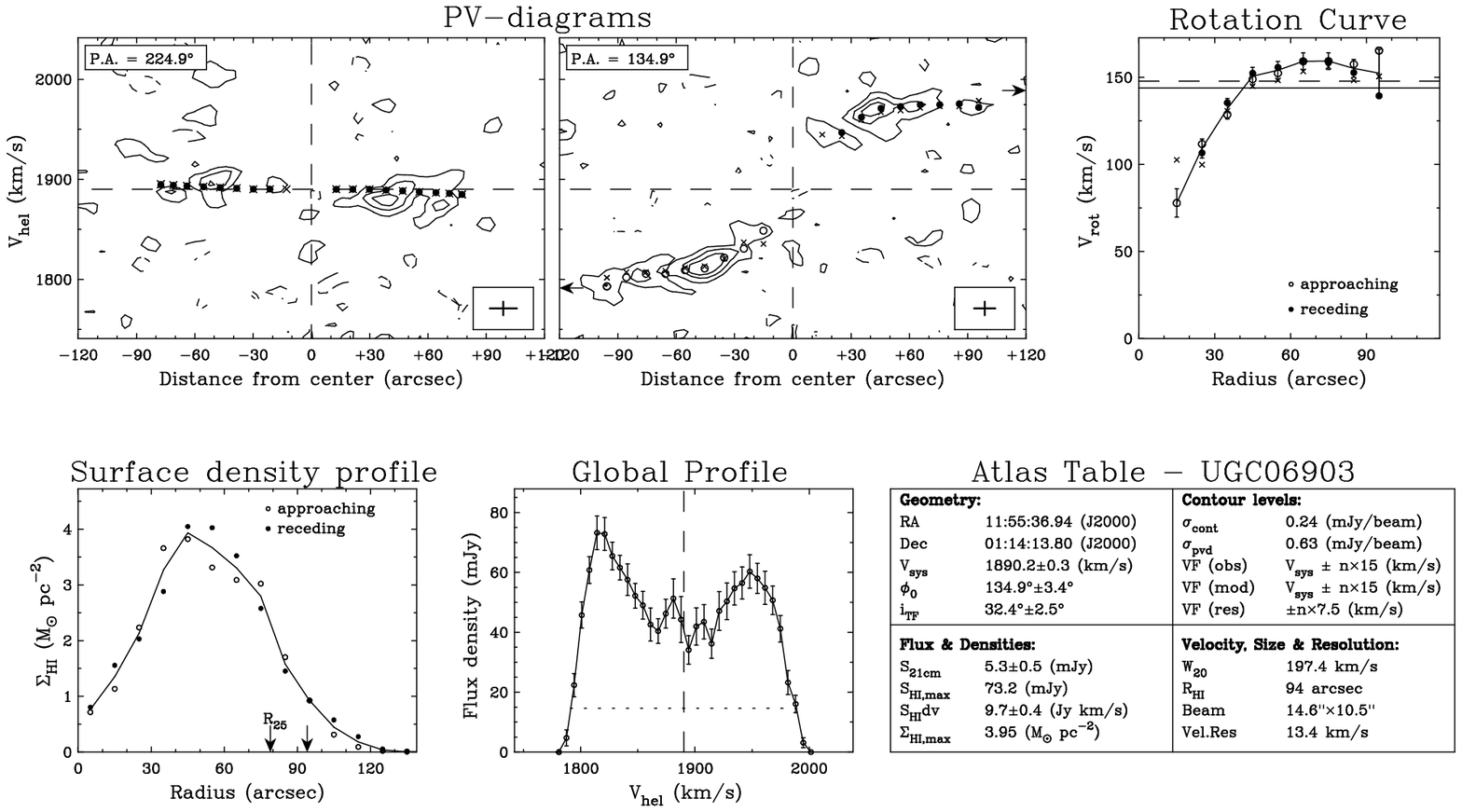}
 \end{figure}

 \begin{figure}
 \centering
 \includegraphics[width=1.0\textwidth]{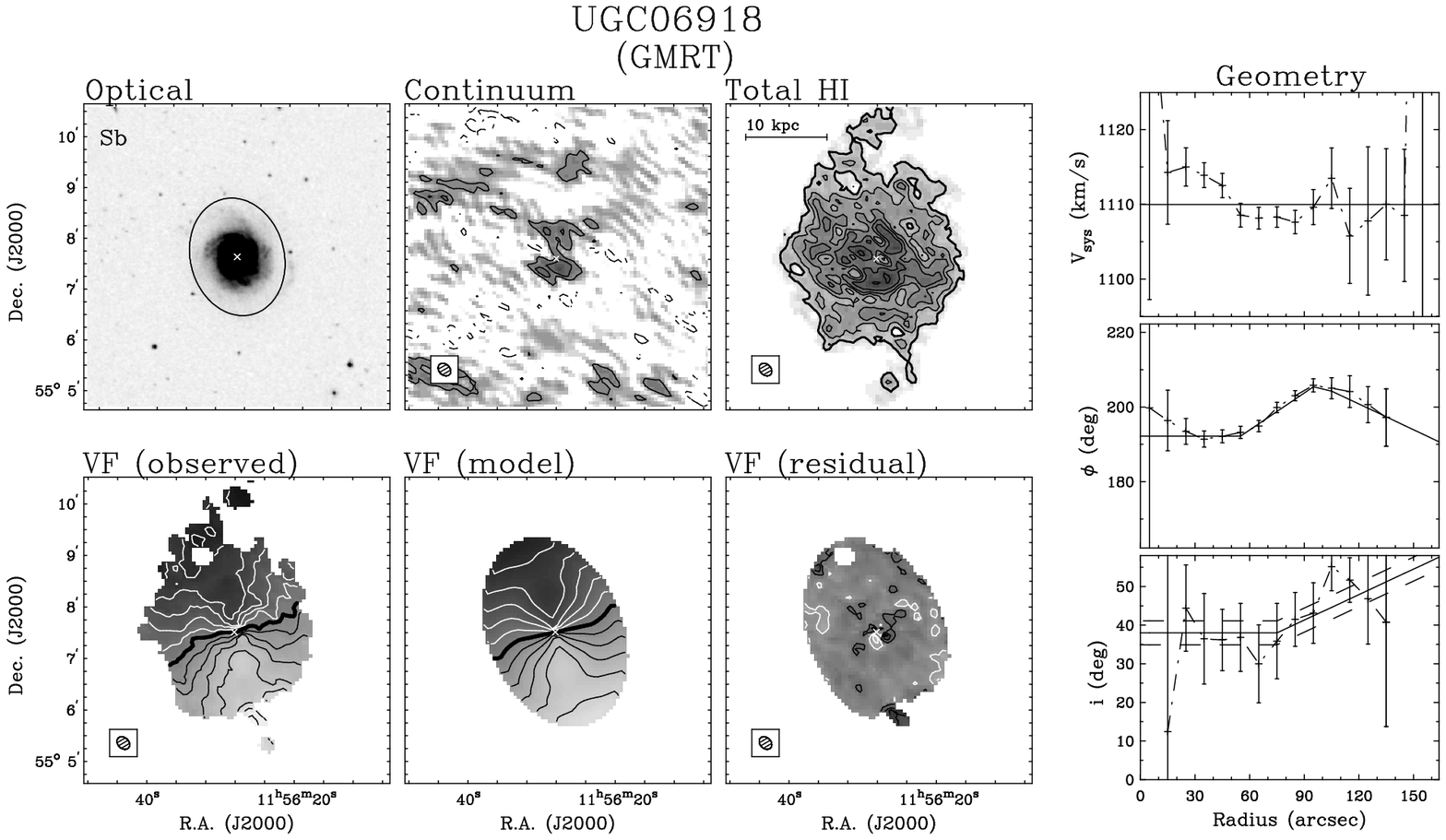}
 \end{figure}

 \begin{figure}
 \centering
 \includegraphics[width=1.0\textwidth]{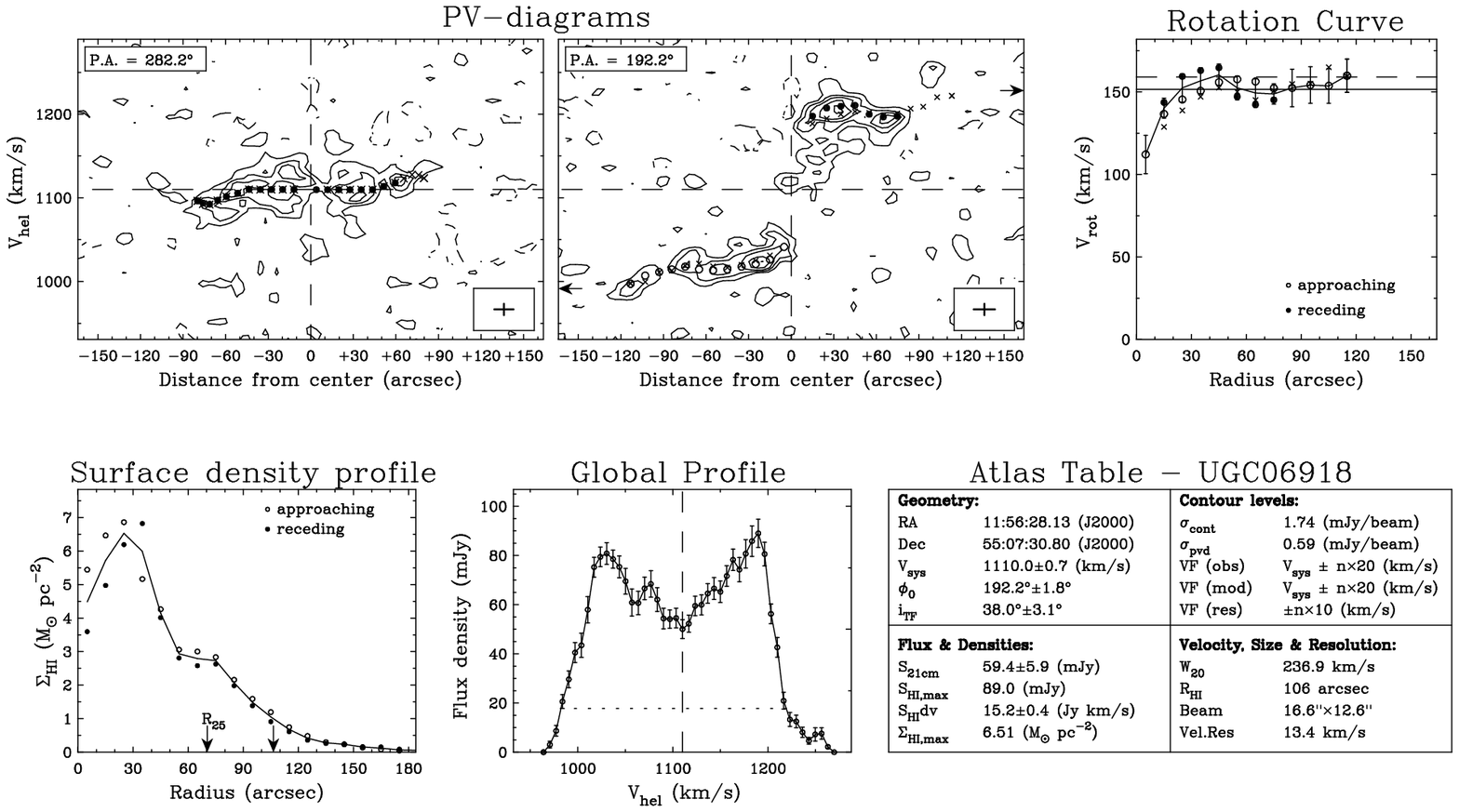}
 \end{figure}

 \begin{figure}
 \centering
 \includegraphics[width=1.0\textwidth]{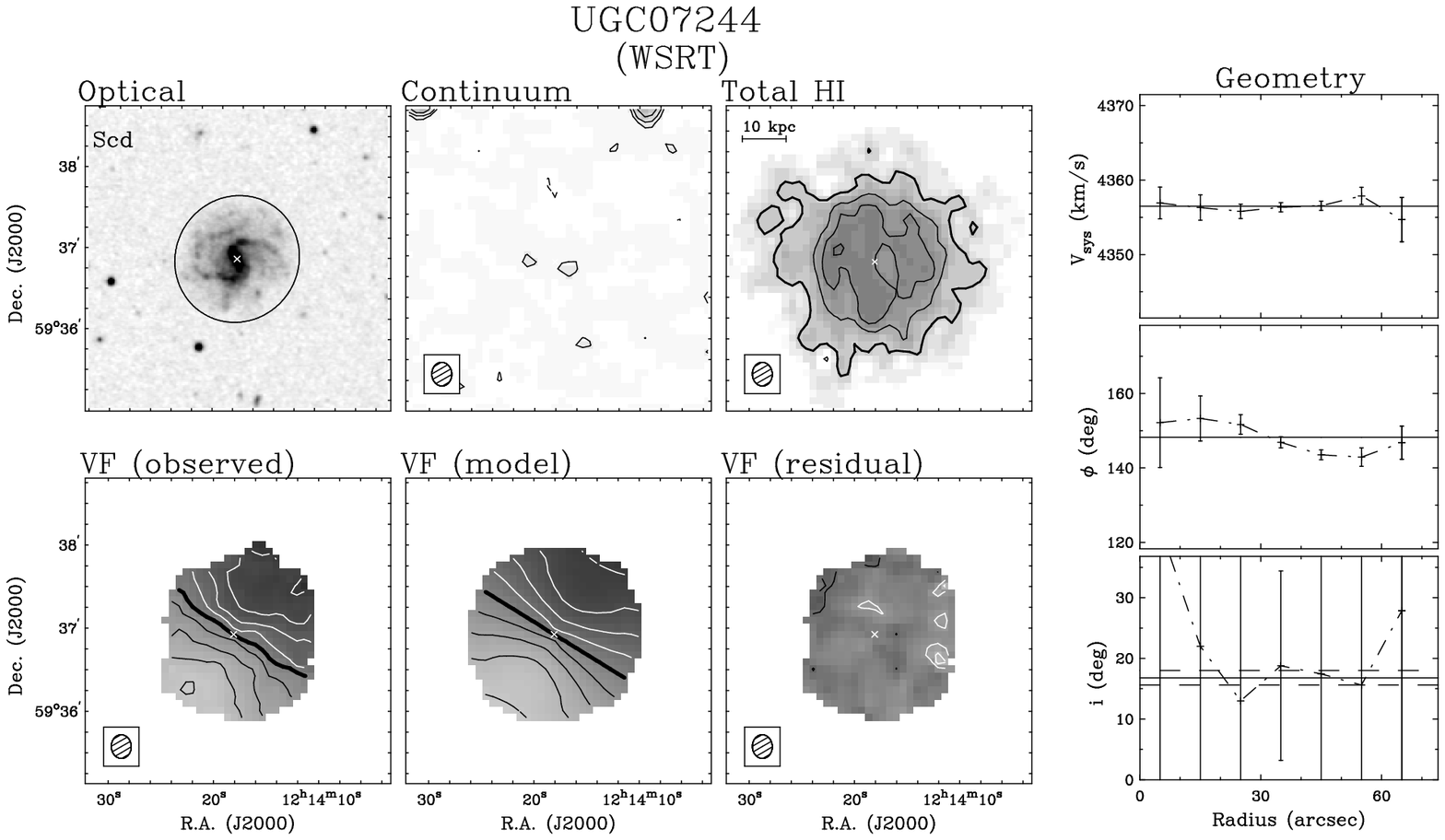}
 \end{figure}

 \begin{figure}
 \centering
 \includegraphics[width=1.0\textwidth]{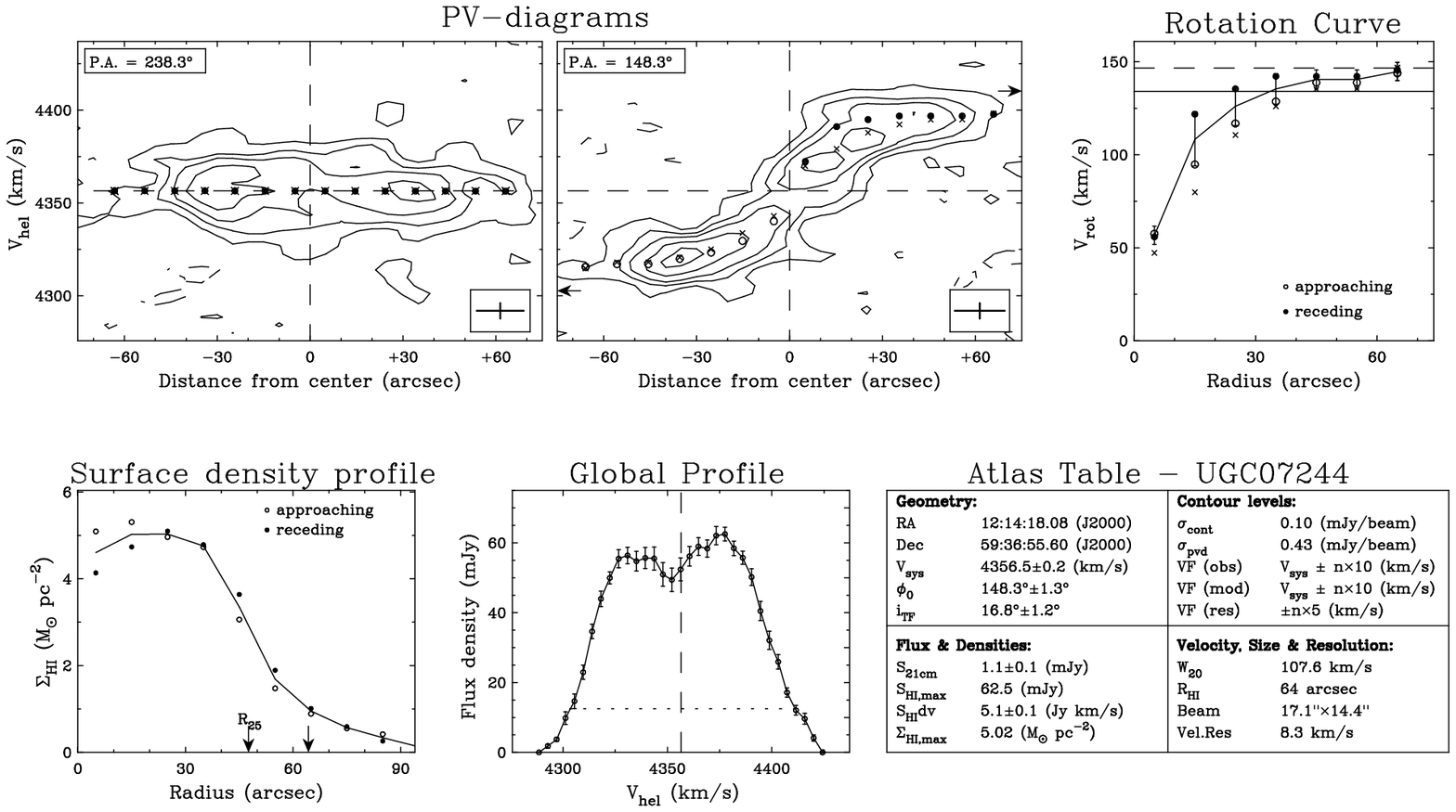}
 \end{figure}

\clearpage

 \begin{figure}
 \centering
 \includegraphics[width=1.0\textwidth]{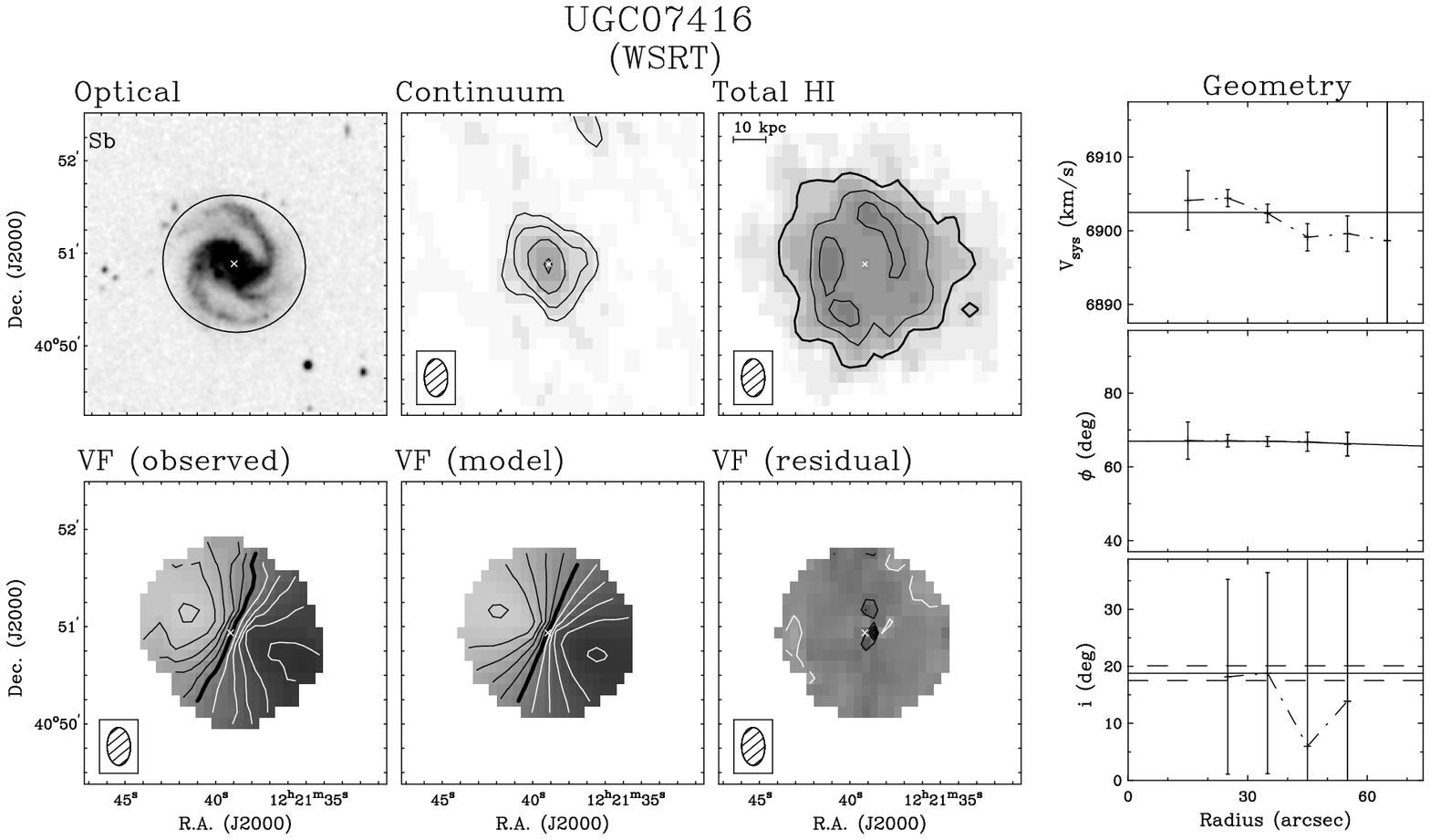}
 \end{figure}

 \begin{figure}
 \centering
 \includegraphics[width=1.0\textwidth]{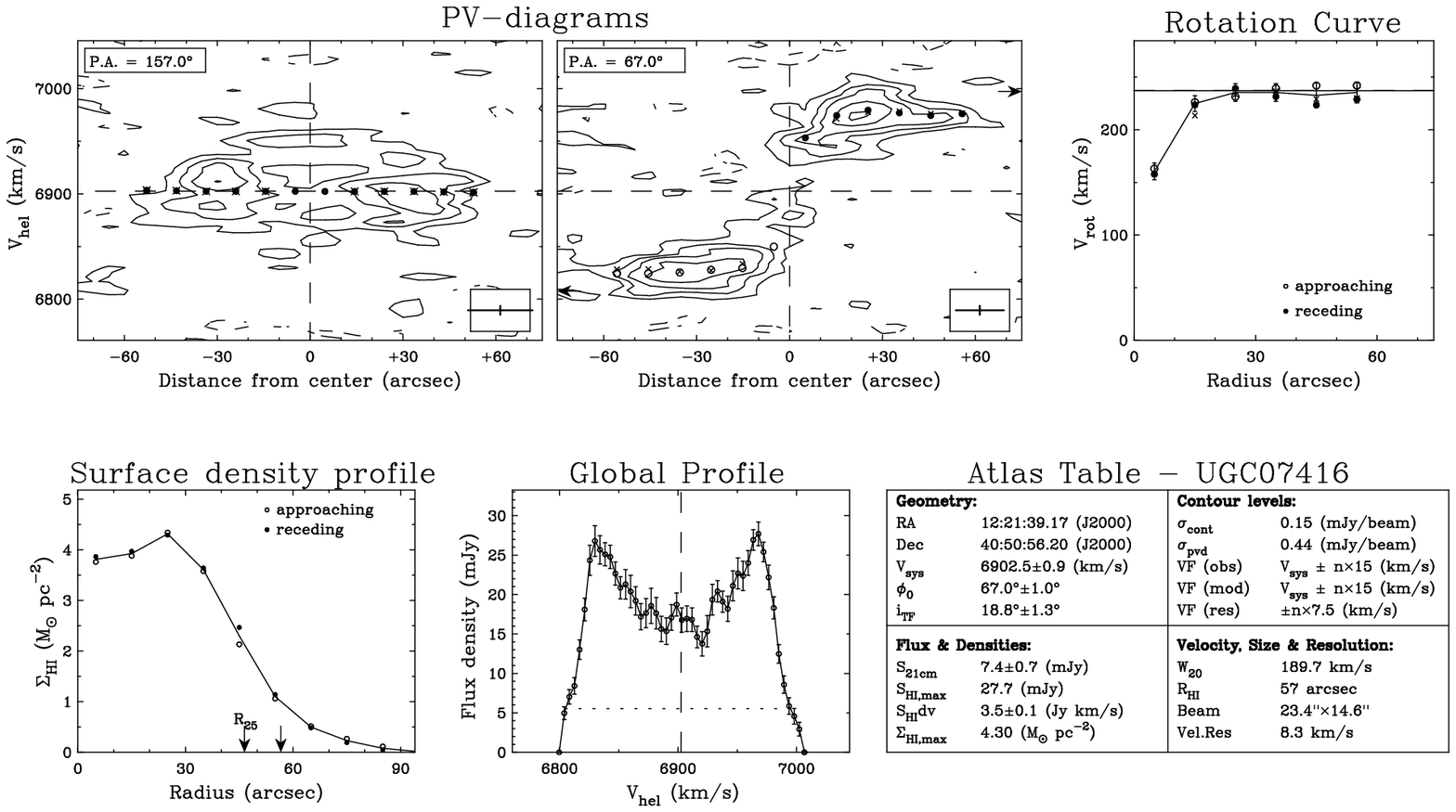}
 \end{figure}

 \begin{figure}
 \centering
 \includegraphics[width=1.0\textwidth]{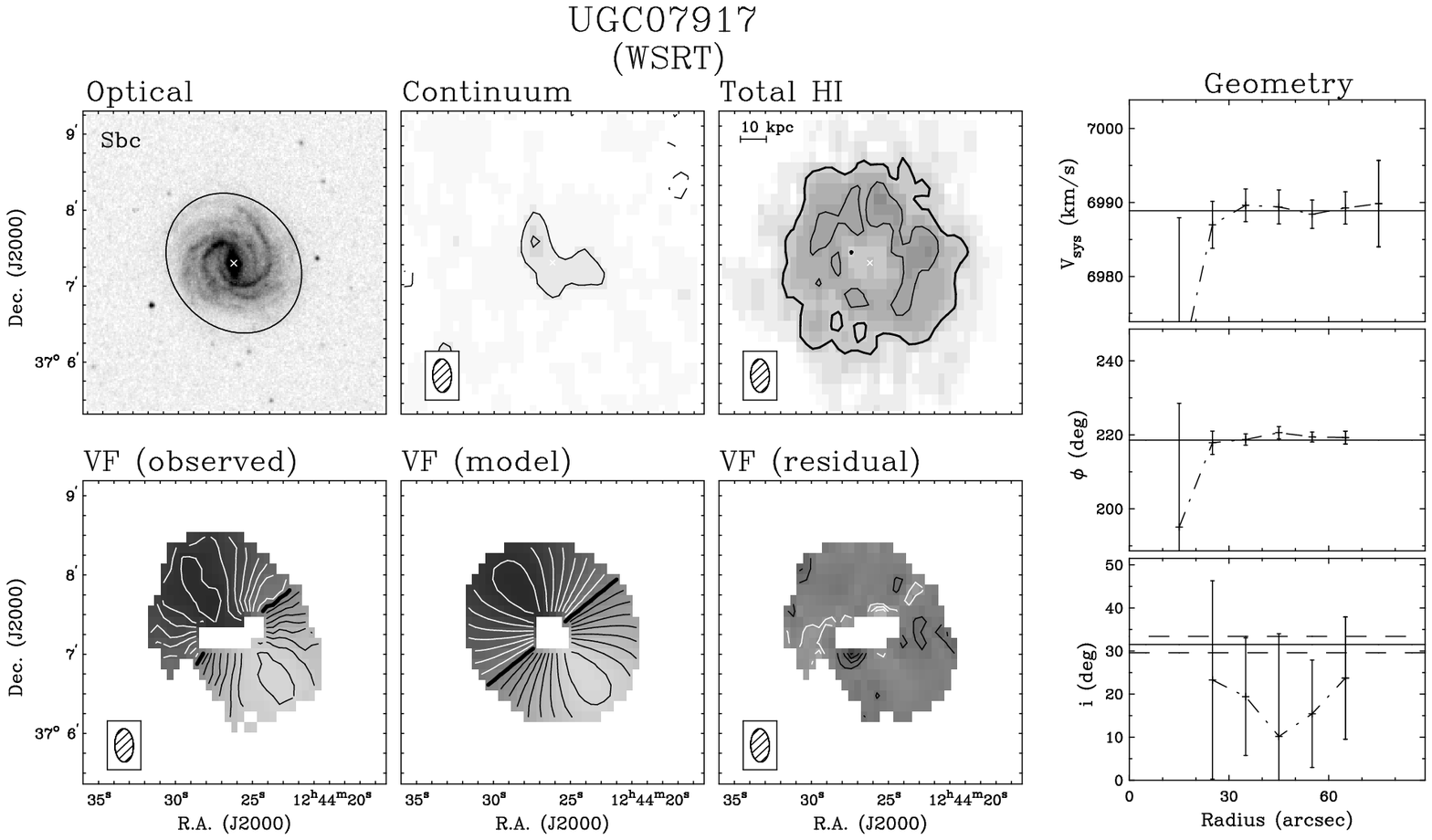}
 \end{figure}

 \begin{figure}
 \centering
 \includegraphics[width=1.0\textwidth]{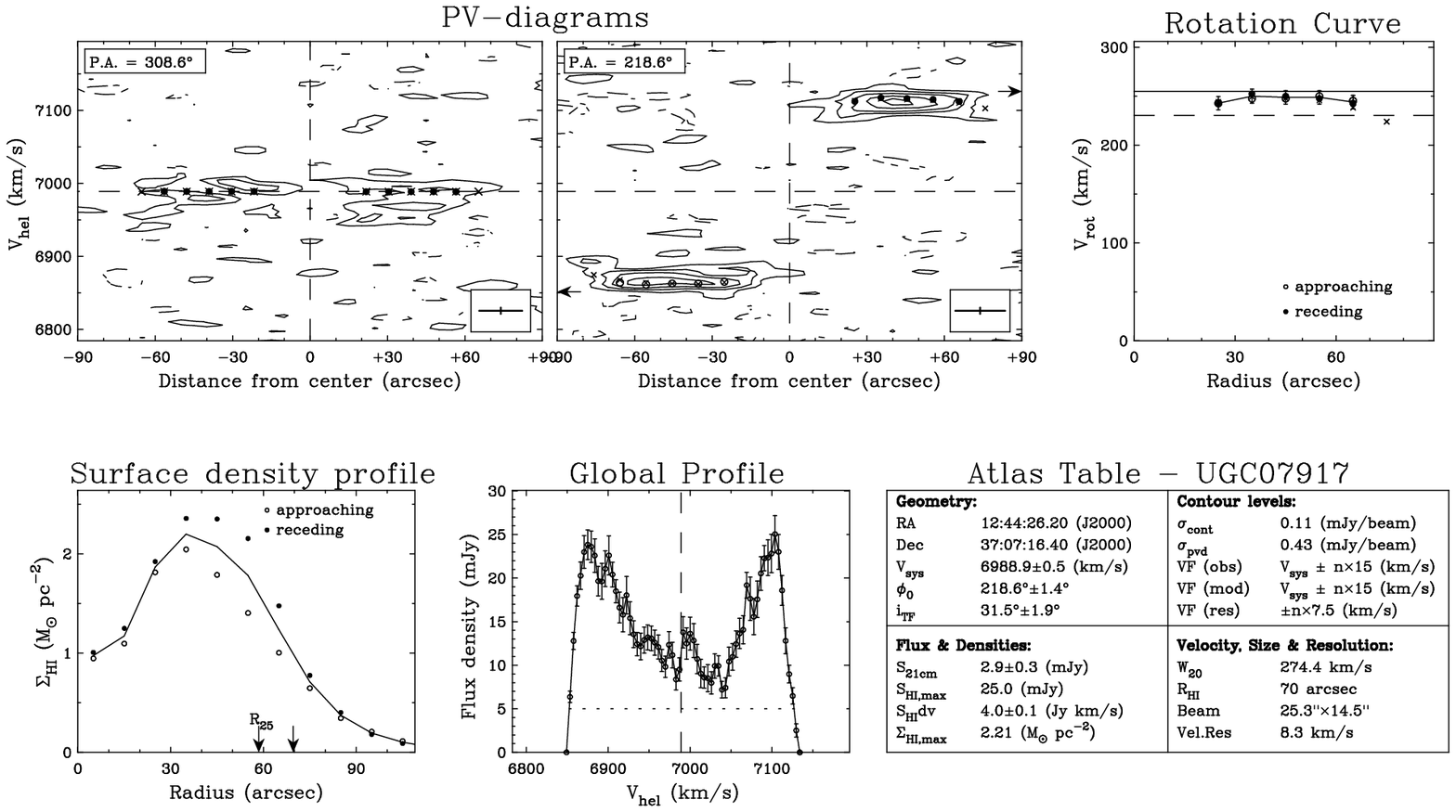}
 \end{figure}

 \begin{figure}
 \centering
 \includegraphics[width=1.0\textwidth]{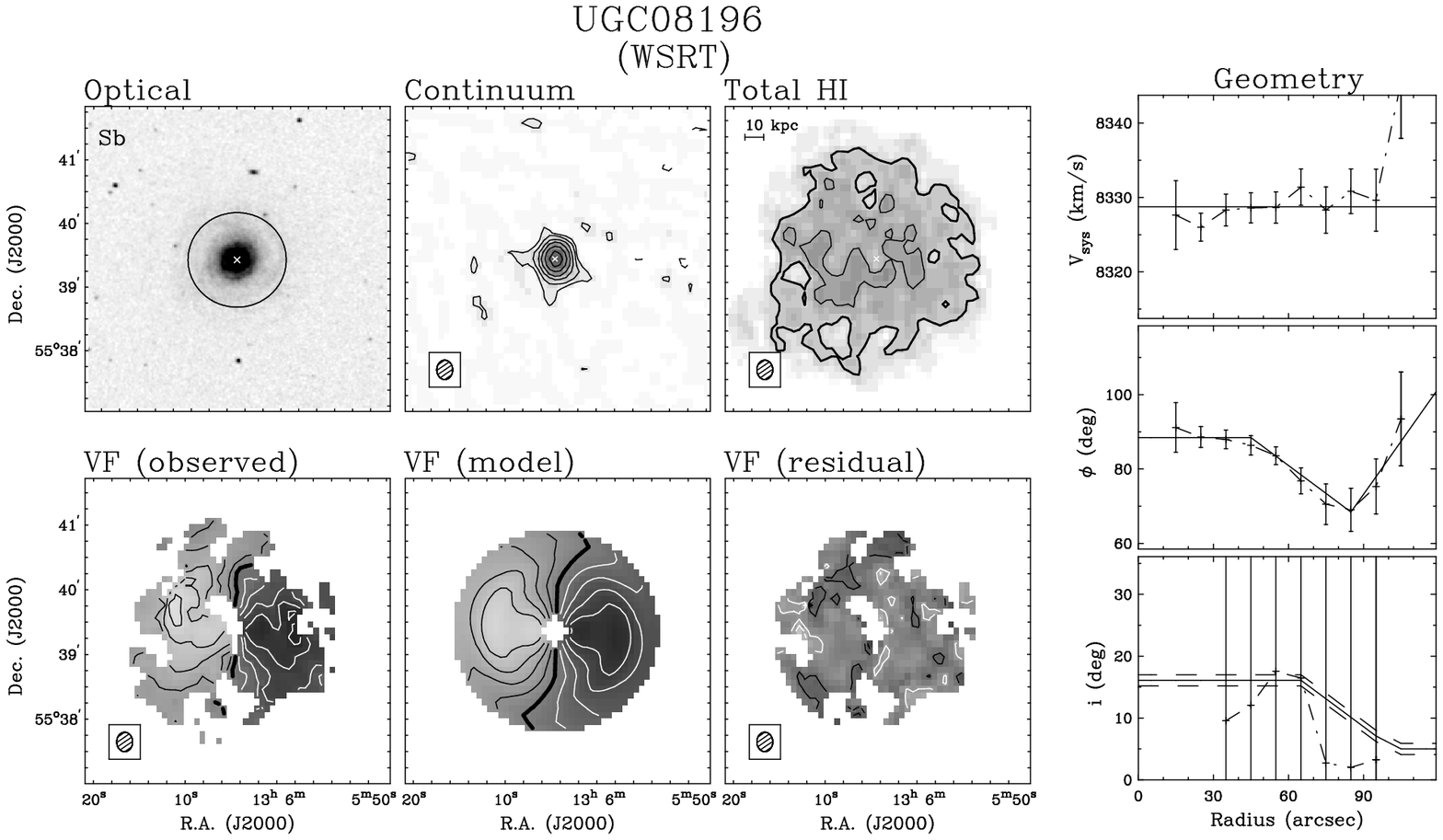}
 \end{figure}

 \begin{figure}
 \centering
 \includegraphics[width=1.0\textwidth]{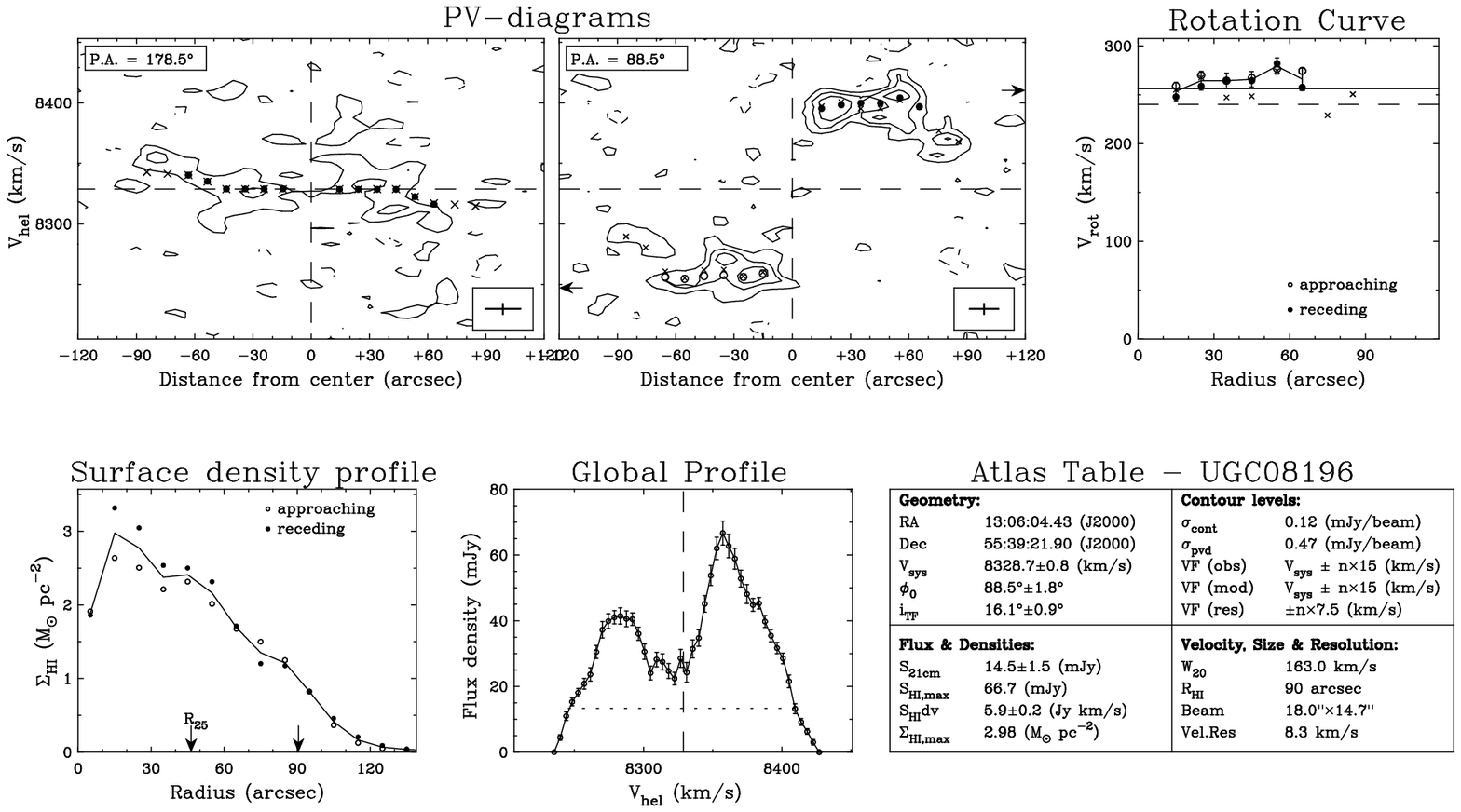}
 \end{figure}

 \begin{figure}
 \centering
 \includegraphics[width=1.0\textwidth]{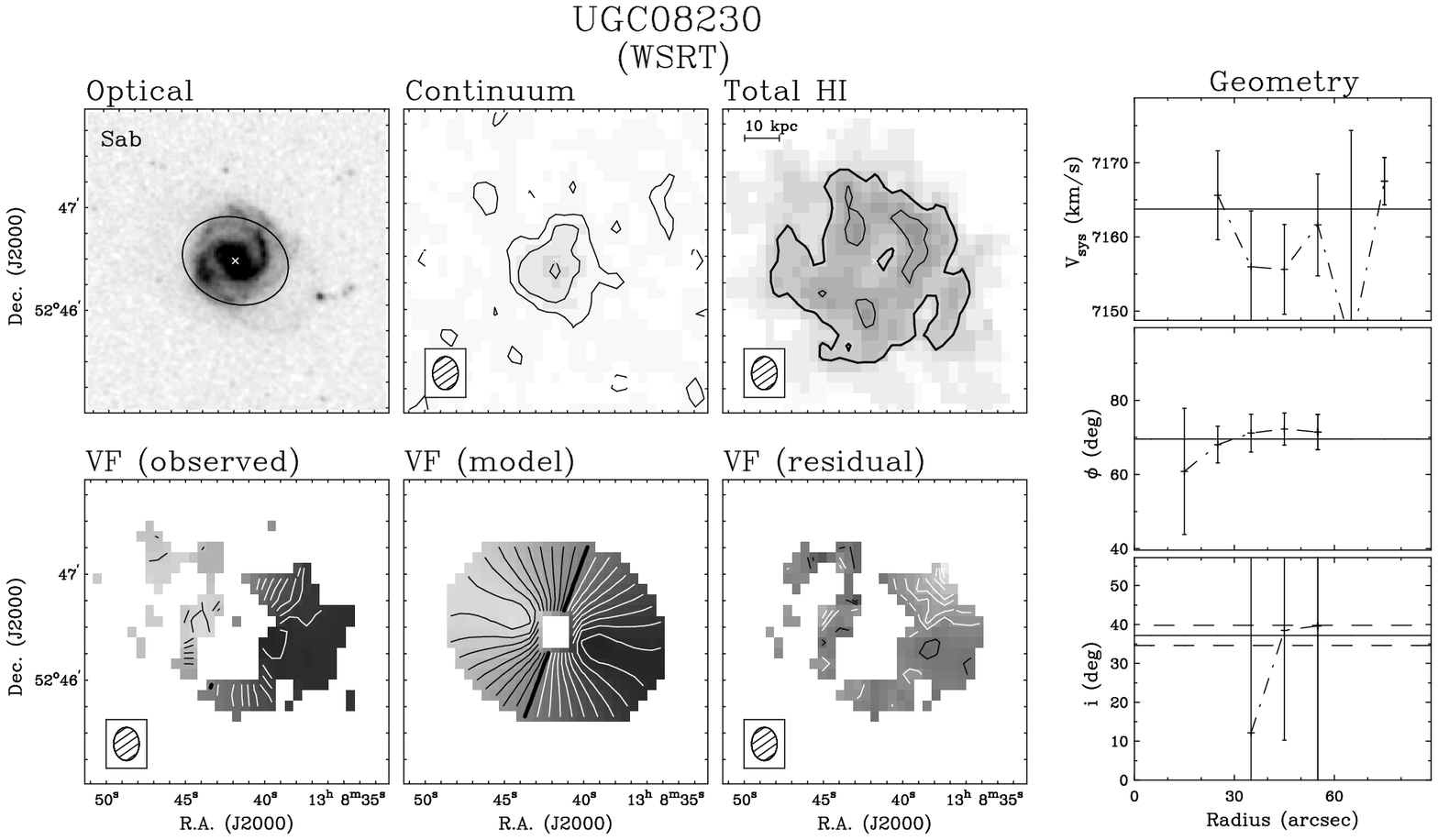}
 \end{figure}

 \begin{figure}
 \centering
 \includegraphics[width=1.0\textwidth]{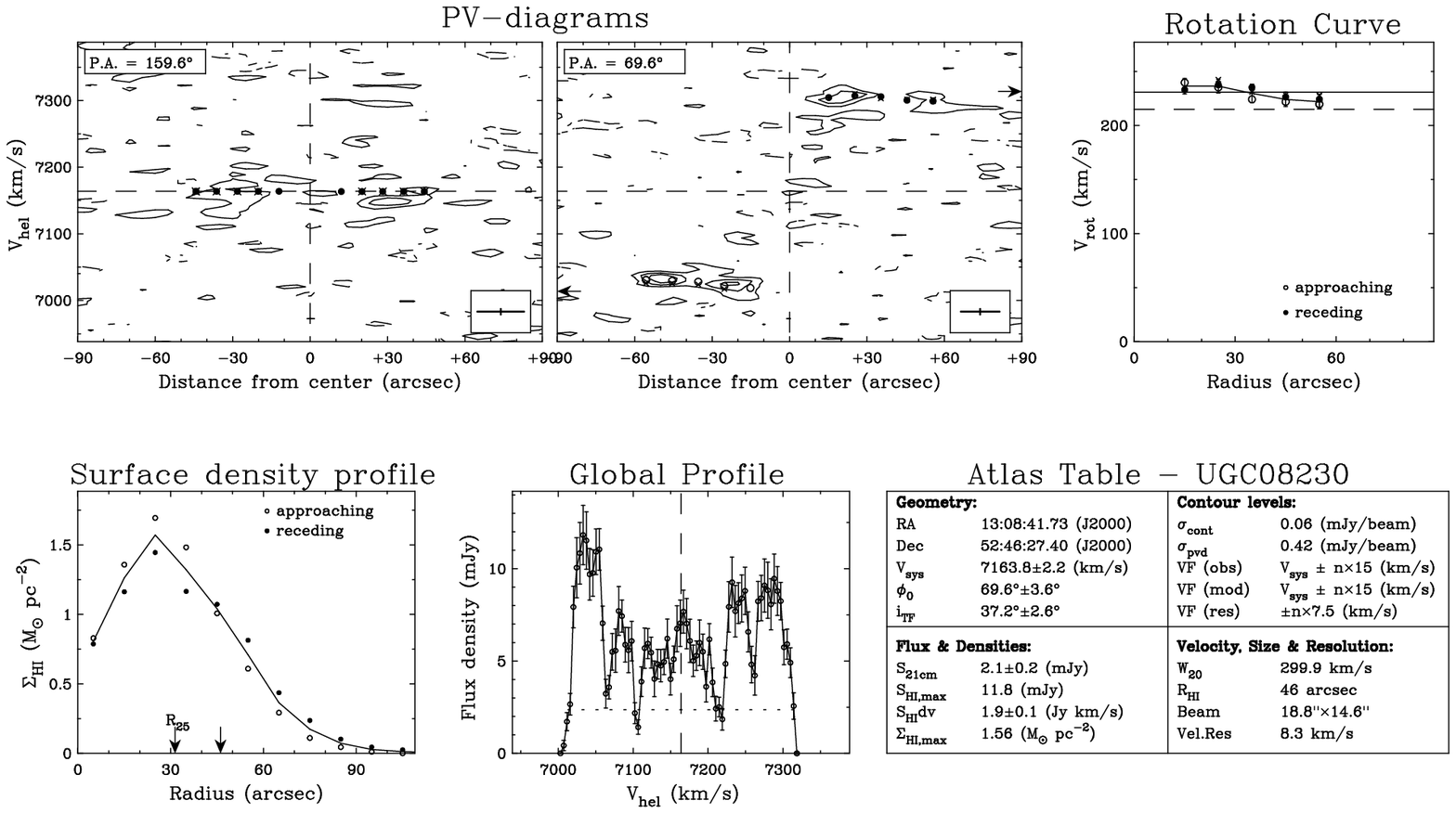}
 \end{figure}

 \begin{figure}
 \centering
 \includegraphics[width=1.0\textwidth]{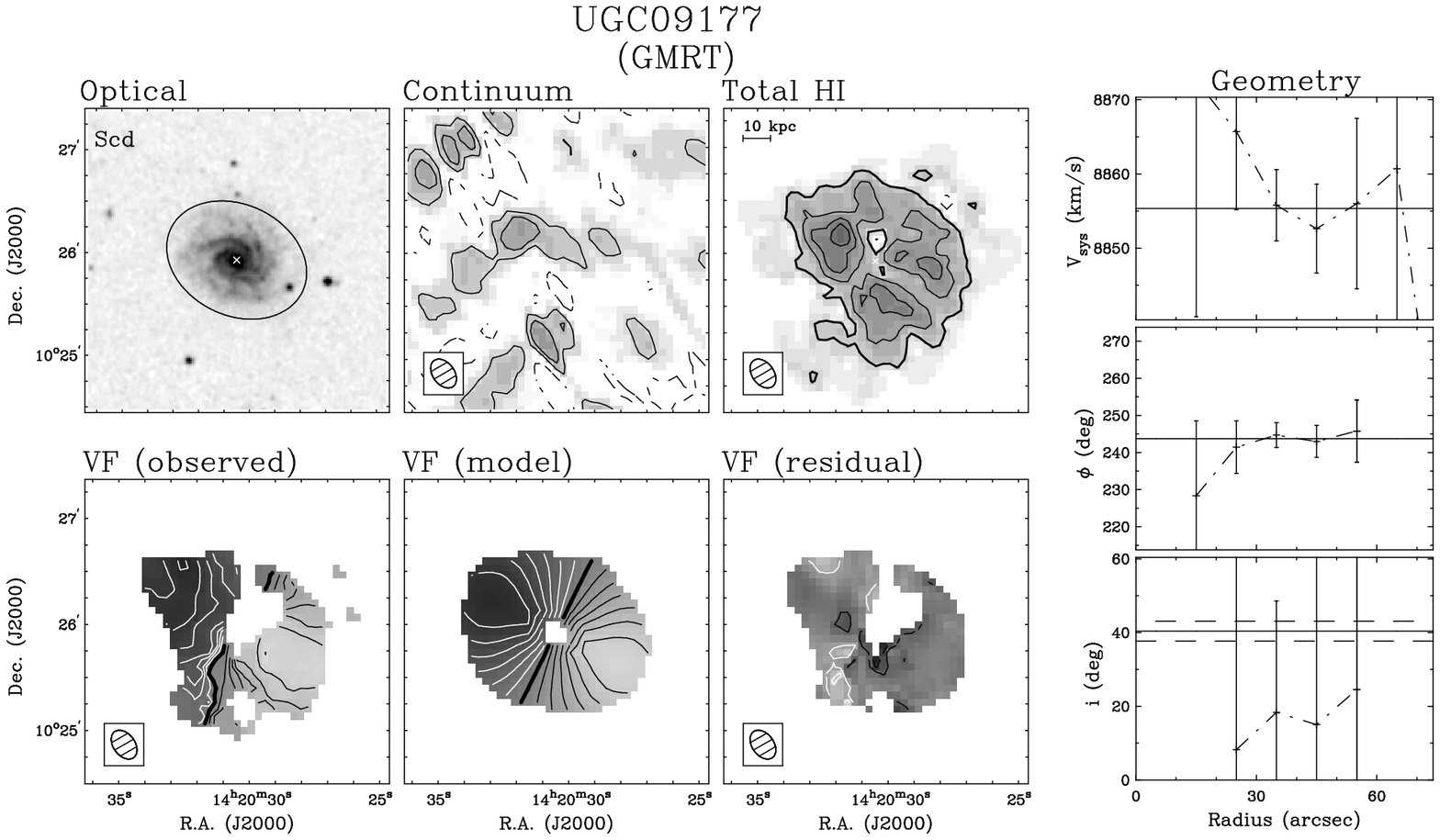}
 \end{figure}

 \begin{figure}
 \centering
 \includegraphics[width=1.0\textwidth]{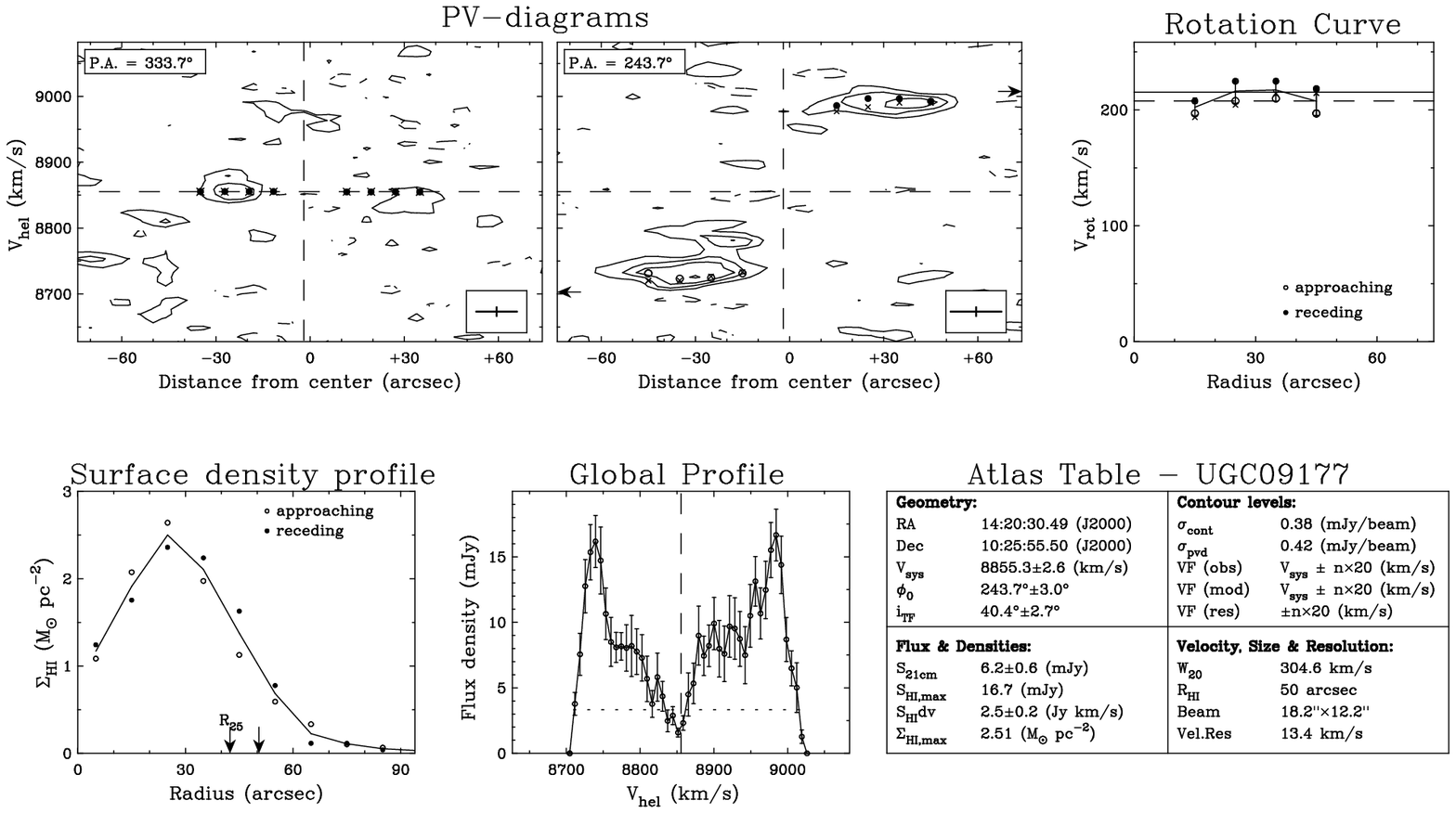}
 \end{figure}

 \begin{figure}
 \centering
 \includegraphics[width=1.0\textwidth]{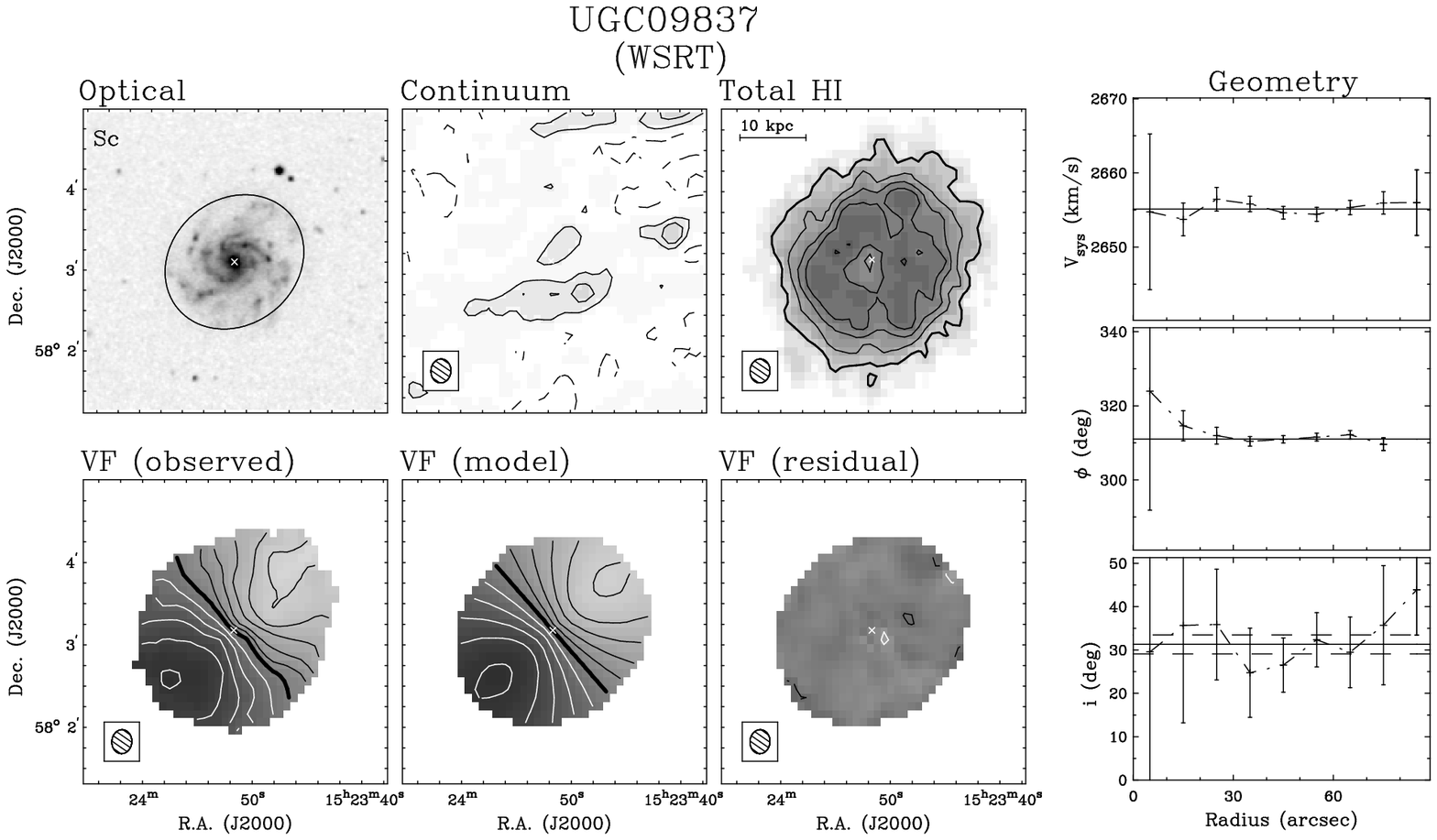}
 \end{figure}

 \begin{figure}
 \centering
 \includegraphics[width=1.0\textwidth]{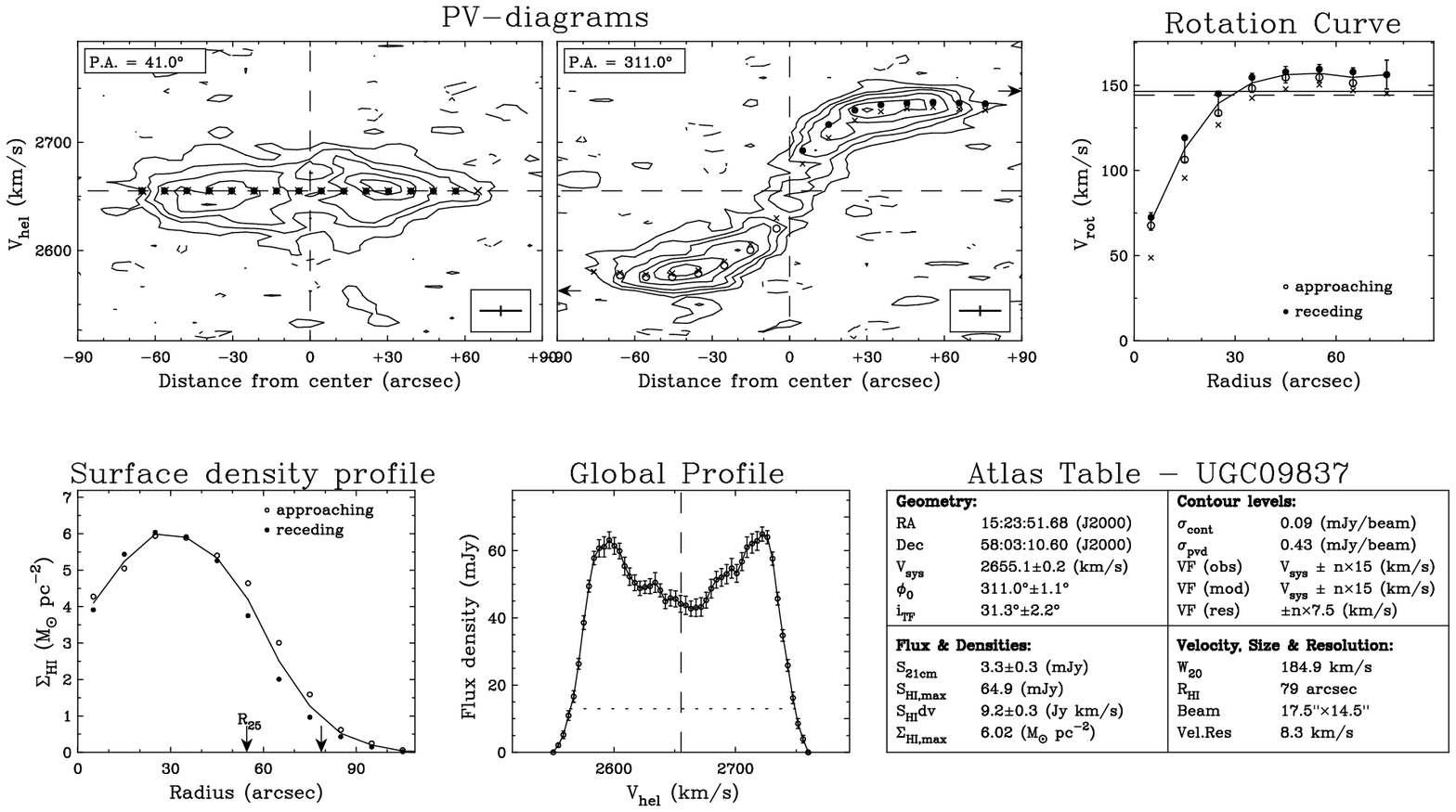}
 \end{figure}

 \begin{figure}
 \centering
 \includegraphics[width=1.0\textwidth]{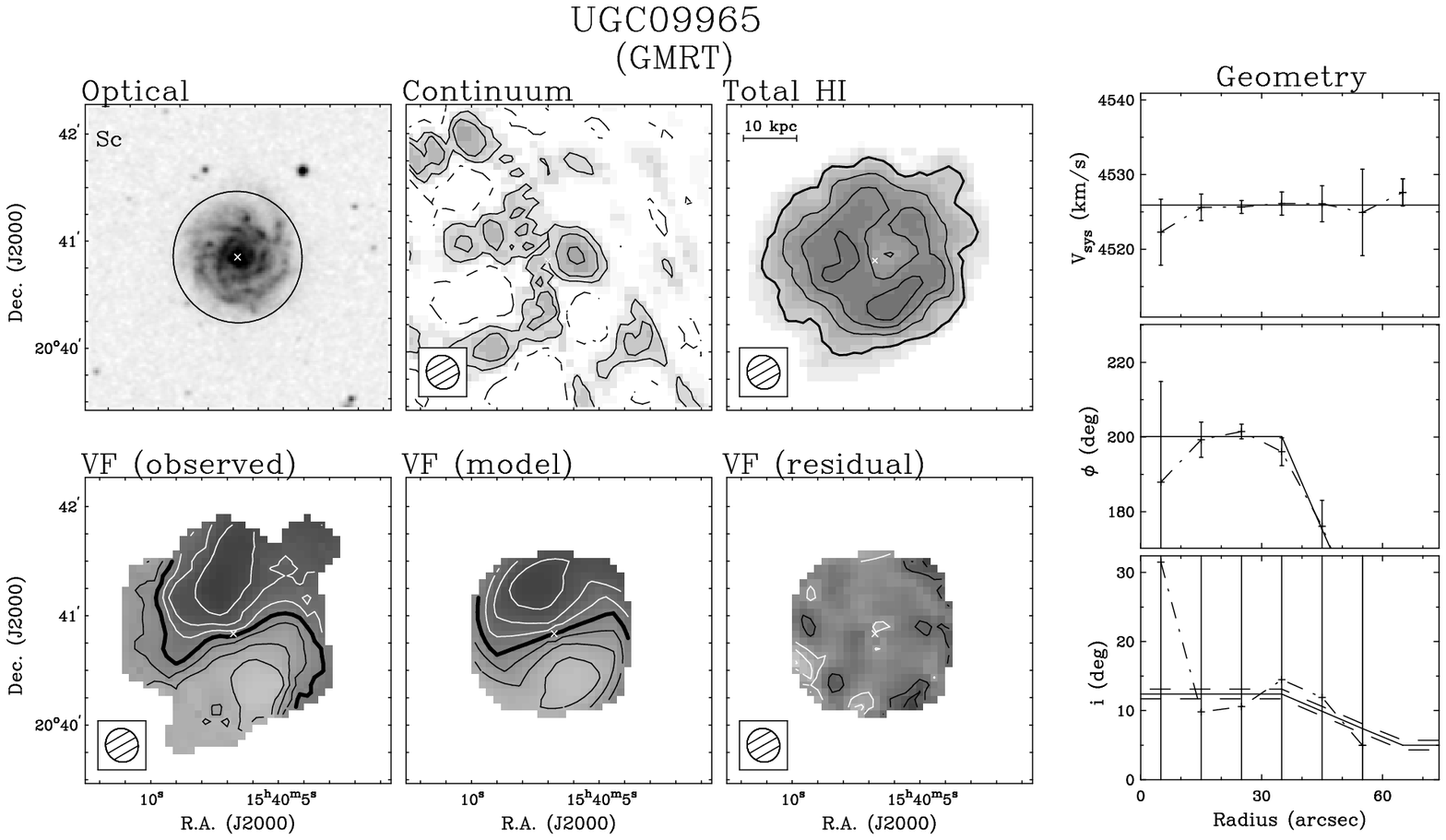}
 \end{figure}

 \begin{figure}
 \centering
 \includegraphics[width=1.0\textwidth]{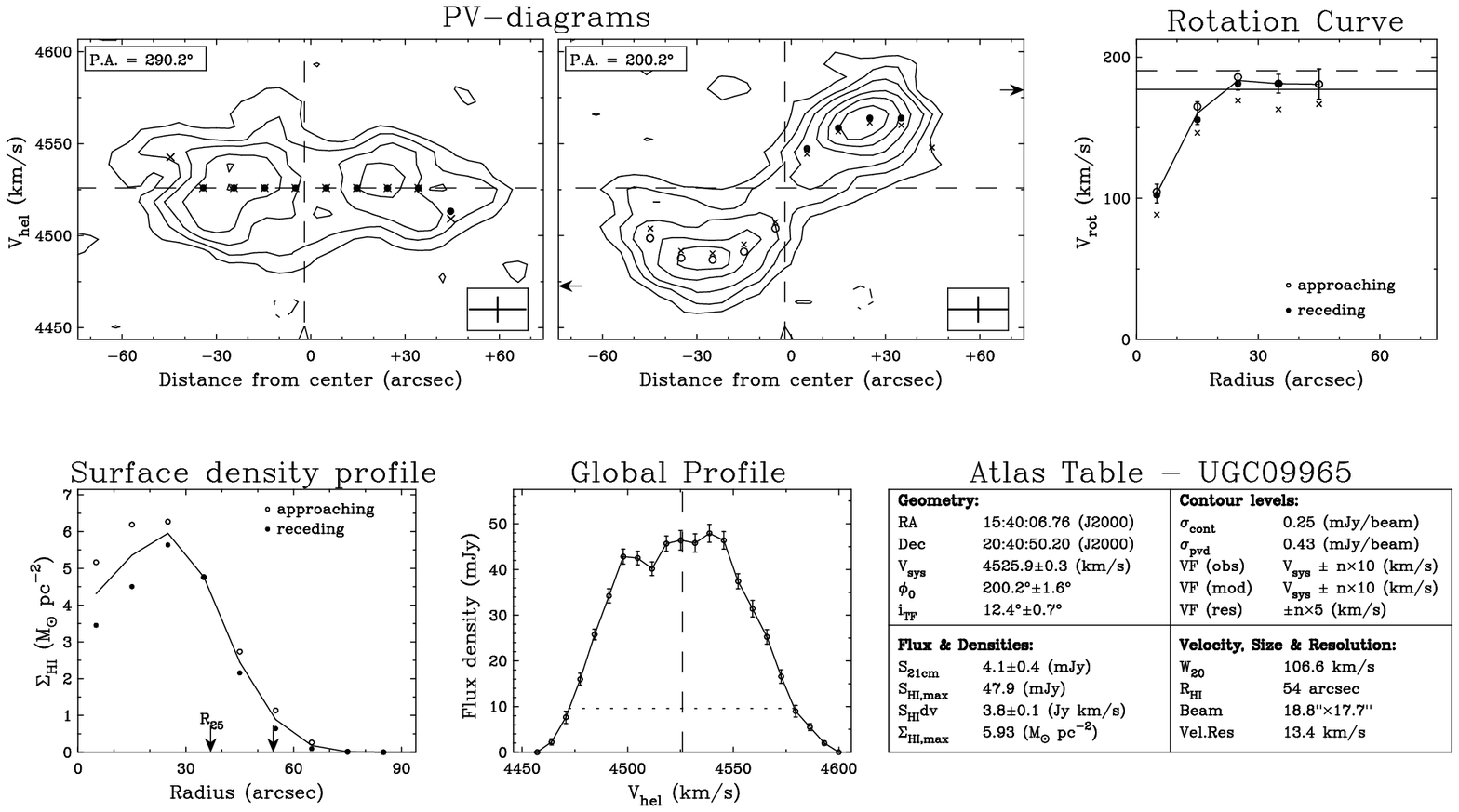}
 \end{figure}

 \begin{figure}
 \centering
 \includegraphics[width=1.0\textwidth]{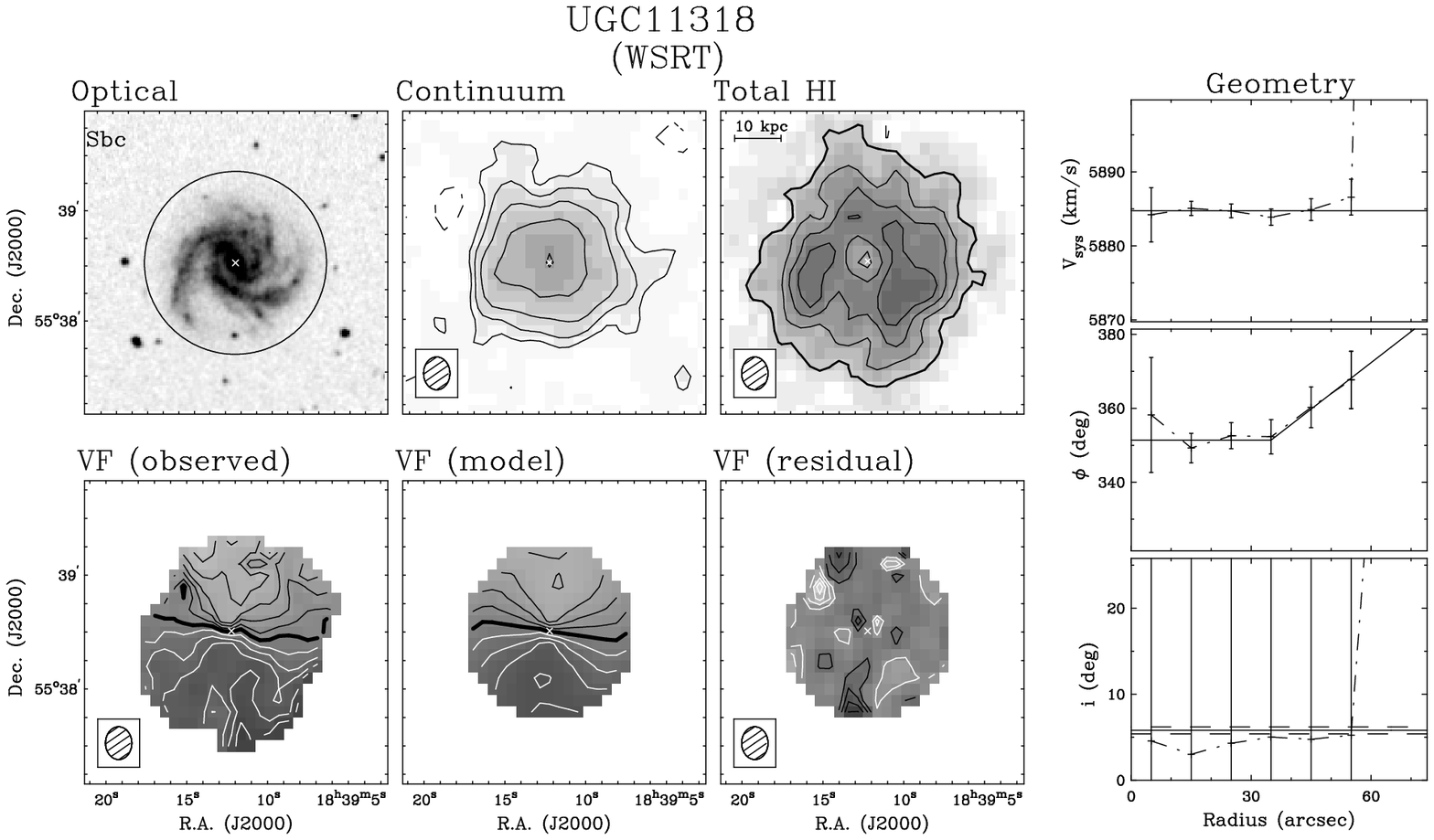}
 \end{figure}

 \begin{figure}
 \centering
 \includegraphics[width=1.0\textwidth]{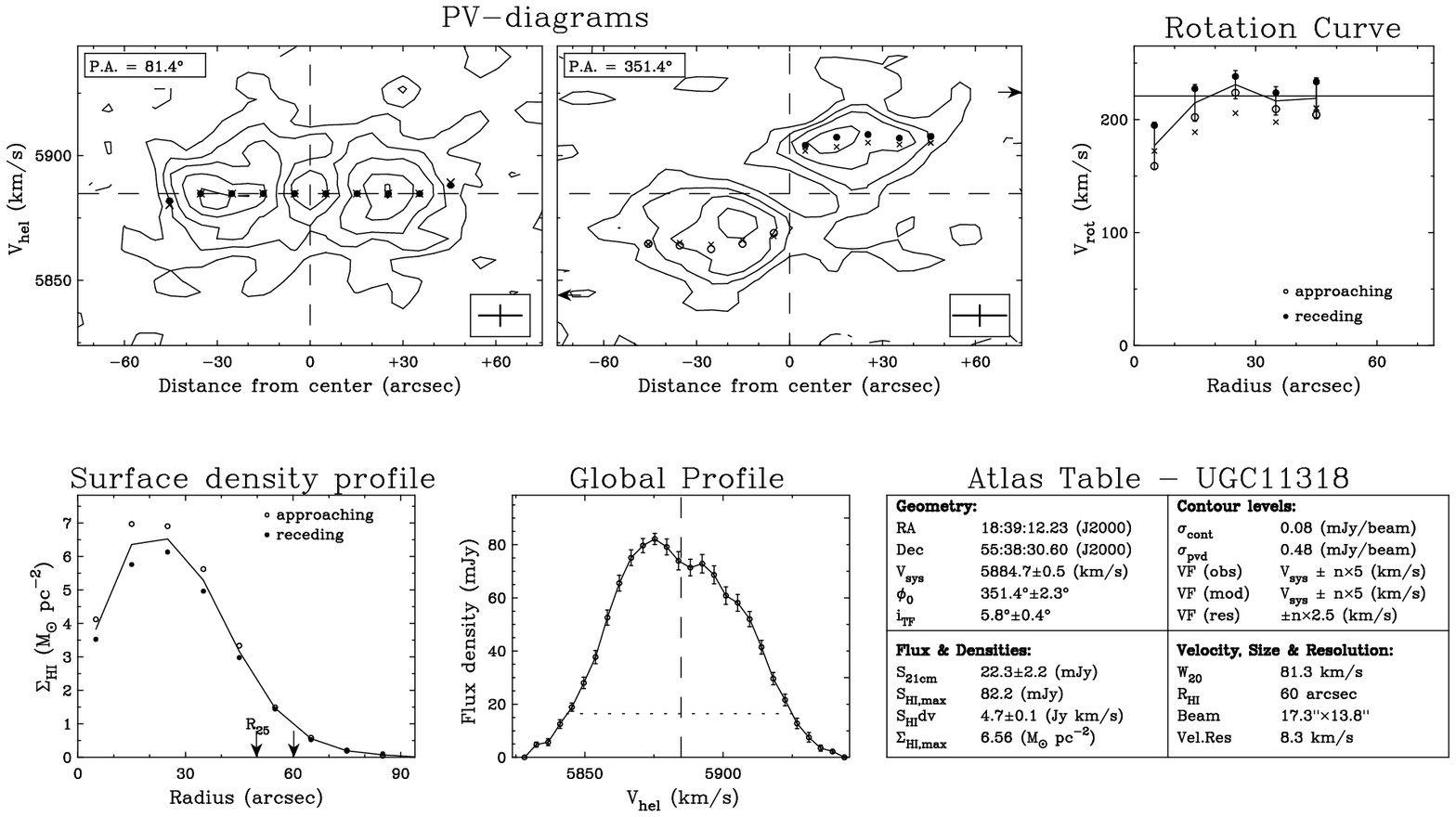}
 \end{figure}

% ========================================================================================

%%%%%%%%%%%%%%%%%%%%%%%%%%%%%%%%%%%%%%%%%%%%%%%%%%%%%%%%%%%%%%%%%%%%%%%%
\end{document}